\documentclass[prx,twocolumn,showpacs,preprintnumbers,amsmath,amssymb]{revtex4-1}

\usepackage{graphicx}
\usepackage{xcolor}

\begin{document}
\title{ Two classes of organization principle: quantum/topological  phase transitions meet complete/in-complete devil staircases
and their experimental realizations }
% in strongly interacting spin-orbit coupled bosons in a square lattice }
%\title{ Unconventional magnetic  phases, novel phase transitions due to in-commensurate magnons,
% In-complete or complete devil staircases and Luttinger liquid Cantor set of
% strongly interacting spin-orbit coupled bosons in a square lattice }
%{\sl   Substantially revised version 2 of arXiv:1603.00451  }
%\title{  Quantum Rotated Ferromagnetic Heisenberg model with generic spin-orbital couplings }
%\title{  Rotated Heisenberg model in a Zeeman field and its applications to cold atoms and materials with spin orbit couplings }
\author{ Fadi Sun$^{1,2}$ and Jinwu Ye $^{1,2}$   }
\affiliation{
%$^{1}$ Department of Physics, Capital Normal University,
%Key Laboratory of Terahertz Optoelectronics, Ministry of Education, and Beijing Advanced innovation Center for Imaging Technology,
%Beijing, 100048, China   \\
$^{1}$Department of Physics and Astronomy, Mississippi State University, MS, 39762, USA \\
$^{2}$ Kavli Institute of Theoretical Physics, University of California, Santa Barbara, Santa Barbara, CA 93106  }
\date{\today }
\begin{abstract}
There exists many quantum or topological phases in Nature. One well known  organization principle
is through various quantum or topological phases transitions between or among these phases.
Another is through either complete or in-complete devil staircases in their quantized forms.
Here, we show that both classes of  organization principle appear in an experimentally accessible system:
strongly interacting spinor bosons  subject to any of the linear combinations of the Rashba and Dresselhaus spin-orbit coupling
(SOC) in the space of
the two SOC parameters $ ( \alpha, \beta) $ in a square lattice.
In the strong coupling limit, it leads to a new quantum spin model called Rotated Ferromagnetic Heisenberg model (RFHM).
%which consists of an anti-ferromagnetic Heisenberg term,
%a ferromagnetic Kitaev term
%and a Dzyaloshinskii-Moriya (DM) term.
The RFHM leads to rich and unconventional magnetic phases even in a bipartite lattice.
They include collinear spin-bond correlated magnetic  Y-x phase,
a non-coplanar $ 3 \times 3 $ Skyrmion crystal phase (SkX), a gapped in-commensurate (IC) non-coplanar IC-SkX-y
phase which reduces to a co-planar IC-XY-y phase when $ \alpha=\beta $,
gapped spiral co-planar commensurate (C) near $ \alpha=\frac{\pi}{N} n, N \geq 3, n \geq 1 $ and gapless IC-YZ-x phases which are
melt into quasi-1d Luttinger liquids (LQ) even at zero temperature by its anisotropic gapless phason modes, so named IC-YZ-x/LQx.
All these phases  are organized by the two different classes of organization principles:  quantum phase transitions or incomplete (when $ \alpha \neq \beta $) and complete  (when $ \alpha=\beta $) devil staircases displaying a fractal structure.
For the first class, we identify a spurious $ U(1) $ symmetry and investigate the order from quantum disorder (OFQD)
phenomenon along the diagonal line slightly away from the Abelian point $ \alpha=\beta=\pi/2 $.
We develop a systematic spin coherent path integral approach to evaluate not only the gap generated, but also the whole spectrum corrected from the OFQD mechanism.  We construct effective low energy actions to describe the 2nd order quantum Lifshitz transition from the Y-x phase to the IC-XY-y phase along the diagonal line. By identifying suitable low energy modes,
we  derive the low energy effective actions
corresponding to C- or IC-magnons away from the diagonal line inside the Y-x phase which lead to quite different
spin-spin correlation functions for the two cases respectively.
We also study the 1st order quantum Lifshitz transition from the Y-x phase to the IC-SkX-y phase from the right and determine the spin-orbital structure of the IC-SkX-y phase.
%The IC-XY-y phase display very similar properties as those observed in some 4d or 5d Kitaev materials.
For the second class, we introduce the topological rational and irrational winding numbers $ W $ to characterize the incomplete or complete devil staircases and also perform their quantizations.  This new organization pattern is beyond any known classification schemes on
quantum/topological phases such as SPT and SET or quantum chaos.
%The $ W $ versus $ \alpha=\beta $ along the diagonal line matches the original simplest Cantor function very well.
The IC-YZ-x/LQx phases form a Cantor set with a fractal dimension along the complete devil staircase. They also take most of measures in the incomplete devil staircases when $ \beta \ll \alpha $.
%1d Luttinger liquids along the complete devil staircase forming a Cantor set with a fractal dimension, etc.
Quantum chaos and quantum information scramblings along the diagonal line $ \alpha=\beta $ are discussed.
%are localized in $ x-$ dimension,
%but free to move in the $ y-$ direction.
%They also display broad distribution of Bragg peaks in $ k_x $ space, so very much resemble the quantum spin liquids in experimental signatures.
%Connections to the classical Frenkel-Kontorowa (FK) model are explored.
%The stripe co-planar ( spiral ) C phases are confined phases with discrete Bragg peaks and gapped magnon excitations, while the IC phases are %deconfined ones with broad distribution of Bragg peaks and gapless phason excitations.
Some possible connections with the topological states in a 1d quasi-crystal, 2d quantum dimer models or
2d deconfined quantum critical point, 2d fractional quantum Hall plateau-plateau transitions and
the 3d cubic code are explored.
Implications on un-conventional magnetic ordered phases detected in the 4d- or 5d-orbital strongly correlated materials with SOC and in the current or near future cold atom systems are presented.
\end{abstract}

\maketitle
\section{ Introduction}

%     There are many previous people's work on
%     the conventional SOC $ \vec{L} \cdot \vec{S} $ which keeps both the Time-reversal and
%     the parity, so is a {\bf scalar} \cite{SLrev1,SLrev2,SLrev3}.

 It was well known that the strong correlations among bosons or fermions lead to many quantum or topological phases
 and phase transitions in materials \cite{scaling,sachdev,aue,wen0,kane,zhang,tenfold,wen}.
 Its combinations with geometric frustrations may lead to new phases of matter such as coplanar spiral magnetic phases, especially topological
 quantum spin liquids (QSL) \cite{frusrev,SLrev1,SLrev2,SLrev3}.
 Its combinations with quenched disorders also lead to new states of matter such as the quantum spin glass or gapless quantum spin liquids
 such as those  in the Sachdev-Ye-Kitaev model (SYK) \cite{SY,kittalk}
 which are closely related to quantum chaos in the black holes through $ AdS/CFT $ correspondences \cite{syk1,syk2,syk3}.
 On the other forefront, Rashba or Dresselhaus spin-orbit coupling (SOC) is ubiquitous in various 2d or layered
 non-centrosymmetric insulators, semi-conductor systems, metals and superconductors \cite{rashba,ahe,socsemi,ahe2,she,niu,aherev,sherev}.
 There are also recent remarkable experimental advances in generating any linear combinations of the 2d Rashba and Dresselhaus SOC for
 both fermions and spinor bosons in both continuum and optical lattices \cite{expk40,expk40zeeman,2dsocbec,clock,clock1,clock2,SDRb,ben}.
 New many body phenomena due to the interplay among strong interactions,
 the SOC and lattice geometries  are being investigated in the current cold atom experiments.
 The 2d Rashba or 3d Weyl SOC $ \vec{k} \cdot \vec{S} $ keeps the Time-reversal, but breaks the parity, so it is a {\bf pseudo-scalar}.
 It is well known that it is this type of SOC which  is responsible for the parity violation in the weak interaction.
%     So our results in Part III may also be useful to lattice QCD calculations with a parity violation.
 In view of its broad impacts in materials, cold atoms and particle physics,
 it becomes urgent, topical and important to investigate what would be the new quantum or topological phenomena due to the
 the interplay between the strong correlations and the ubiquitous Rashba or Dresselhaus SOC on various lattices.

%  My 1999 work with  A. Millis on anomalous Hall effects on CMR materials is due to this type of SOC.

In this work, we address this outstanding problem and discover that the interplay leads to  many novel quantum or
topological phenomena, especially new organization principle of these phases  summarized in the global quantum phase diagram Fig.1
and Fig.\ref{trifeats}.
We establish the Fig.1 and Fig.\ref{trifeats} by the combinations of the approaches from the three directions.

 (1) Extremely anisotropic limit
(solvable line) \cite{rh} $ (\alpha=\pi/2, \beta ) $.  The collinear spin-bond correlated  Y-x phase is the exact
ground state along the line, but becomes just the classical ground state subject to quantum fluctuations away from the line
in the lobe labeled as the Y-x state in Fig.1.
We work out its excitation spectrum and the putative magnon condensation boundaries.
By using the combination of canonical quantization and cohere spin path integral method in the polar coordinate $ ( \eta, \xi) $ of the spin quantization axis along the $ X $ direction, then carefully identifying the low energy mode in the Y-x phase,
we derive the low energy effective actions inside the Y-x phase corresponding to both C- and IC- magnons.
Using these actions, we compute the leading spin-spin correlation functions (SSCFs) corresponding to the two cases respectively
and show they take very different forms.
We push the effective actions to the putative magnon condensation boundary and beyond to study the transition from the Y-x phase to
its neighbouring phases. We find it split into three segments:
the top and the bottom part are pre-emptied by a first order transition and in-complete devil staircase respectively,
%The Y-x state becomes just a metastable state between the first order transition line and the 2nd order magnon condensation boundary.
while the middle part becomes a weakly 1st  order quantum Lifshitz transition from
the Y-x state to a state which is In-commensurate  along $ y $ axis,  non-coplanar  with non-zero Skyrmion density named as IC-SkX-y phase.
We also determine its spin-orbital structure which reduces to the coplanar IC-XY-y phase along the diagonal line
achieved from the approach (2) below.
We contrast this gapped IC-SkX-y phase due to the magnon condensation tuned by the SOC with the gapless  IC-SkX-$\phi$ phase due to the magnon condensation tuned by a longitudinal Zeeman field studied previously in \cite{rhh}.

(2) The isotropic Rashba limit  $ 0 <\alpha=\beta < \pi/2 $.
This line split into two regimes: the quantum phase transition (QPT) regime when $ \alpha_{33} < \alpha < \pi/2  $
and the complete devil staircase regime $ 0 < \alpha < \alpha^{-}_{33} $
separated by the commensurate non-coplanar $ 3 \times 3 $ SkX (hub) phase at $ \alpha^{-}_{33} < \alpha < \alpha_{33} $. In the first regime
 $ \alpha_{33} < \alpha < \pi/2  $, there is a spurious $ U(1) $ symmetry
which leads to a classically infinitely degenerate family of states. By an order from quantum disorder (OFQD) analysis,
we determine the quantum ground
state to be the Y-x state or X-y state which is related to each other by the exact $ [C_4 \times C_4]_D $ symmetry
of the Hamiltonian.
The breaking of the spurious $ U(1) $ symmetry by the Y-x state leads to a spurious Goldstone mode which is nothing but the $ C_{\pi} $ magnons in the Y-x phase from the right studied by the approach (1) above. By choosing the polar coordinate $ (\theta, \phi) $
in the spin quantization axis along the $ Z $ direction,
we construct a spin coherent path integral approach to evaluate the gap generated by the OFQD phenomenon which transfers the spurious Goldstone mode to a gapped pseudo-Goldstone mode.
Then we go further to develop a systematic spin coherent path integral approach to evaluate not only the gap,
but also the whole spectrum corrected from the OFQD mechanism. We construct an effective action to
describe the quantum Lifshitz C-IC transition  from the Y-x  phase to a state which is In-commensurate along $ y $ axis, coplanar  named
as IC-XY-y phase, which is found to be nothing but the planar limit of the non-coplanar IC-SkX-y phase discovered from the approach (1) above.
In the second regime $ 0 < \alpha < \alpha^{-}_{33} $,
we find a $ 3 \times 3 $  non-coplanar Skyrmion crystal ( SkX ) phase, then
successive  principle gapped spiral co-planar Commensurate phases $ N \times 1, N > 4 $ near $ \alpha=\pi/N $ taking most of measures,
then some higher order C-phases near $ \alpha= \frac{\pi}{N} n, n > 1 $ takings small measures, the
spiral gapless IC-YZ-x phases at irrational $ \alpha $ taking zero measures and  forming a Cantor set with a non-integer fractal dimension.
The IC-YZ-x phases support gapless phason mode which, in turn, melts these phases into quasi-1d Luttiger liquid (LQx)
which still break translational symmetry along x-axis. We call these phases as IC-YZ-x/LQx phases.
We introduce the topological rational and irrational winding numbers $ W $ to characterize all these phases
and show that it is topologically equivalent to the original simplest Cantor function in the second regime, but ill defined in the QPT regime.
We investigate the zero and finite temperature phase transitions, especially quantum chaos and information scramblings in both regimes along the diagonal line.

(3) Near the Abelian line $ 0 < \alpha < \pi/2, \beta=0 $, by mapping the system into a classical 1d Frenkel-Kontorowa (FK) model,
 we also find the principle near $ \alpha=\pi/N $, higher order near $ \alpha= \frac{\pi}{N} n, n > 1 $ Commensurate phases and
 the gapless IC-YZ-x/LQs phases. At the classical level, the spiral IC-YZ-x/LQx phases have a broad distribution of  Bragg peaks.
However, the quantum fluctuations due to the gapless phason excitations transfer them into quasi-1d Luttinger liquids.
 They form incomplete devil staircases near the  Abelian line.  We show that the spiral co-planar C phases have gapped magnon excitations and discrete multiple Bragg peaks.
 The topological winding numbers $ W $ can also be used to distinguish the principle spiral co-planar C- phases  near $ \alpha=\pi/N $
 from the higher order ones near $ \alpha= \frac{\pi}{N} n, n > 1 $, despite both have the same symmetry breaking patterns.
 Near the Abelian line $ \beta \ll \alpha $, immersed inside the spiral gapless IC-YZ-x/LQx phases are some small devil staircases with higher topological winding number near $ \alpha= \frac{\pi}{N} n $ displaying a fractal structure.
 When approaching to the diagonal line from $ \beta < \alpha $,
 the principle ones near $ \alpha= \frac{\pi}{N}  $ take more and more measures, the higher orders with $ n > 1 $
 continue to squeeze in, the IC-YZ-x/LQx phases taking smaller and smaller measures. Finally, they turn into a complete devil staircase
 along the diagonal line $ \alpha=\beta $ achieved from the approach (2) above.
 In the  appendix C, we derive the quantization of the 1d FK model which
 in principle, can be used to study all the quantum effects in the fractal structure.

The approaches from the three directions listed as (1) to (3) above
are complementary to each other.
The matching from the three different directions are good check on the consistency of the results achieved
and jointly lead to the global physical picture shown in Fig.1.

%We contrast the two in-commensurate phases IC-XY-y, the spiral IC-YZ-x/LQx and the IC-SkX phase in a longitudinal Zeeman field found in a %previous work \cite{ICSkXh}.
%Especially, the coplanar IC-XY-y phase maybe relevant to the phases already detected in some 4d or 5d Kitaev materials.
%While the co-planar IC phase may be relevant to the materials displaying ¡°Quantum spin liquid " phenomena.
%Some possible far-reaching perspectives are given.
The tight-binding Hamiltonian of (pseudo)-spin $ 1/2 $ bosons  (or fermions) at integer (or half) fillings  hopping in
a two-dimensional square optical lattice subject to any combination of Rashba and Dresselhaus SOC is:
\begin{equation}
	\mathcal{H}_{b/f}= -t\sum_{\langle ij\rangle}(b_{i\sigma}^\dagger U_{ij}^{\sigma\sigma'} b_{j\sigma'}+h.c.)
	+ \frac{U}{2} \sum_i ( n_{i}-N )^2
\label{hubbardint}
\end{equation}
where $ t $ is the hopping amplitude along the nearest neighbors $\langle ij\rangle$,
the non-Abelian gauge fields
$U_{i,i+\hat{x} } =e^{i \alpha \sigma_x}$,
$U_{i,i+\hat{y} }=e^{i \beta \sigma_y}$
are put on the two links in a square lattice.
$ \alpha=\pm \beta $ stands for the Rashba ( Dresselhaus ) case. $ \alpha \neq \beta $ corresponds to any linear combination
of the two.  $U>0$ is the Hubbard onsite interaction.

In the strong coupling limit $ U/t \gg 1 $, to the order $O(t^2/U)$,
we obtain the effective spin $ s=N/2 $ Rotated Heisenberg model:
\begin{equation}
	\mathcal{H}_{RH}  =  -J\sum_i
	[\mathbf{S}_i R(\hat{x},2\alpha)\mathbf{S}_{i+\hat{x}}
	+\mathbf{S}_i R(\hat{y},2\beta)\mathbf{S}_{i+\hat{y}}]
\label{rhgeneral}
\end{equation}
with $J= \pm 4t^2/U > 0 $ for bosons/fermions,
the $R(\hat{x},2\alpha)$, $R(\hat{y},2\beta)$ are the two SO(3) rotation matrices
around the $ X $ and $ Y $ spin axis by angle $2\alpha$, $2\beta$
putting on the two bonds along  $\hat{x} $, $\hat{y} $ respectively.
In this paper, we only focus on spinor bosons which lead to the Rotated ferromagnetic Heisenberg model (RFHM) \cite{rh}.
The fermions which lead to the Rotated ant-ferromagnetic Heisenberg model (RAFHM) \cite{rafhm} will be discussed in a separate publication.

%To understand quantum phenomena in the generic SOC parameter space of Eqn.\ref{rhgeneral}, we take a ``divide and conquer'' strategy.
%We first explore new and rich quantum phenomena along the
%solvable extremely anisotropic limit $\alpha=\pi/2, 0<\beta<\pi/2$.
%Then starting from the deep knowledge along the anisotropic  line,
%we will try to investigate the quantum phenomena at generic $(\alpha,\beta)$.
%The first step has been achieved in \cite{rh}, here, we will achieve the second step.
%This strategy has been successful in solving several strongly coupled models such as Kondo model \cite{kondo0,kondo1,kondo2}
%and quantum dimer model \cite{dimer1,dimer2}.

\begin{figure}[!htb]
	%\centering
\includegraphics[width=0.5\textwidth]{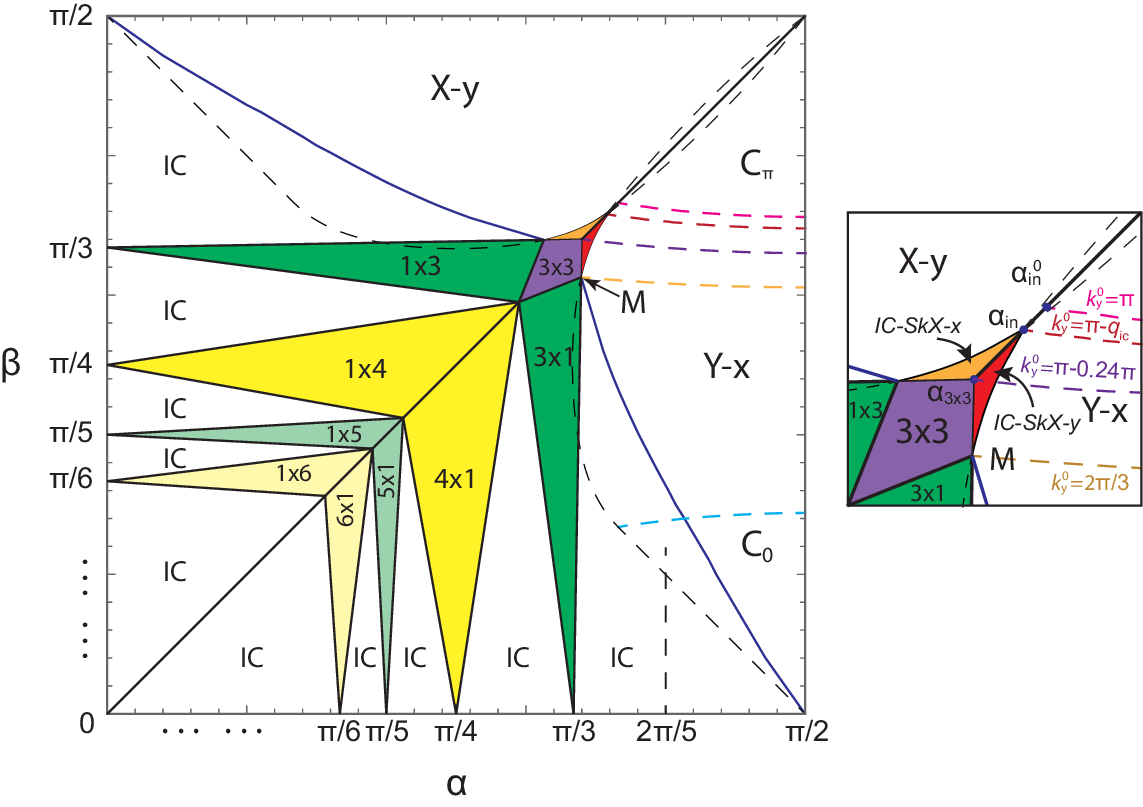}			
\caption{
  The phase diagram of the strongly interacting spinor bosons at a generic SOC $ ( \alpha, \beta) $ in a square lattice.
  The non-coplanar $ 3 \times 3 $ SkX (hub) phase is the only phase respecting the $ [C_4 \times C_4]_D $ symmetry.
 Along the diagonal line $ \alpha=\beta^{+} $, the organization pattern of phases change from
 the complete devil staircase characterized by the topological winding number $ W $ below the hub, to quantum phase transitions
 above the hub: two consecutive 2nd quantum Lifshitz transitions from the hub
 to the in-commensurate (IC-) co-planar IC-XY-y  at $ \alpha= \alpha_{33} $, then to collinear Y-x phase at $ \alpha=\alpha_{in} $.
 Except the hub phase, the diagonal line $ \alpha=\beta $ is a first order transition line between the
  phases along $ \alpha = \beta^{-} $ and those along $ \alpha = \beta^{+} $, both5.03.0 are related by the $ [C_4 \times C_4]_D $ symmetry.
% Along the diagonal line $  \alpha_{in} < \alpha < \pi/2 $, there is a	
% gap opening generated by the order from disorder mechanism.
% the Y-x state becomes metastable between the first order transition
% ( diagonal ) line $  \alpha_{in} < \alpha < \pi/2 $ and the second order transition ( dashed ) line
% driven by the condensations of the $ C-C_{\pi} $ and C-IC with $ \pi- q_{ic} < k^{0}_y < \pi $.
% There is a quantum Lifshitz transition at $ \alpha= \alpha_{in} $ with the dynamic exponent $ z_x=z_y=1 $,
% from the collinear Y-x ( or X-y ) phase to the coplanar IC-XY-y  ( or IC-XY-x ) phase, then a second one from the
% IC-XY-y to the commensurate $ 3 \times 3 $ SkX crystal phase at $ \alpha= \alpha_{33} $ which is a bi-critical point.
% The segment $  \alpha_{33} < \alpha < \alpha_{in} $ is the first order transition line between the IC-XY-y and IC-XY-x.
% It is a bi-critical point  where
% the two second order transition lines from the $ 3 \times 3 $ SkX to the IC-XY-y below the diagonal line
% and to the IC-XY-x above the diagonal line meet the first order transition line between the IC-XY-y and IC-XY-x along the diagonal line.
% The  $ 3 \times 3 $ SkX  and IC-SkX is the only non-coplanar C and IC phase along the diagonal line.
% All the other segments are first order transition lines between $ N \times 1 $ and $ 1 \times N $
% co-planar spiral phase after $ N \geq 4 $.
% The quantum order from disorder phenomenon at $  \alpha_{in} < \alpha < \pi/2 $.
 M is the multicritical point located at $ ( \alpha_M, \beta_M ) $
 where the $ ( 0, \pm 2\pi/3) $ counter line of the Y-x phase from the right hits the corner of the $ 3 \times 3 $ SkX crystal.
 There is also a 1st order quantum Lifshitz transition from
 the Y-x phase on the right to the non-coplanar IC-SkX-y phase driven
 by the condensations of IC- magnons with $ \pi-\pi/3 < k^{0}_y < \pi- q_{ic} $ ( Inset and Fig.\ref{phasesarc} ).
 It reduces to the co-planar IC-XY-y phase when $ \alpha=\beta^{+} $.
 A putative 2nd order transition along the dashed line connecting from $ (\pi/2,0) $ to the $ M $ point
 due to the condensations of the $ C_0 $ and IC-magnons with $ 0 < k^{0}_y < 2\pi/3 $
 is preempted \cite{firstorder} by the ( last)  $ W=1/2 $ segment of the in-complete devil staircase.
 Along the complete devil staircase $ 0< \alpha=\beta^{+} < \alpha^{-}_{33} $, only
 the principle series with $ W=1/N $ is drawn ( see Fig.\ref{cantor},\ref{finiteT} ),
 the IC-YZ-x/LQx form a Cantor set with the fractal dimension.
% The Y-x state becomes metastable between the first order transition line and the second order transition
% ( dashed ) line ( both connect from $ (\pi/2,0) $ to the $ M $ point ).
% Near $ ( \pi/2, \pi/2 ) $, it also becomes metastable between the first order transition line and the second order transition
% ( dashed ) line ( both connect from $ (\pi/2,\pi/2) $ to the $ ( \alpha_{in}, \alpha_{in} ) $ point ).
 The in-complete devil's staircases at a small $ \beta < \alpha $
 consists of commensurate spiral co-planar  phases near $ \alpha=\frac{\pi}{N} n, n \geq 1 $
 with gapped magnons embedded in the sea of the spiral IC-YZ-x/LQx phases with gapless phasons.
% can be obtained by mapping the $ N \times 1 $ ansatz into the FK model Eqn.\ref{fk}.
 %The $ \cdots $ means that this incomplete devil staircases ends to some IC phases near the origin $ \alpha=\beta=0 $.
 Immersed inside the spiral IC-YZ-x/LQx phases are some small devil staircases
 with higher topological winding number $ W= n/N $
 ( such as at $ 3/7, 2/5, 2/7...... $ ) displaying fractal structures.
% For example, the $ \alpha= 2\pi/5 $ state is stable only when $ \beta $ is small,
% there are always some IC phases intervening between the $ 2\pi/5 $ state and the Y-x phase.
% There is a spurious Mirror symmetry about $ \beta=\pi/4 $ at the LSW order, but it was spoiled by higher order terms in $1/S $ expansion.
% There are order from disorder and mass generation phenomena along $ \alpha =\beta $, but not along its mirror image $ \alpha= \pi/2- \beta $.
% The in-complete devil staircases become complete ones  along the diagonal line.
% The Y-x ( X-y ) is the only collinear, the $ 3 \times 3 $ SkX ( IC-XY-y or IC-XY-x ) is the only C ( IC ) non-coplanar phase,
% all the other phases are stripe coplanar spiral C or IC phases. In fact, all the spiral states at $ \alpha=\frac{n}{N} \pi $
% is the same state as that at $ \alpha=\pi/N $.
 The relevant numbers are $ \alpha^{0}_{in} \sim 0.3611 \pi, \alpha_{ic} \sim 0.3526 \pi, \alpha_{33} \sim 0.3402 \pi, (\alpha_M,\beta_M) \sim(0.33952\pi,0.31284\pi)  $ and $ q_{ic} \sim 0.18 \pi $. The inset shows the quantum phase transitions
 from the Y-x phases ( Fig.\ref{phasesarc} ). For the three building blocks, see Fig.2. }
\label{phasedia}
\end{figure}

\begin{figure}[!htb]
	%\centering
\includegraphics[width=0.47\textwidth]{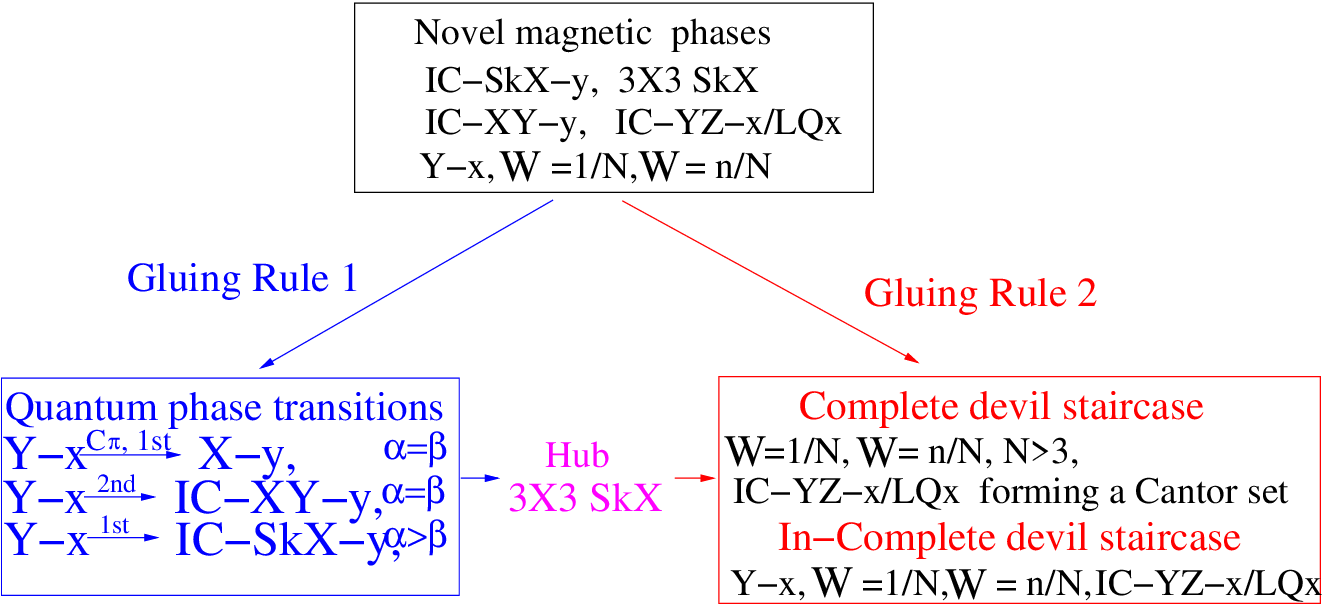}			
\caption{ The three building blocks in Fig.1: Unconventional magnetic phases which can be organized either in novel quantum phase transitions
 or complete/In-complete devil staircases. The non-coplanar $ 3 \times 3 $ SkX phase is the only phase respecting the $ [C_4 \times C_4]_D $ symmetry along the diagonal line and acts as
 the hub ( central node ) phase dividing the two different organization principles. Y-x is the only collinear phase.
 The quantum Lifshitz transition from the Y-x to the IC-XY-y along the diagonal line need to be studied by involving the OFQD \cite{NOFQD}.
 The IC-SkX-y away from the diagonal line maybe relevant to 4d/5d Kitaev materials.
 Despite $ W=1/N $ and $ W=n/N, n> 1 $ have the same symmetry breaking patterns, they can still be distinguished by
 $ W $ which characterize the complete/In-complete devil staircases.
 Along the diagonal line, the IC-YZ-x/LQx forms a Cantor set with a fractal dimension  along the complete devil staircase. Away from the diagonal line, they take a finite measure. The putative transition from the Y-x driven by $ C_0 $ and part of IC- magnons Eq.\ref{eta33} below the M point
 is pre-emptied by the $ W=1/2 $
 plateau of the in-complete devil staircase, so belong to the right box.
 The gluing rule 1 can be described by various effective quantum field theories.
 While the gluing rule 2 defies any effective quantum field theories. }
\label{trifeats}
\end{figure}

The RFHM Eq.\ref{rhgeneral} at a generic $ (\alpha, \beta ) $ has the translational,
the time reversal $ {\cal T} $ symmetry.
Along the extremely anisotropic limit $\alpha=\pi/2, 0<\beta<\pi/2$, there are three spin-orbital coupled $ Z_2 $
symmetries $ {\cal P}_x, {\cal P}_y, {\cal P}_z $ \cite{rh}. Most importantly,
there is
a hidden spin-orbital coupled $ U(1) $ symmetry generated by
$ U_1(\phi)=e^{ i \phi \sum_{i} (-1)^x S^{y}_i } $ and also the Mirror symmetry $ {\cal M} $: under the local rotation
$\tilde{\mathbf{S}}_{i} =R(\hat{x},\pi ) R(\hat{y},\pi n_2) \mathbf{S}_{i}$, $ \beta \rightarrow  \pi/2 - \beta $.
The middle point $ \beta=\pi/4 $ respects the Mirror symmetry and is also the most frustrated point.
However, any deviation from the  extremely anisotropic line $ \alpha \neq \pi/2 $ spoils the $ U(1) $ and Mirror symmetry.
Along the isotropic Rashba limit $ \alpha=\beta $, the $ {\cal P}_z $ symmetry along the anisotropic limit is enlarged to the spin-orbital coupled $ [C_4 \times C_4]_D $ symmetry
around the $ z $ axis. Of course, along the bottom Abelian line $ 0 < \alpha < \pi/2, \beta=0 $, it has
the $ \tilde{SU}(2) $ symmetry in the $ \tilde{SU}(2) $ basis  $ \tilde{\mathbf{S}}_n= R(\hat{x}, 2 \alpha n )\mathbf{S}_n $.
Because $ \beta < \alpha $ lower-half is related to the $ \beta > \alpha $ upper half in Fig.\ref{phasedia}
by the $ [C_4 \times C_4]_D $ transformation, so in the following, we mainly focus on the lower half.
%We will approach the global phase diagram Fig.1 from all the three lines: the solvable line $ \alpha=\pi/2, 0<\beta<\pi/2$,
%the Abelian line $ 0 < \alpha < \pi/2, \beta=0 $ and the diagonal line  $ 0 < \alpha=\beta < \pi/2 $.

% Some applications of the RFHM to these materials have been discussed in \cite{rhh} and will be discussed further near to the end of this paper.

The rest of the paper is organized as follows.
In Sec.II-VI, we explore all these novel quantum phases and the quantum/topological phase transitions among them.
In Sec.VII-VIII, we investigate the second organization principle of these phases
 in-complete or complete devil staircase displaying fractals, especially their quantization ( Fig.\ref{trifeats} ).
This gluing pattern presents a complete new class of organization principle which is beyond any known classification schemes
in symmetry breaking, topological phases or quantum chaos.
In Sec. IX, We also contrast these novel phases and their gluing patterns with those due to geometric frustrations/quenched disorders,
1d Aubry-Andre (AA) model/Hofstadter butterfly, in 2d quantum dimer models,
2d quantum Hall plateau-plateau transitions/2d quantum Hall edge states and 3d cubic code with a fractal structure.
In Sec. X, we discuss the experimental implications. We find that the IC-SkX-y phase in Fig.1 display very similar properties as those un-conventionally in-commensurate ordered magnetic phases observed in some 4d or 5d Kitaev materials.
We stress the important roles due to the DM term which breaks the parity.
In Sec. XI, we summarize our results, compare with an exact theorem on possible topological states in the presence of SOC
and outline some future perspectives.
In several appendices, we perform specific calculations by canonical quantization, path integral, especially the shift between the two approaches to support  the new and important concepts made in the main text.

%are localized in one dimension, but free to move in the other.
%They also display broad distribution of Bragg peaks in $ k_x $ direction which very much resemble the quantum spin liquids in experimental %signatures, so may be relevant to the materials displaying ¡°Quantum spin liquid "

\section{ C and IC Magnons in the Y-x state, their condensations and putative 2nd order transitions }
% Quantum fluctuations along the dashed line $ ( \alpha=\pi/2, \alpha < \beta ) $ have been studied in \cite{rh}.
 The firmly established results and physical insights \cite{rh} achieved on the extremely anisotropic line
 $ ( \alpha=\pi/2, \alpha < \beta ) $ pave the way to study the physics
 at generic $ ( \alpha, \beta ) $ in Fig.\ref{phasedia}. Especially, we will follow how the three kinds of magnons
 response and evolve when moving away from the line.

Making a globe rotation $R_x(\pi/2)$ to align spin along the Z-axis and then introducing Holstein-Primakoff bosons $ a $ and $ b $ for
the two sublattice, we can expand the Hamiltonian in the powers of $1/\sqrt{S}$,
\begin{align}
	H=E_0+2JS\Big[H_2
	+\Big(\frac{1}{\sqrt{S}}\Big)H_3
	+\Big(\frac{1}{\sqrt{S}}\Big)^2H_4+\cdots\Big]
\label{swcubic}
\end{align}
where the symbol $H_n$ denotes the $n$-th polynomial of the boson operators,
$E_0=-2NJS^2\sin^2\alpha$ is the classical ground state energy of the Y-x state.
   Performing a unitary transformation, then a Bogoliubov transformation on $ H_2$, one can diagonize $ H_2 $ as ( appendix A ):
\begin{align}
	H_2= E_2
	+2\sum_k(\omega_k^+\alpha_k^\dagger\alpha_k
		+\omega_k^-\beta_k^\dagger \beta_k)
\label{h2e2}
\end{align}
where $ E_2=\sum_k(\omega_k^++\omega_k^--2\sin^2\alpha) $ is the quantum correction to the ground state energy at the LSW order,
$\omega_k^{\pm}=\sqrt{(\lambda_k^{\pm})^2-\chi_k^2}$,
$\lambda_k^{\pm}=\sin^2\alpha-\frac{1}{2}\cos2\beta\cos k_y
\pm\frac{1}{2}\sqrt{\sin^4\alpha\cos^2k_x+\sin^22\beta\sin^2 k_y}, \chi_k=\frac{1}{2}\cos^2\alpha\cos k_x $.
Obviously, $\omega_k^{\pm}=\omega_{-k}^{\pm}$ which is dictated by the symmetries of the Hamiltonian and the Y-x state.
Note that to the LSW order, the dispersion still has the Mirror symmetry under the $ \beta \rightarrow \pi/2-\beta $.
However, the mirror symmetry will be spoiled by the higher order terms starting at $ H_3 $.

As shown in \cite{rh}, at $ \alpha=\pi/2 $, the Y-x state is the exact ground state, $ \chi_k=0 $, there is no need for the extra
Bogoliubov transformation, the spin wave dispersion reduces to $ \omega_k^{\pm}=\lambda_k^{\pm} $.
As shown in \cite{rhtran}, any transverse field $ h_x $ or $ h_z $ transfers the Y-x state into a co-planar canted state.
In a sharp contrast, here, under $ \pi/2 - \alpha \neq 0 $, the Y-x state remains the classical state,
but not the exact eigenstate anymore due to the quantum fluctuations introduced by $ \alpha \neq \pi/2 $.
%DO NOT BE CONFUSED between operator $\alpha_k$ and parameter $\alpha$.
 From $ \omega_k^{-} $, one can identify the minimum $ (0,k_y^0)$ of spin-wave dispersion
 corresponding to the magnons C-$C_0$, IC, C-$C_{\pi} $ respectively ( See appendix A ).
% Setting $  \lambda=\frac{\cos(2\beta)}{\sin(2\beta)}
%	\sqrt{\sin^4(\alpha)+\sin^2(2\beta)} $, then when $ \lambda \in (-\infty,-1),[-1,1],(1, \infty) $, $ k_y^0=\pi,
%	\arccos\left[\lambda \right ],0 $ corresponding to the magnons C-$C_0$, C-IC, C-$C_{\pi} $ respectively.
 Near $ (0,k_y^0)$, their dispersions take the relativistic form:
\begin{equation}
 \omega_{-}(q) = \sqrt{ \Delta^2 + v^{2}_x q^{2}_x + v^{2}_y q^{2}_y }
\label{relagap}
\end{equation}
  The gap and the two velocities are given in the appendix A.

   The Staggered magnetization and specific heat of the Y-x phase at $ T \ll \Delta $ are:
\begin{eqnarray}
	M(T) & \sim & M(T=0)- \frac{T\Delta}{2\pi v_xv_y}\sqrt{1+\frac{\cos^4\alpha}{4\Delta^2}}e^{-\Delta/T}   \nonumber  \\
    C(T) & \sim  & \frac{1}{2\pi v_xv_y}\frac{\Delta^3}{T}e^{-\Delta/T}
\label{mc}
\end{eqnarray}
  where $ M(T=0)= S-\frac{1}{N}
	\sum_k(\frac{\lambda_k^+}{2\omega_k^+}
	       +\frac{\lambda_k^-}{2\omega_k^-}-1) $ is the $ T=0 $ magnetization.
  At $ \alpha=\pi/2 $, replacing $v_x$ by $\sqrt{\Delta/m_x}$ and $v_y$ by $\sqrt{\Delta/m_y}$, Eqn.\ref{mc}
  gives back to those along the solvable line in \cite{rh}.
  The spin-spin correlations functions (SSCFs ) and structure factors have been evaluated in appendix E for both C- and IC- magnons.

  Solving $\Delta=0$ leads to the 3 segments of their condensation boundary:
\begin{align}
\alpha=
\begin{cases}
	&\pi/2-\beta, \\
	& \arcsin\left[
		\frac{\sqrt{6}\sin2\beta}{\sqrt{9\sin^22\beta-1}}
	\right],	\\
	& \beta,
\end{cases}
\label{segments}
\end{align}
    for $ 0\leq\beta\leq\pi/2-\arccos(1/\sqrt{6}), \pi/2-\arccos(1/\sqrt{6})\leq\beta\leq\arccos(1/\sqrt{6}) $ and
    $ \arccos(1/\sqrt{6})\leq\beta\leq\pi/2 $ respectively.
    At the LSW order, it still has the mirror symmetry under $ \beta \rightarrow \pi/2-\beta $.

\subsection{ Spurious Goldstone mode along the diagonal line near $ \alpha=\beta=\pi/2 $. }

    The C$_\pi$ magnons condense along the diagonal line $ \arccos(1/\sqrt{6})\leq \beta \leq \pi/2 $ with the gapless relativistic dispersion:
\begin{align}
	\omega_-(q)=\sqrt{v_x^2 q_x^2+v_y^2 q_y^2}
\label{gapless}
\end{align}
   where $ v_x=\cos(\alpha)/2,~
	v_y=\cos(\alpha)\sqrt{1-6\cos^2(\alpha)}/2 $.
 As to be shown in the next section, it is a spurious Goldstone mode due to the breaking of a
 spurious $ U(1) $ symmetry at the classical level. $ v_y $ also vanishes at the boundary.

Obviously, both velocities vanish at the Abelian point $ \alpha=\pi/2, \beta=\pi/2 $
dictated by the enlarged $ SU(2) $ symmetry. Moving away from the Abelian point $ \alpha=\beta=\pi/2 $, $ v_x$ keeps increasing,
but $ v_y$ increases first, reaches a maximum, then decreases and vanishes at the boundary between C-$ C_{\pi} $ and IC- magnons
$ \alpha^{0}_{ic}= arccos(1/\sqrt{6} ) \sim 0.36614 \pi $.
When pushing to higher orders, $ \omega_-(q)=\sqrt{v_x^2 q_x^2+v_y^2 q_y^2 + u^2 q_y^4+ \cdots } $,
we find it is a putative $ ( z_x=1, z_y=2 ) $ quantum Lifshitz transition from the Y-x state to
an incommensurate state ( Fig.4a).
However, as to be shown in the following section, the gapless mode along the diagonal line and the mirror symmetry $ {\cal M} $ under $ \beta \rightarrow \pi/2 - \beta $ are just spurious facts of the LSW approximation.
%However, the  quantum Lifshitz transition remains, but with a different dynamic exponent than
%$ ( z_x=1, z_y=2 ) $.
%When going beyond the LSW, there is a gap opening along the diagonal line, the mirror symmetry should also be spoiled.

\begin{figure}
\includegraphics[width=\linewidth]{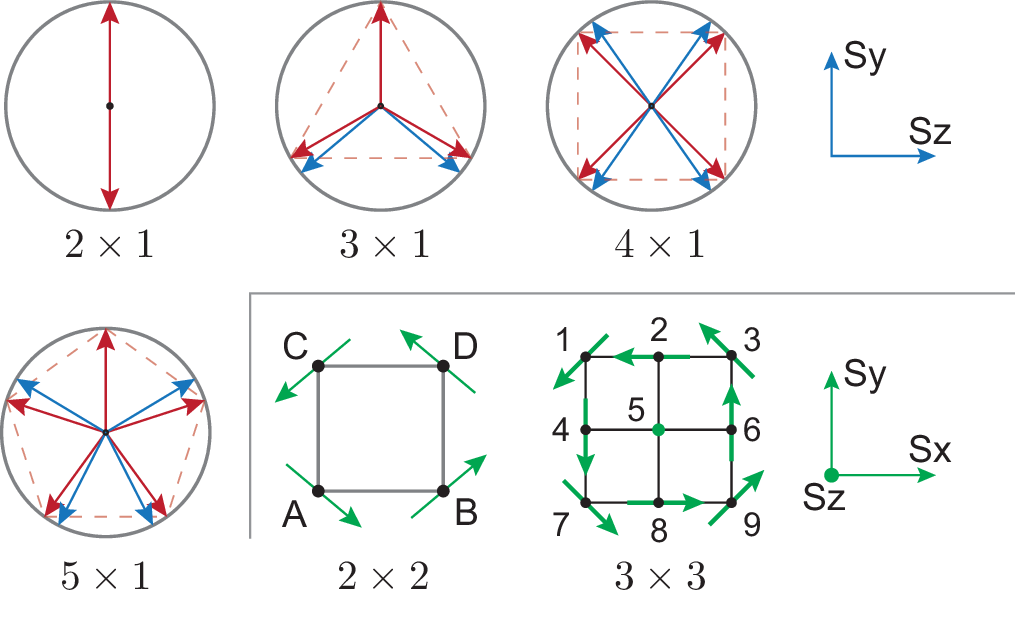}
\caption{
 Some most robust Collinear, spiral, vortex and non-coplanar states in Fig.\ref{phasedia}.
 Top layer:  the $ 2 \times 1 $  (Y-x) state $ S^{y}= (-1)^x $ is the exact quantum  ground state \cite{rh} at $ \alpha=\pi/2 $.
 It is the only collinear phase. All the others with $ N \times 1, N \geq 2 $ are spiral co-planar phase in the YZ plane.
% {\sl The first capital letter indicates spin polarization along $ Y $ direction,
%    the second small letter indicates the orbital ordering along the $ x $ bond. We will use this notation to label all the spin-orbital correlated phases
%    in this proposal. }
 When $ \beta $ is small, the $ 3 \times 1 $ spiral state is close to be a FM state in the rotated basis
 $ \tilde{\mathbf{S}}_n = R(\hat{x}, 2 \alpha n )\mathbf{S}_n $.
 All the red arrows in $ 120^{\circ} $ structure (connected by the dashed line) will be transformed to a FM state in the rotated basis, the blue arrows
 are actual spiral spin orientations which only deviate slightly from the red arrows. The deviation angles increases as $ \beta $ increases in the
 $ 3 \times 1 $ staircase.
 The spiral states at  $ 4 \times 1 $, $ 5 \times 1 $ and other devil's staircases (not shown) can be similarly constructed.
 The inset show the spin axis for the collinear and spiral states.
 The degeneracy is $ 2 N $ for odd $ N $ and $ N $ for even $ N $.
 There is a small magnetization for $ N $ odd, but exactly zero for $ N $ even.
 Bottom layer: the classically degenerate $ 2 \times 2 $ vortex state along the diagonal line
 $ \alpha=\beta $ is simply a FM state in the XY plane
 in the rotated basis $  \tilde{\mathbf{S}}_n= R(\hat{x},\pi n_1)  R(\hat{y},\pi n_2)\mathbf{S}_n  $.
%In fact, the classical $ Z-AFM $ state $ S^{z}=(-1)^{(x+y)} $ found in \cite{classdm2} is just the FM state along
%$ (0,0,1) $ direction in the rotated basis with a 4 sublattice structure.
%It is degenerate with the $ 2 \times 2 $ vortex state  at the center $ \alpha=\beta=\pi/2 $, but has a  higher energy at any  $ \alpha=\beta < \pi/2 $.
The $ 3 \times 3 $ skyrmion crystal (non-coplanar) state with non-vanishing Skyrmion density
$ \mathbf{S}_i \cdot  \mathbf{S}_j \times \mathbf{S}_{k} \neq 0 $
happens near $ \alpha=\beta=\pi/3 $ which is the most frustrated regime in the Wilson loop \cite{rh}.
The inset show the spin axis for the $ 2 \times 2 $ vortex and $ 3 \times 3 $ SkX states. }
\label{allphases}
\end{figure}

\section{  Order from disorder along the diagonal line near the $ \alpha=\beta=\pi/2 $ Abelian point. }
% The QFH model  along the dashed line  $ ( \alpha=\pi/2, \alpha < \beta ) $ have been throughly  studied in \cite{rh}.
% The $ 2 \times 1 $ ( $ Y-x $ state ) was proved to be the exact quantum ground state, so no quantum fluctuations at $ T=0 $.
In this section, we will the following 3 effects of the order from quantum disorder mechanism
(1) determine the true quantum ground state along the diagonal line near the Abelian point
 $ \alpha=\beta=\pi/2 $
(2)  by using the spin coherent state path integral method, determine the magnon gap at $ \vec{q}=0 $
     generated by the order from quantum disorder mechanism.
(3) push method much further to evaluate its correction to the whole spectrum instead of just at $ \vec{q}=0 $.

\subsection{  The selection of the quantum ground state by the order from quantum disorder mechanism  }
\begin{figure}[!htb]
	%\centering
\includegraphics[width=0.5\textwidth]{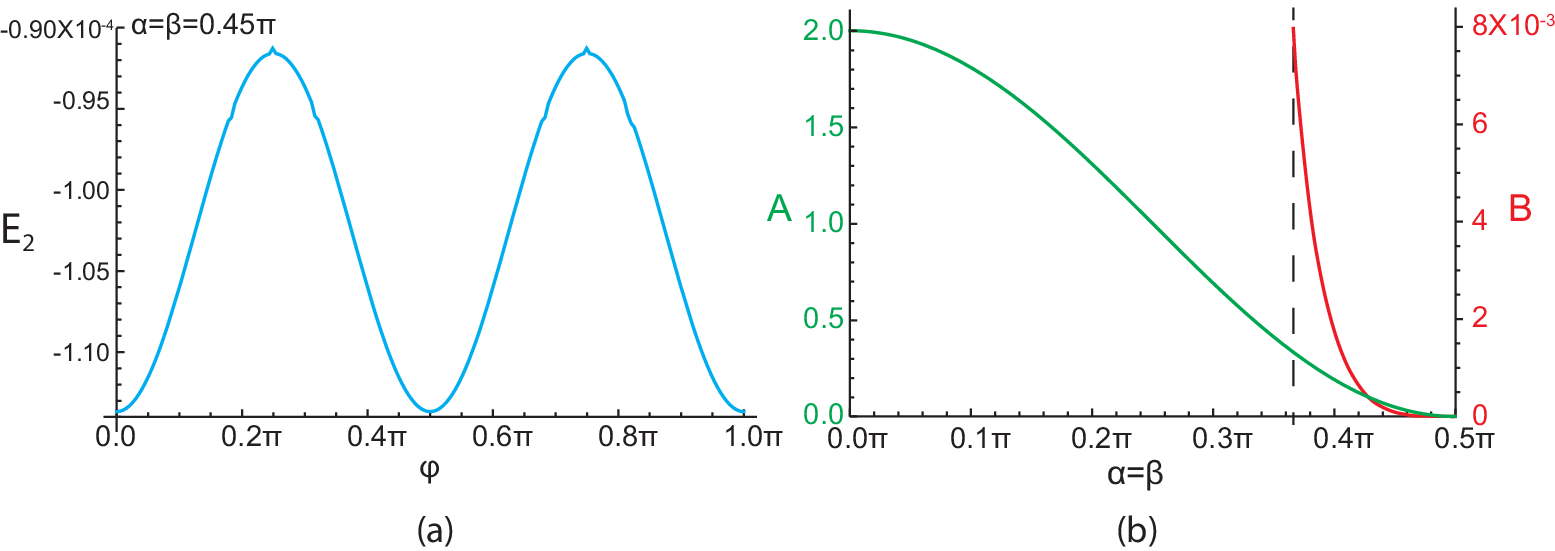}	
%\quad
%\includegraphics[width=0.2\textwidth]{KC.eps}	
	\caption{ The effective potential generated by the OFQD and the gap opening
    on the spurious gapless mode along the diagonal line in Fig.\ref{phasedia}.
   (a) The quantum correction to the ground-state energy from the LSW.
	$\phi=0$ corresponds to X-y state and $\phi=\pi/2$ corresponds to Y-x state.
    So the quantum fluctuations pick up Y-x or X-y as the ground state which is related to each other
    by the $ [C_4 \times C_4]_D $ symmetry.
    (b) The classical coefficient $ A( \alpha)/J $ on the left axis and
    the quantum one $ B(\alpha )/J $ on the right axis. Both vanish at
    the Abelian point $ \alpha=\beta =\pi/2 $  as $ \sim ( \pi/2- \alpha)^2 $ and are monotonically increasing function
    when moving away from the Abelian point. Note that $ A \gg B $.
    The Dashed line is located at
    $ \alpha^{0}_{in} \sim 0.3661 \pi $ where the Y-x state becomes unstable at the LSW order.
    After incorporating the gap opening, the $ \alpha^{0}_{in} $ is shifted to a  smaller value $ \alpha_{in} \sim 0.3526 \pi $. }
\label{orderdis}
\end{figure}

 The first thing to do is to find  what is the true quantum ground state along the diagonal line near the Abelian point
 $ \alpha=\beta=\pi/2 $.
 At the classical level, the $ 2 \times 1 $  Y-x  stripy state $ S^y=(-1)^x $ is degenerate with the $ 1 \times 2 $  X-y stripy state $ S^x=(-1)^y $.
 In fact, we find there is a family of states called $ 2 \times 2 $ vortex states in Fig.\ref{allphases}:
\begin{equation}
  \mathbf{S}_i=
( (-1)^{i_y}\cos\phi,(-1)^{i_x}\sin \phi,0 )
\label{2times2}
\end{equation}
  which are  degenerate at the classical level.
 In general, this family breaks the $ [C_4 \times C_4]_D $  symmetry except at $ \phi= \pm \pi/4, \pm 3\pi/4 $.
 When $ \phi=0, \pi/2 $, it recovers to the X-y and Y-x state respectively.
 Quantum fluctuations ( "order from disorder" mechanism ) are needed to find the unique quantum ground state upto
 the  $ [C_4 \times C_4]_D $  symmetry in this regime.
 To perform a LSW calculation, one need to introduce a 4 sublattice structure $ A, B, C, D $ shown in Fig.\ref{allphases}.
 After making suitable rotations to align the spin quantization axis along the Z axis, we introduce  4 HP bosons $ a, b, c, d $
 to perform a systematic $ 1/ S $ expansion shown in Eqn.\ref{swcubic} where
 $E_0=-2NJS^2(1-\cos2\alpha\sin^2\phi-\cos2\beta\cos^2\phi)$ is the classical ground state energy,
 $ H_2 $ can be diagonized by a unitary transformation, then followed by a Bogoliubov transformation as:
\begin{align}
	H_2= E_2 + 2\sum_{n,k} \omega_n (k)\alpha_{n,k}^\dagger \alpha_{n,k}
\label{fournk}
\end{align}
    where $ n=1,2,3, 4 $ is the sum over the 4 branches of spin wave spectrum in the
    Reduced Brillouin Zone (RBZ) $ -\pi/2 < k_x, k_y < \pi/2 $,
    $ E_2 $ is the $ 1/S $ quantum correction to the ground-state energy $ E_0 $:
\begin{align}
	E_2 =\sum_{k,n}[\omega_n(k)
	-(1-\cos2\alpha\sin^2\phi-\cos2\beta\cos^2\phi)/2]
\label{quantumphi}
\end{align}
   Obviously, near the Abelian point $ \alpha=\beta=\pi/2 $, if $ \alpha > \beta $, it picks the Y-x state with $ \phi=\pi/2 $.
   If $ \alpha < \beta $, it picks the X-y state with $ \phi= 0 $.
   Setting $ \alpha=\beta $, the $ E_0 =-2NJS^2(1-\cos2\alpha) $ becomes $ \phi $ independent, indicating
   the classical degenerate family of states characterized by the angle $ \phi $ along the whole diagonal line $ \alpha=\beta $. Fortunately,
   the quantum correction $ E_2(\phi)=\sum_{k,n}[\omega_n(k,\phi)-\sin^2\alpha] $ does depend on $ \phi $.
As shown in Fig.\ref{orderdis}a, we find that $E_2(\phi)$ reach its minimum at $\phi=0$
( X-y  state with the degeneracy $ d=2 $ ) or $\phi=\pi/2$ ( Y-x state with the degeneracy $ d=2 $ ) which is
related to each other by the $ [C_4 \times C_4]_D $ symmetry which dictates:
\begin{equation}
  E_2(\phi)=E_2^0- B \cos 4 \phi  + \cdots
\end{equation}
 where $ \cdots $ means higher order harmonics in $ \cos 4 \phi $.
 Obviously, the global form of $ E_2(\phi) $ keeps the $ [C_4 \times C_4]_D $ symmetry of the Hamiltonian,
 its four minima at $ \phi=0, \pi/2, \pi, 3 \pi/2 $ lead to the  X-y state with the degeneracy $ d=2 $ and
 the Y-x state with the degeneracy $ d=2 $ respectively.

If picking one of the 4 minima such as $\phi=0$, then one breaks the $ [C_4 \times C_4]_D $ symmetry.
Expanding $E_2(\phi)$ ( in unit of $ 2 J S $ ) around one of its minima  $\phi=0$:
\begin{equation}
  E_2(\phi)=E_2^0+ \frac{1}{2} B(\alpha) \phi^2+ \kappa \phi^4 + \cdots
\label{Bkappa}
\end{equation}
where one can identify the coefficient $ B(\alpha )$ as plotted in the Fig.\ref{orderdis}b (right axis).

The quantum order from disorder selection of the Y-x or X-y state along the diagonal line shows that there is a direct first order transition from
the Y-x state to the X-y state along the diagonal line in Fig.1. So along the diagonal line, there is any mixture of the Y-x and X-y state.
Similar first order transition between vacancy induced supersolid (SS-v) and interstitial induced supersolid (SS-i)
and any mixtures of the two along the particle-hole symmetric line at the half filling in a triangular lattice were discussed in \cite{dual1,dual2,dual3}.

% However, we expect that the quantum fluctuations ( "order from disorder" ) encoded in the RFHM  will pick the
% $ 2 \times 2 $ vortex state as the ground state. Following the SWE developed in \cite{rh,rhh,rhht,rafh,sw1,sw2,sw3}, we will
% confirm this expectation and also determine the excitation spectra above this classical vortex state.

\subsection{  The magnon gap generated by the order from disorder mechanism: Pseudo-Goldstone mode }
  The gapless nature of the spin wave spectrum Eqn.\ref{gapless} along the diagonal line is just a
  spurious fact of the LSW approximation.
%  It will be gapped out by the higher order
%  terms in the $1/S $ expansion Eqn.\ref{swcubic}. As shown in the next section,
%  the quantum Lifshitz transition remains, but with a different dynamic exponent $ z=1 $
%  than that $ ( z_x=1, z_y=2 ) $ got within the LSW.
   By using the spin coherent state path integral formulation \cite{aue,swgap},
   we will evaluate  the leading order corrections to the gap at the minimum $ (\pi,0) $ of the
   C-$ C_{\pi} $ magnons.
  A general uniform state at $ \vec{q}=0 $ can be taken as a FM state with the polar angle $ (\theta, \phi) $
  in the $ \tilde{\tilde{SU}}(2) $ basis with $\tilde{\tilde{\mathbf{S}}}_{i}
  = R(\hat{x},\pi n_1)  R(\hat{y},\pi n_2) \mathbf{S}_{i} $ at the  $ \alpha=\beta=\pi/2 $ Abelian point.
  After transforming back to the original basis by using
$\tilde{\tilde{S}}_1=R_z(\pi)S_1$,
$\tilde{\tilde{S}}_2=R_y(\pi)S_2$,
$\tilde{\tilde{S}}_3=R_x(\pi)S_3$,
$\tilde{\tilde{S}}_4=S_4$,
it leads to a  $ 2 \times 2 $ state characterized by the two angles $ \theta $ and $ \phi $.
%\begin{eqnarray}
%    S_1 & = & (-\cos\phi\sin\theta,-\sin\phi\sin\theta,\cos\theta), \nonumber   \\
%	S_2 & = & (-\cos\phi\sin\theta,\sin\phi\sin\theta,-\cos\theta),  \nonumber    \\
%	S_3 & = & (\cos\phi\sin\theta,-\sin\phi\sin\theta,-\cos\theta),  \nonumber    \\
%	S_4 & = & (\cos\phi\sin\theta,\sin\phi\sin\theta,\cos\theta)
%\label{22vortex}
%\end{eqnarray}
   Along the diagonal line, its classical energy becomes
\begin{align}
	E_0=J[-2 \sin^2 \alpha - 2\cos^2 \alpha \sin^2 \theta ]
\label{classtheta}
\end{align}
  which is, as expected, $ \phi $ in-dependent. But one can see any deviation from the Abelian point picks up the XY plane
  with $ \theta=\pi/2 $. So it reduces to the $ 2\times 2 $ vortex state in Fig.\ref{allphases} used in the " order from  disorder "
  analysis in the last section.
  Expanding around the minimum $ E_0 =J[-2\sin^2\alpha+ 2\cos^2\alpha (\theta-\frac{\pi}{2})^2+\cdots] $ gives
  the stiffness $ A =2J\cos^2\alpha$ shown in Fig.\ref{orderdis}b (left axis ). Using the spin coherent state analysis,
  we can write down the quantum spin action at $ \vec{q}=0 $:
\begin{equation}
  {\cal L}( \vec{q}=0 )= i S \cos \theta \partial_{\tau} \phi + S^2 A ( \theta-\pi/2)^2 +  S B \phi^2
\label{action00}
\end{equation}
  where we put back the spin $ S $, the first term is the spin Berry phase term, the second
  $ A \sim (\pi/2-\alpha)^2 $ and the third $ B \sim (\pi/2-\alpha)^2 $ are from the classical analysis
  in Eqn.\ref{classtheta} and the  order from the quantum disorder (OFQD) analysis to LSW order in Eqn.\ref{quantumphi} respectively.
  Eqn.\ref{action00} leads to the gap:
\begin{equation}
   \Delta_B= 2\sqrt{S A B} \propto \sqrt{S}
\label{diagap}
\end{equation}
  In fact, there are also corrections from the cubic $ H_3 $ and quartic $ H_4 $ terms in Eqn.\ref{swcubic},
  but they only contribute to order of $ 1 $ which is subleading to the $ \sqrt{S} $ order in the $ 1/S $ expansion \cite{sw1,japan}.
  As shown in Fig.\ref{orderdis}b, because both $ A $ and $ B $ are monotonically increasing along the diagonal line,
  so the gap also increase. Plugging their values at
  $ \alpha=\alpha^{0}_{in} = \arccos(1/\sqrt{6})\sim 0.3611 \pi $,
  one can see $ A $ is 3 orders of magnitude larger than $ B $: $A/J=1/3, B/J \approx0.008 $.
  Putting $S=1/2$, we find the maximum gap near the quantum Lifshitz transition $ \Delta_B/J \sim 0.036  $.

  In short, there is a spurious Goldstone phase mode $ \phi $ due to the spontaneous breaking of the spurious $ U(1) $ symmetry
  in the XY plane Eq.\ref{gapless}. Then the order from quantum disorder (OFQD) phenomenon generates
  a gap $ \Delta_B $ to this spurious Goldstone mode
  and transfers it into a pseudo-Goldstone mode Eq.\ref{gapspectrum}. In this process, the coefficient $ B $ is generated by the quantum fluctuations, while $ A $ is due to classical, so $ A \gg B $. In fact, there are 3 orders of magnitude differences between the two.
  This can be contrasted with a spurious $ SU(2) $ symmetry broken to $ U(1) $ resulting a spurious quadratic ferromagetic mode case presented
  in \cite{NOFQD} where both coefficients $ A $ and $ B $ are generated by  quantum fluctuations.

\subsection{ The spectrum corrected by the order from quantum disorder phenomena:
  a spin coherent state approach }

In the previous section, we derived the gap $ \Delta_B $ Eq.\ref{diagap} generated by order from disorder.
It is just a correction to the spectrum at $ k=0 $.
In this subsection, we will derive the correction to the whole spectrum, especially in the long wavelength limit

In the spin-coherent state path integral, the action takes form
\begin{align}
    \mathcal{A}
	=\int d\tau\big(\sum_i iS\cos\theta_i(\tau)\partial_\tau\phi_i(\tau)
	    +\mathcal{H}[\theta,\phi]\big)
\label{oldaction}
\end{align}
where $\mathcal{H}$ stands for the rotated Heisenberg model Eq.\ref{rhgeneral}
expressed in terms of polar angle $\theta$ and azimuthal angle $\phi$ in the polar coordinate with the spin quantization
axis along the $ Z- $ axis.
%and $\mathbf{S}_i=S(\sin\theta_i\cos\phi_i,\sin\theta_i\sin\phi_i,\cos\theta_i)$.

The classic ground-state near $\alpha=\beta\approx\pi/2$, the $2\times2$ vortex state, has the following 4-sublattice structure:
$A$-sublatice $(\theta_i,\phi_i)=(\pi/2,\phi_0)$,
$B$-sublatice $(\theta_{i+\hat{x}},\phi_{i+\hat{x}})=(\pi/2,-\phi_0)$,
$C$-sublatice $(\theta_{i+\hat{y}},\phi_{i+\hat{y}})=(\pi/2,\pi-\phi_0)$, and
$D$-sublatice $(\theta_{i+\hat{x}+\hat{y}},\phi_{i+\hat{x}+\hat{y}})=(\pi/2,\pi+\phi_0)$.
The 4-sublatice structure can be also written as $(\theta_i,\phi_i)=(\pi/2,i_y\pi+(-1)^{i_x+i_y}\phi_0)$,
where $\phi_0$ is an arbitrary angle due to the spurious U(1) symmetry.
Expanding the Hamiltonian around the classic ground-state with a general $\phi_0$,
and retaining up to second order in the fluctuations lead to:
\begin{widetext}
\begin{align}
    \mathcal{H}[\theta,\phi]
	=&-2NJS^2\sin^2\alpha+
	2JS^2\sum_{k\in RBZ}\big[\sum_{\Gamma=A,B,C,D}
	\sin^2\alpha
	(\delta\theta_{\Gamma,k}\delta\theta_{\Gamma,-k}+\delta\phi_{\Gamma,k}\delta\phi_{\Gamma,-k})   \nonumber\\
	&-\cos2\alpha\cos k_x(\delta\theta_{A,k}\delta\theta_{B,-k}+\delta\theta_{C,k}\delta\theta_{D,-k})
	-\cos2\alpha\cos k_y(\delta\theta_{A,k}\delta\theta_{C,-k}+\delta\theta_{B,k}\delta\theta_{D,-k})\nonumber\\
	&+(\sin^2\phi_0-\cos2\alpha\cos^2\phi_0)\cos k_x(\delta\phi_{A,k}\delta\phi_{B,-k}+\delta\phi_{D,k}\delta\phi_{C,-k})\nonumber\\
	&+(\cos^2\phi_0-\cos2\alpha\sin^2\phi_0)\cos k_y(\delta\phi_{A,k}\delta\phi_{C,-k}+\delta\phi_{D,k}\delta\phi_{B,-k})\nonumber\\
	&-i\sin2\alpha\cos\phi_0\sin k_x
	(\delta\theta_{A,k}\delta\phi_{B,-k}+\delta\theta_{B,k}\delta\phi_{A,-k}
	-\delta\theta_{D,k}\delta\phi_{C,-k}-\delta\theta_{C,k}\delta\phi_{D,-k}) \nonumber\\
	&-i\sin2\alpha\sin\phi_0\sin k_y
	(\delta\theta_{A,k}\delta\phi_{C,-k}+\delta\theta_{C,k}\delta\phi_{A,-k}
	-\delta\theta_{D,k}\delta\phi_{B,-k}-\delta\theta_{B,k}\delta\phi_{D,-k})\big]
\end{align}
\end{widetext}
The similar expansion of the Berry phase term leads to $-iS\int d\tau\sum_k\sum_\Gamma \delta\theta_{\Gamma,k}\partial_\tau\delta\phi_{\Gamma,-k}$,
which dictates $-S\delta\theta_{\Gamma}$ is conjugate to $\delta\phi_{\Gamma}$, namely, $ [ -S\delta\theta_{\Gamma}, \delta\phi_{\Gamma}]=i \hbar $( See also the Eq.\ref{qpi} in a different polar coordinate $ (\eta, \xi) $  with the spin quantization axis along the $ X- $ axis.  ).

From the Hamiltonian, one can extract the 4 eigenmodes $4JS\,\omega_{1,2,3,4}(\mathbf{k})$,
which are also obtained in Eq.\ref{fournk}.
Because the $\omega(\mathbf{k})$ depend on $\phi_0$,
thus the quantum fluctuations contribute an effective potential $E_\text{ofd}(\phi_0)=2JSE_2(\phi_0)$
which is nothing but the quantum correction to the ground-state energy  Eq.\ref{quantumphi}.
It is $E_\text{ofd}(\phi_0)$ which determines the quantum ground-state
to be X-y state $(\phi_0=0)$ or Y-x state $(\phi_0=\pi/2)$ ( see Fig.3).

Picking the ground state to be the Y-x state. Then it simplifies to a 2 sub-lattice structure.
Here we still stick to the 4 sub-lattice structure ( also used in appendix B )  which is related to the 2 sub-lattice structure
by just folding or unfolding the Brillouin Zone (BZ) $ 0 < k_x < \pi, -\pi < k_y < \pi $.

Expanding $E_\text{ofd}(\phi_0)$ around $\phi_0=\pi/2+\delta\phi_0$ leads to:
\begin{align}
	E_\text{ofd}(\phi_0)& =E_{\min}+NJSB(\delta\phi_0)^2+\cdots   \nonumber\\
	 & \approx E_{\min}+JSB\sum_{k\in \text{RBZ}}\sum_{\Gamma} \delta\phi_{\Gamma,k}\delta\phi_{\Gamma,-k}
\label{B2}
\end{align}
where the coefficient $B$ is listed in Eq.\ref{Bkappa} and  given in Fig.3b.

By adding $E_\text{ofd}$ back to the action Eq.\ref{oldaction}, one reaches the corrected action:
\begin{align}
    \mathcal{A}_\text{ofd} = \mathcal{A} +E_\text{ofd}(\phi_0)
\end{align}
 which gives  the corrected 4 eigen-modes. In the long wave-length limit, in the unit of $4JS$, the lowest one leads to the
 corrected spectrum which stands for the pseudo-Goldstone mode:
\begin{equation}
    \omega_{-}(q)=\sqrt{\Delta^2_B+ v_x^2 q_x^2+ v_y^2 q_y^2 }
\label{gapspectrum}
\end{equation}
  where the gap $ \Delta_B $ is given in Eq.\ref{diagap} and  the two velocities $ v_x, v_y $ also receive some corrections shown in Eq.\ref{vxvy}.  It will be used in the next section to derive the quantum Lifshitz transition.

 In the appendix B, we will re-derive Eq.\ref{gapspectrum} from the canonical quantization approach which is
 complementary to the spin-coherent path integral approach used here.
 Using two different but complementary approaches to derive the same result may lead to additional insights on
 the new physics of corrected spectrum due to the order from quantum disorder phenomenon.

\section{  Quantum Lifshitz transition from the Y-x  to IC-XY-y  state along $ \alpha=\beta^{+} $. }
As shown in the last section, there is a gap $ \Delta_B $ opening at $ \vec{q}=0 $ along the diagonal line,
so the  quantum Lifshitz transition point mentioned at Sec.II will shift to a smaller value of $ \alpha $.
% One can derive the  by incorporating the site dependence...it is not known how to incorporate the site dependence into
% the term generated from "order from disorder " mechanism...
%Obviously, the magnons always take the relativistic form Eqn.\ref{relagap} to any order of $1/S $, so
%it is justified to incorporate the gap $ \Delta_B $  into the spin-wave dispersion $\omega_k$ in Eqn.\ref{gapless} at the LSW order.

\begin{figure}[!htb]
    \centering
    \includegraphics[width=0.48\textwidth]{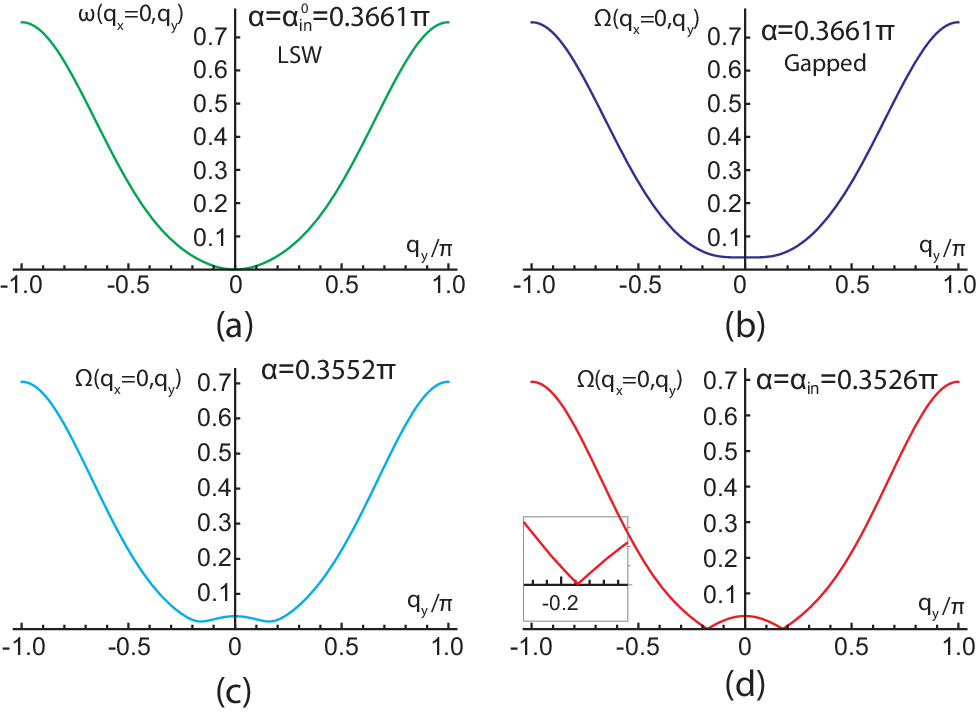}
    \caption{ The quantum Lifshitz C-IC transition from the Y-x state to the IC-XY-y state along the diagonal line $ \alpha=\beta $.
    The momentum is expanded near $ \vec{k} = (0, \pi ) + \vec{q} $.
    (a) The transition happens at $\alpha=\alpha_{\rm in}^0=0.3661\pi$ at the LSW order with the dynamic exponent $ ( z_x=1, z_y=2 ) $.
    (b) Order from disorder mechanism generates a gap $ \Delta_B $ to the spin wave spectrum at $\alpha=\alpha_{\rm in}^0=0.3661\pi$.
    (c) As $ \alpha $ decreases further, the Y-x state supports the IC- magnons at $ ( 0, k^{0}_y ) $.
    (d) The C-IC transition due to the condensations of the IC- magnons
        at $\alpha=\alpha_{\rm in}=0.3552\pi$ with the onset in-commensurate order $ q_{ic}=\pm (\Delta_B/u)^{1/2} \sim 0.18 \pi $ and
        the dynamic exponent $ (z_x=1, z_y=1 ) $ as shown in the inset.  }
\label{cictransition}
\end{figure}

%     whose evolution is shown in Fig.\ref{cictransition}
%    ( because $ B \ll A $, so we ignore its small contribution to $ \omega_q $ ).
Because the spectrum along $ q_x $ is non-critical, so one can just put $ q_x=0 $ in Eq.\ref{gapspectrum}:
 \begin{align}
	\omega_{-}(q_x=0, q_y)=\sqrt{ \Delta^2_B+v_y^2 q_y^2 + u^{2} q_y^4 + \cdots }
\label{gap}
\end{align}
    where $ v^{2}_y= a ( \alpha -\alpha^{0}_{in}  ) $ changes sign at $ \alpha = \alpha^{0}_{in} \sim 0.3611 \pi $ ( Fig.4a ).
    Its evolution is shown in Fig.\ref{cictransition}.
%    ( because $ B \ll A $ as shown in Fig.3b, so we ignore its small contribution to $ \omega_q $ ).

    From the gap vanishing condition \cite{loff} at the IC wave-vectors $ q_{ic}= \pm (\Delta_B/u)^{1/2} $, one can see
    the quantum Lifshitz transition is shifted to $ \alpha_{ic} = \alpha^{0}_{in}- 2 u \Delta_B/a  $.
    Plugging in the values of $ \Delta_B $ and $ u $,
    we find  the onset orbital order wavevector $ q_{ic} \sim 0.18 \pi $ (  Fig.\ref{cictransition}d ) and the shift is so small that $ \alpha_{ic} \sim 0.3526 \pi $
    remains larger than $\alpha_{33} \sim 0.3402 \pi $ ( to be defined in the next section )  as shown in Fig.\ref{phasedia}.
    So there must be an In-commensurate phase intervening between the Y-x state and the  $ 3 \times 3 $ state
    when $ \alpha_{33} < \alpha < \alpha_{ic} $ in Fig.1.

    The transition from the Y-x to the In-commensurate phase is a quantum Lifshitz transition with the dynamic exponent $ z_x=z_y=1 $ ( Fig.4d)
    instead of the one with $ ( z_x=1, z_y=2 ) $ at the LSW order in Fig.4a. The IC phase has the
    4 orbital order wave-vectors $ ( 0, \pm ( \pi-q^{0}_y ) ) $ and $ ( \pi, \pm ( \pi-q^{0}_y ) ) $
    with $ q^{0}_{y} \geq  q_{ic} $.
The spin structure of this IC phase will be determined in the following from the effective action Eq.\ref{Yxp}.
It is found to be an in-commensurate coplanar phase in the XY plane with the in-commensurate momentum along the $ k_y $ direction
%with a non-vanishing Skyrmion density $ \vec{S}_i \cdot (\vec{S}_j \times \vec{S}_k ) \neq 0 $
which we name as IC-XY-y phase \cite{spinICXYy}. The Y-x state has the C-$ C_{\pi} $ magnons when $ \alpha^{0}_{in} < \alpha < \pi/2 $,
    the IC- magnons at  the two  minima $ ( 0, \pm k^{0}_y) $ with  $  \pi- q_{ic} < k^{0}_y < \pi $
    when $ \alpha_{in} < \alpha <  \alpha^{0}_{in} $ as shown in Fig.4c.

 Now we construct a GL action in terms of the pseudo-Goldstone mode $ \phi $ to describe the quantum Lifshitz transition.
 This is a symmetry based phenomenological theory which is independent of the  $ 1/S $ expansion Eq.\ref{swcubic}.
 However, the phenomenological parameters in the effective GL action can be evaluated by the $1/S $ spin-wave expansion
 which is a microscopic calculation. Of course, the microscopic calculation in the previous sections can guide us to
 construct the phenomenological effective GL action  consistent with all the symmetries of
 the microscopic Hamiltonian Eq.2

 Inside the Y-x phase along the diagonal line $ \alpha=\beta $,
 after integrating out the massive conjugate variable $ \theta- \pi/2 $, adding the effective potential Eq.\ref{B2} generated  from
 the OFQD mechanism, we reach
 the following effective GL action in the continuum limit:
\begin{eqnarray}
{\cal L}_{Y-x}[ \phi ] & = & \frac{1}{2A} ( \partial_\tau \phi )^2  + v^2_x ( \partial_x \phi)^2
    + v^2_y ( \partial_y \phi)^2 + u^2 ( \partial^2_y \phi)^2  \nonumber  \\
    & + & \frac{1}{2} B \phi^2 + \kappa \phi^4 + \cdots
\label{Yx}
\end{eqnarray}
 where all the phenomenological parameters can be evaluated by the microscopic $ 1/S $ expansion.
 For example,  $ A $ is from a classical contribution, $ B $ and $ \kappa $ are from the OFQD,
 all were evaluated in Sec.III and shown in Fig.3b.

  Note that despite the $ [C_4 \times C_4]_D $ symmetry of the Hamiltonian along the diagonal SOC line $ \alpha= \beta $,
  the Y-x state breaks this symmetry, so the effective action Eq.\ref{Yx} inside the Y-x state also breaks the $ [C_4 \times C_4]_D $ symmetry.

   As shown in Sec.III-A, when moving away from the Abelian point $ \alpha=\beta=\pi/2 $, the coefficient $ B $ increases,
   the coefficient $  v^2_y= a ( \alpha - \alpha^0_{in}  ) $ changes sign at $ \alpha= \alpha^0_{in}\sim 0.3611 \pi $.
   However, due to the gap term $ \frac{1}{2} B \phi^2 $ generated from the OFQD mechanism,
   there is a quantum Lifshitz transition at the two In-commensurate wavevectors at $ \pm q_{ic} $ where the gap vanishes.
   It is physically more transparent to re-write Eq.\ref{Yx} in the momentum space:
\begin{eqnarray}
{\cal L}[ \phi ]_{Y-x,D} & = &  \phi(- \omega_n, -q_x, -q_y)
[\omega^2_{n}/A +  v^2_x q^2_x + u^2 ( q^2_y - q^2_{ic} )^2          \nonumber  \\
 & +  & \Delta ] \phi( \omega_n, q_x, q_y) + \kappa \phi^4 + \cdots
\label{Yxp}
\end{eqnarray}
 where $ -\pi/2 < q_x, q_y < \pi/2 $ is in the Reduced Brillouin Zone (RBZ) and
 $ \Delta= \Delta^2_B- \frac{a^2}{4u^2} ( \alpha-\alpha^0_{in} )^2 $ is the tuning parameter of the transition.

  The spin can be expressed in terms of the order parameter $ \phi $  when using the shift $ \phi \rightarrow \phi+ \pi/2 $ in Eq.\ref{2times2}.
\begin{equation}
  \mathbf{S}_i=
(- (-1)^{i_y}\sin\phi,(-1)^{i_x}\cos \phi,0 )
\label{2times2order}
\end{equation}
  So we conclude that when $ \Delta >0 $, then $ \langle \phi \rangle =0  $ in Eq.\ref{2times2order} shows it is inside the Y-x phase,
  when $ \Delta < 0 $,  then
\begin{equation}
   \langle \phi \rangle = P_0 \cos ( q_{ic} y + \phi_0 )
\end{equation}
  which has a modulation wavevector $ q_{ic} $. $ P_0, \phi_0 $ need to be fixed by the 4th order term.
  Substituting it into Eq.\ref{2times2order} shows that the system is in the IC-XY-y phase.
  The $ \theta $ has been fixed to be at its classical value $ \theta_0=\pi/2 $.
  Its quantum fluctuation $ S(\theta- \theta_0) $ plays the conjugate variable to the pseudo-Goldstone mode $ \phi $
  ( see a similar Eq.\ref{qpi} in the 2d quantization of the 1d FK model. ).

It is easy to see that when scaling to the two bosonic "Dirac" points $ (q_x=0, q_y= \pm q_{ic} ) $, the dynamic exponent is
$ z_x=1, z_y=1 $.  So the upper critical dimension is $ d=3 $. Obviously, the $ \kappa $ term is relevant at the two bosonic Dirac points.
By using $ 1/N $ expansion or $ 4- \epsilon $ RG method with $ \epsilon =1 $, one can determine the QC scaling
functions of the out of time ordered spin correlation function Eq.\ref{spinotoc}.

   Note that although the IC-XY-y phase breaks the crystal translational symmetry along the $ x $ axis
   only to two sites per unit cell, it completely breaks the crystal translational symmetry along the $ y $ axis.
   It is infinitely degenerate, but discrete and countable. So its excitation spectrum should still have a gap.
   Because the crystal momentum $ k_y $ is not a good quantum number anymore, so there maybe dis-commensurations or domain walls along the $ y $ axis.
   It remains interesting to determine the distributions of these dis-commensurations and their repulsive interactions in the
   IC-XY-y phase. Its finite temperature properties will be discussed in Sec.VIII.

 Similarly, starting form the X-y phase, one can reach the IC-XY-x phase with
the 4 orbital order wave-vectors $ ( \pm ( \pi-q^{0}_x ), 0 ) $ and $ ( \pm ( \pi-q^{0}_x ), \pi ) $.
So along the diagonal line $ \alpha_{ic}  < \alpha <  \pi/2 $ (  $ \alpha_{33} < \alpha < \alpha_{ic} $ ),
there must be co-existence of the Y-x and X-y ( IC-XY-y and IC-XY-x ) phases with any ratios ( Fig.\ref{phasedia} and its inset ).
This physical picture will be substantiated further from the anisotropic line $ ( \alpha=\pi/2, \beta ) $
approached from the right. Indeed, as to be shown in Sec.V-B,
there is also a transition from the Y-x state on the right to the IC-XY-y
due to the condensations of the IC- magnons with $ \pi- \pi/3 < k^{0}_y < \pi -0.18 \pi $
( or equivalently $   0.18\pi < q^{0}_y <  \pi/3  $ ) in Fig.1.

%   its excitation spectrum $ E( k_x, y ) $  and the spin-spin correlation
%   functions $ \langle S_{\alpha} ( k_x, y, \omega ) S_{\beta} ( -k_x, y^{\prime}, -\omega ) \rangle $.

\subsection{ Contrast to the quantum Lifshitz transition in quantum dimer model }

%  It may also be instructive to compare all the phases in Fig.1 with those in the Quantum dimer model (QDM) \cite{dimer,dimer1}
%  in a bipartite lattice. Of course, our RFHM is a quantum spin model subject to a SOC in a bipartite lattice,
%  while the QDM in a bipartite lattice is not a quantum spin model.

It is well known that  Rokhsha-Kivelson Quantum Dimer (QD) model \cite{dimer,dimer1,dimer2} in a bi-partite lattice such as
a square or a honeycomb lattice can be described by a
  quantum Lifshitz action with the dynamic exponent $ z=2 $.
   In the height $ \chi $  representation, the $ 2 + 1 $ QD model can be written as:
\begin{eqnarray}
{\cal L}_{QD} & = & \kappa ( \partial_\tau \chi )^2 + \rho_s ( \nabla \chi)^2 + K ( \nabla^2 \chi)^2
       +  u ( \nabla \chi)^4     \nonumber  \\
       & + & \lambda \cos 2 \pi\chi   + \cdots
\label{qd}
\end{eqnarray}
   At the QCP $  \rho_s=0 $, there is a line of fixed point controlled by the parameter $ K $ with the dynamic exponent $ z=2 $.
   The monopole term $ \lambda $ is irrelevant in some ranges of $ K $, but $ u $ is marginally irrelevant.
   At $ \rho_s > 0 $, it is in the columnar  VBS, while in $ \rho_s < 0 $, it is in the staggered ( tilted ) VBS phase.
   The authors in \cite{dimer1,dimer2} pointed out that there could be many commensurate
   and in-commensurate ( quasi-periodicity  ) VBS phases intervening between the column VBS and the staggered  VBS phase. They may form
   fractals and in-complete devil staircases between the two ending commensurate phases.

  There are several crucial differences between the QLCP in the QD in Eq.\ref{qd}  and the QLCP in Eq.\ref{Yx},
 (1) the dynamic exponent is $ z=2 $, here it is $ z_x=z_y=1 $.
 (2) the physical quantity in the QD is the dimer ( or VBS) density
  $  n \sim e^{i 2 \pi \chi } + h.c. $
  which can be compared to the quantum spin Eq.\ref{2times2order}.
 (3) The two phases on both side of the QLCP are very much different.

  In fact, the Abelian line $ ( 0< \alpha < \pi/2, \beta=0) $ in Fig.1 resemble the Rokhsar-Kivelson (RK) point in the QDM. Indeed, as said in
  the introduction, all the phases along the  Abelian line can be transformed into a FM state in the rotated $ \tilde{SU}(2) $ basis  $ \tilde{\mathbf{S}}_n= R(\hat{x}, 2 \alpha n )\mathbf{S}_n $. Any small $ \beta >0 $ turns it into an in-complete devil staircase.
  It was known that the wavefunction at the RK point can also be written as a FM state along the $ \hat{x} $ direction. So the C and IC magnetic phases in the in-complete devil staircases at a small $ \beta $
  near the Abelian line in Fig.1 ( see Sec.VII ) can be contrasted to those C- and IC- VBS near the RK point \cite{dimer1,dimer2}
  which also form in-complete devil staircases.
  We expect they may melt into quantum spin liquids (QSLs) in a frustrated lattice such as a
  triangular lattice \cite{dimer0,dimer00}.

%    This is one of the central results achieved in this paper. It will be confirmed further
%    when approaching the $ 3 \times 3 $ SkX state from below the diagonal line $ \beta < \alpha $.

\section{  Non-coplanar $ 3 \times 3 $ Skyrmion Crystal phase and Co-planar spiral phases along the diagonal
line $ \alpha=\beta $ away from the $ \alpha=\beta=\pi/2 $ Abelian point. }

In this section, we will first discuss the $ 3 \times 3 $ non-coplanar Skyrmion Crystal phase, then its connection to its
two neighbouring phases: IC-XY-y and Y-x phase. We will also study the transition from the Y-x to the IC-XY-y
driven from the right solvable line $ (\alpha=\pi/2, \beta=0 ) $, also that from the Y-x to the $ 3 \times 3 $ through the
 Multi-critical point (M) in the contour line $ (0, \pm 2 \pi/3 ) $.  This approach from the right solvable line confirm and strengthen the results achieved
in Sec.III and IV along the diagonal line. It also provides additional insights on all the phases around the Multi-critical point (M) in Fig.1.
Finally we explore the other co-planar states at $ \alpha=\beta=\pi/N $ shown in Fig.1 whose details were shown in Fig.5

\subsection{ $ 3 \times 3 $ non-coplanar Skyrmion Crystal phase (SkX): the hub phase in Fig.\ref{trifeats}.  }
%and the in-commensurate SkX phase intervening   between the Y-x phase and the  $ 3 \times 3 $ SkX phase. }
% As discussed in the previous sections, near $ \alpha=\beta=\pi/2 $, we take $ 2 \times 2 $ ansatz.
Near $ \alpha=\beta=\pi/3 $, it is natural to take a $3\times3$ ansatz: $S_{(i_x,i_y)}=S_{(i_x+3m,i_y+3n)}$ with $m,n\in \mathbb{Z}$.
We  estimate its classical ground-state energy by minimizing
$ E_{3\times3}(\{\phi_i,\theta_i\}_{0\leq i\leq 9})$ over its 18 variables.
Along  the diagonal line ($\alpha=\beta$), as long as $\alpha$ is not too small, the minimization of $E_{3\times3}$
always leads to a $ [C_4 \times C_4]_D $ symmetric $ 3 \times 3 $ SkX state shown in Fig.\ref{allphases}.
This is in sharp contrast to the case near $ \alpha=\beta=\pi/2 $ where the classical analysis only leads to the degenerate
family of $ 2 \times 2 $ vortex states shown in Fig.\ref{allphases}.
A quantum " order from disorder " analysis in Sec.III-A is needed to show the $ 2 \times 2 $ vortex state phase separates into
any mixtures of the Y-x state and X-y state along the diagnose line.

Comparing the classical ground energy of the $ 3 \times 3 $ SkX with that of the  Y-x state
$E_{Y-x}=-2J\sin^2\alpha$ leads to a putative first order transition between the two states at
$\alpha_{33} \approx0.3402 \pi $ which is smaller than  $ \alpha_{ic} \sim 0.3526 \pi $.
So a putative direct first order transition between the Y-x state and the $ 3 \times 3 $ SkX splits into
2 second order quantum Lifshitz transitions with $ z=1 $ with the IC-XY-y phase intervening between them.
In fact, $ \alpha_{33} $ also shifts to a smaller value due to the intervening of the IC-XY-y phase, but
for simplicity, we still use the same symbol.  The point $  \alpha= \alpha_{33} $ in Fig.\ref{phasedia} is a bi-critical point
which means two 2nd order transition lines meet one 1st order transition line.
%Similarly, by LSW, we can determine the excitations spectra above the $ 3 \times 3 $ SkX.
%We expect the transition from the IC-SKY/Y-x state to the $ 3 \times 3 $ SkX is also a quantum Lifshitz transition with $ z=1 $.

  We will determine the 9 ordering wave-vectors of the $ 3 \times 3 $ SkX which can be directly detected by Bragg spectroscopies in cold atoms
  or neutron scattering in materials.
  For simplicity, we only explicitly determine the spin-orbital configuration of the $ 3 \times 3 $ SkX along the diagonal line
  in both real space and momentum space.
  In fact, it is the only state respecting the $ [C_4 \times C_4]_D $ symmetry along the diagonal line.
  However, as shown in Fig.3, it remains stable in a regime around the diagonal line. Because it always have the same symmetry,
  so should have the same 9 ordering wave-vectors as the one along the diagonal line.

%{\sl 1. The structure of the $ 3 \times 3 $ SkX in real space }

   The $ 3 \times 3 $ SkX along the diagonal line shown in Fig.\ref{trifeats} and \ref{diagstair} respects the $ [C_4 \times C_4]_D $ symmetry.
   The spin in the center is along $ z $ axis. Due to the $ [C_4 \times C_4]_D $ symmetry, there are only two pairs of
   independent angles $ (\eta_1, \xi_1) $ and $ (\eta_2, \xi_2) $ characterizing the set $ ( S_1, S_3, S_7, S_9 ) $  and
   $ ( S_2, S_4, S_6, S_8 ) $ in the classical state respectively:
\begin{eqnarray}
	S_5 & = & (0,0,1),  \nonumber   \\
	S_1 & = & (-\cos\xi_1\sin\eta_1,-\sin\xi_1\sin\eta_1,\cos\eta_1),   \nonumber   \\
	S_3 &=&(-\sin\xi_1\sin\eta_1,\cos\xi_1\sin\eta_1,\cos\eta_1),   \nonumber   \\
	S_7 & = & (\sin\xi_1\sin\eta_1,-\cos\xi_1\sin\eta_1,\cos\eta_1),   \nonumber   \\
	S_9 &= &(\cos\xi_1\sin\eta_1,\sin\xi_1\sin\eta_1,\cos\eta_1),  \nonumber   \\
	S_2 & = & (-\cos\xi_2\sin\eta_2,-\sin\xi_2\sin\eta_2,\cos\eta_2),  \nonumber   \\
    S_4 & = & (-\sin\xi_2\sin\eta_2,\cos\xi_2\sin\eta_2,\cos\eta_2),  \nonumber   \\
	S_6 & = & (\sin\xi_2\sin\eta_2,-\cos\xi_2\sin\eta_2,\cos\eta_2),	 \nonumber   \\
    S_8 &= & (\cos\xi_2\sin\eta_2,\sin\xi_2\sin\eta_2,\cos\eta_2),
\label{xieta}
\end{eqnarray}

For $\alpha=\beta=\pi/3$, the ground state energy per site is $E_{\rm GS}=-1.53608 J $
and the two pairs of angles are $ (\eta_1, \xi_1)= ( 0.59 \pi, \pi/4 ) $ and $ (\eta_2, \xi_2)= ( 0.49 \pi, 0 ) $ leading to the total spin:
\begin{align}
	S_{\rm tot}=\sum_i S_i=(0,0,0.004088)
\end{align}
   which has exact vanishing $ S_x, S_y $ components, but still a small non-vanishing $ S_z $ component justifying the name SkX.

%{\sl 2. The structure the $ 3 \times 3 $ SkX in momentum space }

For general $ 3 \times 3 $ SkX, we can always expand it in terms of its 9 ordering wavevectors $ \frac{2\pi}{3}(m, n) $
\begin{align}
	S^{\alpha}(x,y)=\sum_{m,n=0,1,2}\rho^{\alpha}_{mn}e^{i\phi_{mn}}e^{i\frac{2\pi}{3}(m x+ n y)}
\label{mn}
\end{align}
   where $ \alpha=X,Y,Z$ are the spin's three components. Here, we take the $ Z $ component as an illustration.

Instead of using the real spin-orbital configuration in Eqn.\ref{xieta},
to make the expression simple,
we just use the following simplest spin-orbital configuration which has the same symmetry as  Eqn.\ref{xieta}:
the spin at the center has $S^z=1$ and all other spins in Fig.\ref{diagstair}b in the XY plane, so
having no $S^z$ components: $ S^z(2,2)=1, S^z(i,j)=0, \textmd{ for } i\neq2 \textmd{ or } j\neq 2 $ which is more like a meron.

 The components in Eqn.\ref{mn} are $ \rho_{m,n}=1/9 $ and
\begin{align}
	\phi=
	\begin{pmatrix}
	  \phi_{11} &\phi_{12} &\phi_{13}\\
	  \phi_{21} &\phi_{22} &\phi_{23}\\
	  \phi_{31} &\phi_{32} &\phi_{33}\\
	\end{pmatrix}
	=
	\frac{2\pi}{3}
	\begin{pmatrix}
	  -1 &0  &1\\
	   0 &1 &-1\\
	   1 &-1 &0\\
	\end{pmatrix}
\end{align}
  which leads to a very simple expression:
\begin{align}
	S^z(x,y)=\frac{1}{9}\sum_{m,n=0,1,2} e^{i\frac{2\pi}{3}[m (x+1)+ n (y+1)]}
\end{align}
   The $ S^{x} $ and $ S^{y} $ components can be similarly constructed.
 The real spin configuration in Eqn.\ref{xieta} can be similarly computed, but with a more complicated expression.
 Of course, the 9 ordering wavevectors stay the same.

 Note that the $ 2 \times 2 $ vortex state in Fig.\ref{allphases} has only two ordering wavevectors $ ( 0, \pi) $ and $ ( \pi,0) $,
 the other two $ (0,0) $ and $ ( \pi,\pi) $ are excluded due to the fact the  $ 2 \times 2 $ vortex state is a co-planar state
 instead of a non-coplanar one.

 In the Fig.1, from the right solvable line $ (\alpha=\pi/2, \beta=0 ) $,
 there is a direct transition from the Y-x state to the $ 3 \times 3 $ SkX state at the M point through the contour $ ( 0 , \pm 2\pi/3 ) $.
 The transition indicates the orbital orderings $  ( 0 , \pm 2\pi/3 ) $ and
   $ ( \pi,0 ) + ( 0 , \pm 2\pi/3 )= ( \pi, \pm 2\pi/3 ) $.
  The former does belong to the 9 ordering wavevectors of  the $ 3 \times 3 $ SkX state,
 but the latter does not. Similarly, approaching from the X-y  state leads to $ ( \pm 2\pi/3, 0 ) $ ordering wavevectors.

 So far, the discussions are classical. By using the spin wave calculations,
 one can incorporate the quantum fluctuations and  find its excitation spectra.
 Due to the $ 3 \times 3 $ structure, there should be 9 branches, the lowest of which should take
 the same form as Eq.\ref{gapspectrum}. Then by using the combinations of
 the spin wave analysis and the coherent spin path integral method developed in appendix C-F,
 it is possible to construct a GL action similar to Eq.\ref{Yx}
 to describe the quantum Lifshitz transition near the Bi-critical point $ \alpha_{33} $ in Fig.1 from the $ 3 \times 3 $ SkX to the IC-XY-y along the diagonal line $ \alpha=\beta^- $ and IC-SkX-y away from it.

\begin{figure}[!htb]
\includegraphics[width=4.5cm]{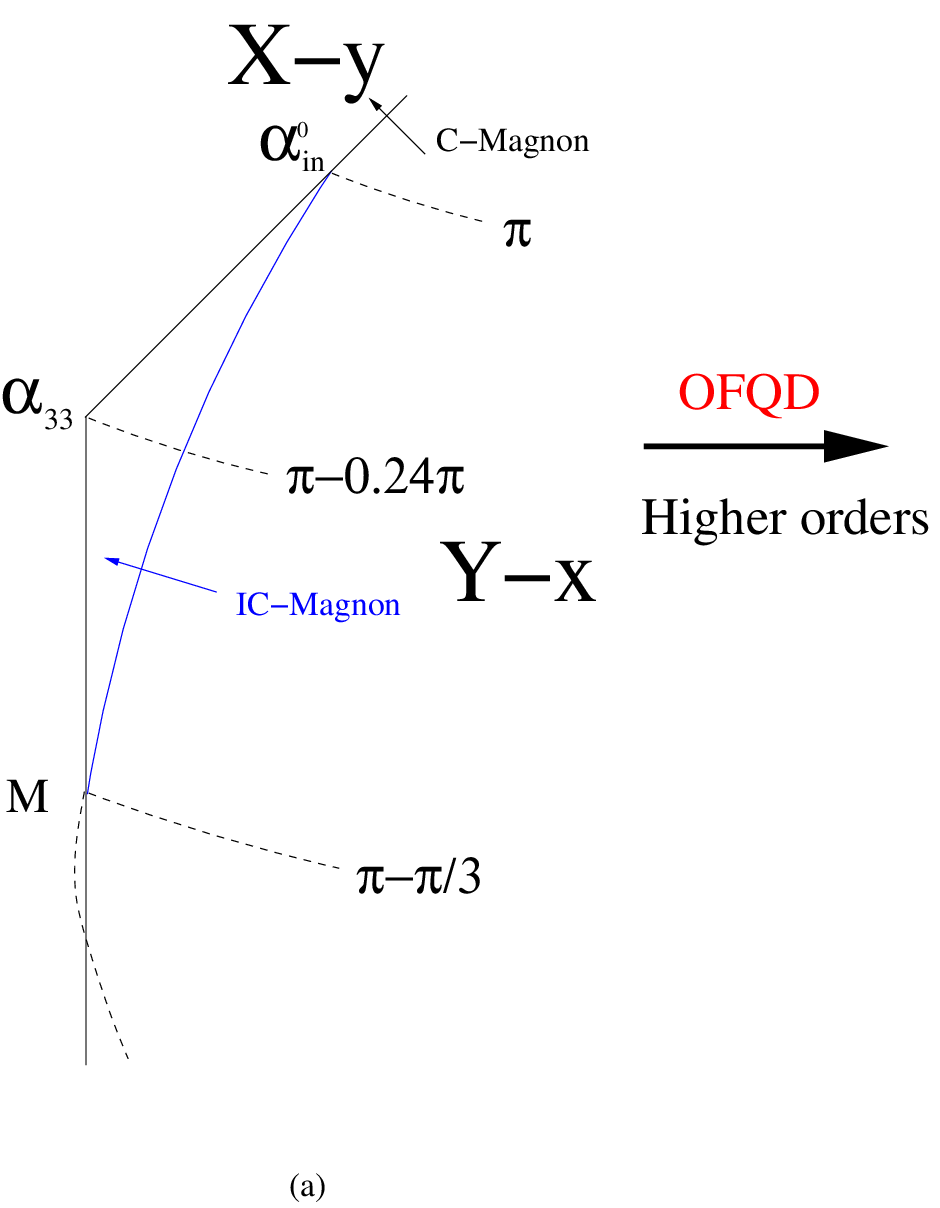}
\hspace{0.4cm}
\includegraphics[width=3.5cm]{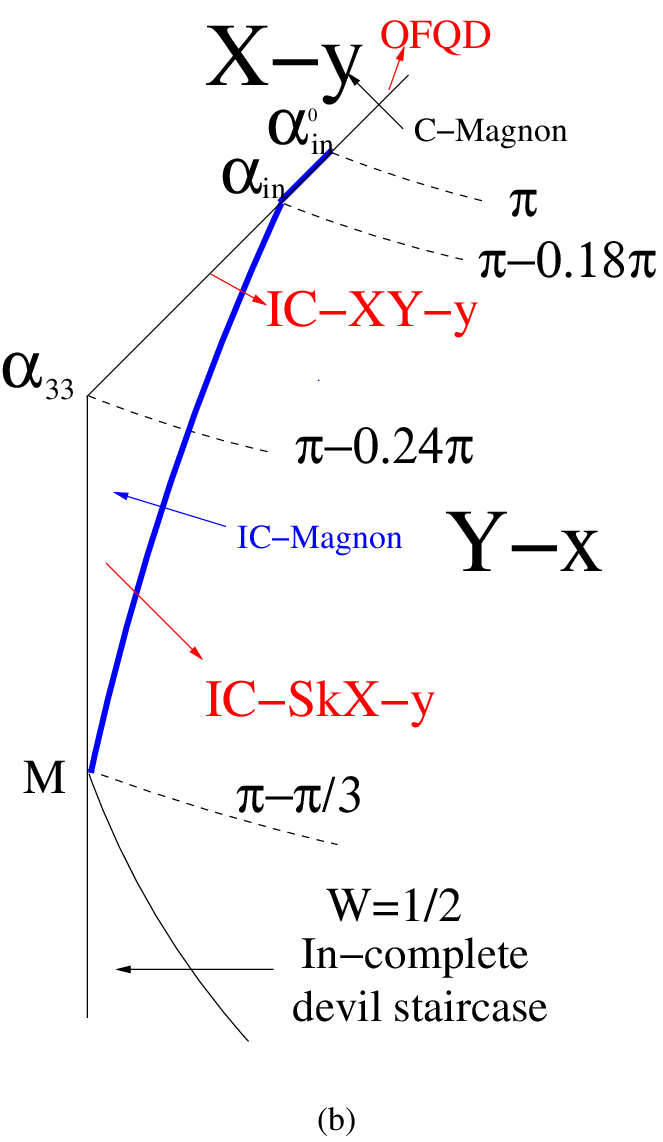}
%\includegraphics[width=4.25cm]{Fig2-3.eps}
%\hspace{-0.5cm}
%\includegraphics[width=4.0cm]{globalphaseM.eps}
%\hspace{0.2cm}
%\includegraphics[width=4.5cm]{contourline.eps}
%\hspace{0.2cm}
%\includegraphics[width=6.5cm]{contours2.eps}
%\hspace{0.2cm}
%\includegraphics[width=4.0cm]{Min0.eps}
\caption{ The fate of the three segments of the magnon condensation boundary Eq.\ref{segments}  from the Y-x phase on the right.
The quantum Lifshitz action from the Y-x phase on the right to the X-y phase
by C$_\pi$ magnons and to the IC-SkX phase by the IC-magnons. It match the results achieved along the diagonal line in Sec.III and IV.
(a) At the quadratic ( or Gaussian ) level. There is a putative 2nd order transition
from the Y-x to the X-y driven by the C$_\pi$ magnons and to the IC-SkX by the IC-  magnons $  \pi-\pi/3 < Q < \pi $ respectively.
The lower part below $ M $ corresponds to the the putative 2nd order transition from the Y-x phase driven by condensations of
IC- magnons $ 0 < Q < \pi-\pi/3 $ and C$_0$ magnons respectively.
(b)  The leading higher order effects  along the diagonal line are due to the OFQD
which can be evaluated in the $ 1/S $ expansion in Sec.III and IV.
They shift the magnon condensation boundary in (a).
The IC magnons from the right are confined to $ \pi-\pi/3 < Q < \pi -q_{ic}, q_{ic}=0.18 \pi $.
While the $ \pi -q_{ic} < Q < \pi $  IC -magnons remains gapped and placed along the diagonal line
$ \alpha_{in} < \alpha < \alpha^0_{in} $. The 1st order transition line from the Y-x to the IC-SkX-y ends at the
2nd order transition point  $ \alpha= \alpha_{in} $ which is a bi-critical point.
The putative 2nd order transition line below the M point in (a) is changed to the $ W=1/2 $ plateau of the in-complete devil staircase.
So the organization principle changes from the segment of an in-complete devil staircase to a 1st order QPT at the M point.
See also the inset in Fig.1. }
\label{phasesarc}
\end{figure}

\subsection{ The quantum Lifshitz transition from the Y-x state on the right $ \beta < \alpha $  to the IC-SkX-y state
 through the condensations of IC- magnons  }
  As shown in the inset of Fig.\ref{phasedia}, approaching from the right  solvable line $ (\alpha=\pi/2, \beta=0 ) $ in the Y-x phase,
  the crossing point between the $ ( 0, \pm 2\pi/3) $ counter line of the Y-x phase and
  the C-IC condensation boundary ( dashed line ) just hits the corner of the $ 3 \times 3 $ SkX crystal
  at the multicritical M  point located at $ ( \alpha_M, \beta_M )=(0.33952\pi, 0.31284 \pi) $.
  So there should be a quantum Lifshitz phase transition Y-x to the IC-SkX-y
  due to the condensations of IC- magnons with the ordering wavevectors $ \pi-\pi/3 < k^{0}_{y} < \pi- 0.18 \pi $.

  In the appendix F, we construct an effective GL action  to describe such a transition driven by
  the condensation of the IC-magnons from the right in the Y-x lobe in Fig.1 ( dropping $ \delta $ for the notational convenience ):
\begin{eqnarray}
	{\cal L}[\eta ]_{Y-x,IC} & = & \eta(-k,-i\omega_n) [ \omega_n^2  + v^2_x k^2_x + v^2_y (k^2_y-Q^2)^2      \nonumber   \\
     & + & \Delta^2 ] \eta(k,i\omega_n)  + \lambda \eta^3 + \kappa \eta^4 + \cdots
\label{etaeta3}
\end{eqnarray}
 where  $ 0 < k_x < \pi, -\pi < k_y < \pi $ is in the BZ and
 $ \pi-\pi/3 < Q < \pi-q_{ic} $. It takes the same form as Eq.\ref{Yxp} achieved along the diagonal line in the polar coordinate $ (\theta,\phi ) $, except the existence of the cubic term away from the diagonal line.
 It is instructive and important to reach this form from the right in a different  polar coordinate $ (\eta,\xi ) $.
 It can be taken as generalizing the non-linear Sigma model \cite{scaling} such as Eq.\ref{nafm}  to  the presence of SOC
  to study the transition from the Y-x to the IC-SkX
  state due to the IC- magnon condensation in the momentum range $ \pi-\pi/3 < k^{0}_{y} < \pi-q_{ic} $.

   When getting into the IC-SkX-y phase where $ \Delta^2 < 0 $ , its mean field solution can be written as:
\begin{eqnarray}
  S \eta_A &=& S \eta_B = P_0 \cos ( Q y + \phi_0 ),~~~~~  \nonumber   \\
  S \xi_A  &=& - S \xi_B= - \frac{ P_0 \cos \theta_0 }{ 1+ \sin \theta_0 }  \sin ( Q y + \phi_0 )
\label{SkXAB}
\end{eqnarray}
   which shows there is always a phase $ \pi/2 $ difference between $ \eta $ and $ \xi $.

   Plugging Eq.\ref{SkXAB} into Eq.\ref{spinYx} displays the spin-orbital structure of the IC-SkX-y phase shown in Fig.1.
   To the linear order, we obtain explicitly the spin-orbital structure of the IC-SkX-y phase:
\begin{eqnarray}
  S^x_A & = & S^x_B= P_0 \cos ( Q y + \phi_0 ),~~~~~  \nonumber   \\
  S^z_A & = & S^z_B  = - \frac{ P_0 \cos \theta_0 }{ 1+ \sin \theta_0 }  \sin ( Q y + \phi_0 )
\label{SkXABlinear}
\end{eqnarray}
   where $ P_0, \phi_0 $ need to be fixed
   by the cubic or 4th order term. It gives the two transverse spin components in the two sublattices A and B.
   Their maximum magnitude ratio may also be fixed by the in-commensurate momentum $ Q $ as shown in Eq.\ref{Qvalue}.
   The longitudinal $ Y $  component can be taken as  $ S^{y}=(-1)^{x} $ to the linear order.
   Obviously, it has a non-vanishing Skyrmion density $ \vec{S}_i \cdot (\vec{S}_j \times \vec{S}_k ) \neq 0 $.
   It is constructive to compare with the IC-SkY-$\phi $ induced by a Zeeman field Eq.\ref{icskxh}
   where the $ U(1)_{soc} $ dictates the two transverse components must have the same magnitude.
   Along the diagonal line $ \alpha=\beta $, the IC-SkX-y phase reduces to the IC-XY-y phase
   where  $ S^z_A= S^z_B=0 $. If setting $ Q= \pi + q_{ic} $, it is identical to the IC-XY-y phase listed in Eq.\ref{2times2order}.

\begin{figure}[!htb]
\centering
\includegraphics[width=0.9\linewidth]{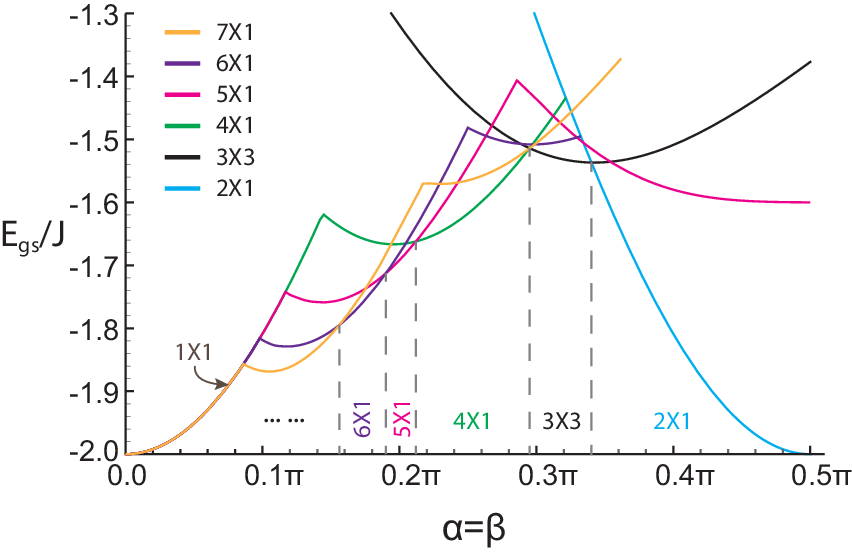}
%\quad
%\includegraphics[width=0.25\textwidth]{33SkX}
    \caption{ Classical Energy competition from different spiral C-phases with $ W=n/N $ along the diagonal line $ \alpha=\beta $.
    The only collinear state is the $ 2 \times 1 $ Y-x state.
    The only non-coplanar state is the $3\times3$ SkX phase which has lower energy than the $3 \times 1$ coplanar state.
    As shown in Fig.1, there is an IC-XY-y phase with a finite measure intervening between the $3 \times 3 $ SkX and the Y-x phase.
    All the other $  N \times 1, N \geq 4 $ states are co-planar states in the $ YZ $ plane.
    All the energy level crossings can be read from the figure and shown along the diagonal line in Fig.1.
%    Inset: the spin-configuration of the $ 3 \times 3 $ SkX in real space.
    Near the Abelian line $ \beta \ll 1 $, the $ N \times 1 $ ansatz means the optimized energy from the $N\times1$ coplanar state
    $E_{\rm gs}=\min\limits_{\{\xi_n\}} E_{N\times1}$ defined in the Frenkel-Kontorova Model Eqn.\ref{fk} and shown in Fig.2.
    In fact, there are always some higher order IC spiral phases intervening between the two principle
    $  N \times 1, N \geq 4 $ and $  (N+1) \times 1, N \geq 4 $ states. For example, as explained in Sec.VII-B, there is
    a $ W=2/7 $ phase intervening between $ W=1/4 $ and the $ 3\times 3 $ SkX phase. This suggests there is a complete devil staircase
    along the diagonal line. See Fig.\ref{finiteT}.  }
\label{diagstair}
\end{figure}

   Recall that Eq.\ref{Yxp} holds along the diagonal line $ \alpha=\beta $ where the cubic term
   vanishes  $ \lambda=0 $. There is also a spurious symmetry, the OFQD analysis generates the effective potential Eq.\ref{Bkappa}
   which leads to the specific value of the quartic term $ \kappa $.
   It describes a transition from the Y-x phase to the co-planar IC-XY-y phase with the dynamic exponent $ z_x=z_y=1 $.
   Here it is away from the diagonal line and approaching from the right, so the cubic term exists in general. However, if it is close to the diagonal line as shown
   in Fig.1, the cubic term is small. So it describes a transition from the Y-x phase to the non-coplanar IC-SkX-y phase which could be a weakly first-order transition, so it maybe approximately taken as  a second order transition with the dynamic exponent $ z_x=z_y=1 $.
%   If it turns out to be a second order transition, then it has the dynamic exponent $ z_x=z_y=1 $.
   Combing the results achieved along the diagonal line in Sec.IV and those achieved here from the right, we expect that the IC-SkX-y phase
   away from the diagonal line   reduces to the IC-XY-y phase along it.
   The former is non-coplanar with non-vanishing Skyrmion density, the latter is coplanar with vanishing Skyrmion density,
   but both phases have the same symmetry breaking patterns ( Fig.\ref{phasesarc} ).
   It is remarkable that one reach the same physics from the right and along the diagonal line.

  Putting $  \alpha=\beta= \alpha_{33} $ into the formula for the constant contour at $ (0, k^{0}_{y} ) $ listed in the appendix A,
  one gets $ k^{0}_{y} \sim \pi-0.24 \pi $. So one can see that along the diagonal line
  $  \alpha_{33} < \alpha < \alpha_{ic} $, the ordering wavevector of the IC-XY-y is $ 0.18 \pi < q^{0}_{y} < 0.24 \pi  $.
  While the transition from the $ 3 \times 3 $ SkX to the IC-SkX-y on the right is through the condensations of IC- magnons
  with  $ 0.24 \pi < q^{0}_{y} < \pi/3  $.
  Of course, the coplanar IC-XY-x phase with the ordering wavevectors $ ( \pm k^{0}_{x}, 0 ) $ and $ ( \pm k^{0}_{x}, \pi ) $
  is related to IC-XY-y by the $ [C_4 \times C_4]_D $ rotation.

\subsection{ Principle Co-planar spiral states near $ \alpha=\beta=\pi/N $ and the first order transition line
between $ \alpha=\beta^{+} $ and $ \alpha=\beta^{-} $. }
We find even at the classical level, there is a first order transition from the $ 4 \times 1 $ state to the $ 1 \times 4 $ state
along the diagonal line ( Fig.\ref{diagstair} ). While one need resorts " order from quantum disorder" mechanism in Sec.III
 to select out Y-x and X-y  state
as the quantum ground state  near  $ \alpha=\pi/2 $. This may be due to the fact that only near $ \alpha=\pi/2 $,
the Y-x and  X-y are collinear states and orthogonal to each other, while
all the other commensurate  states near $ \alpha=\pi/N, N > 2 $ are non-collinear ( but co-planar in the YZ plane )
spiral phases and not orthogonal to each other.
It turns out the $ 3 \times 3 $ SkX is the only commensurate non-coplanar state along the diagonal line ( see  Fig.\ref{diagstair}   )
which respects the $ [C_4 \times C_4]_D $ symmetry. There is a 1st order transition from it to $ 3 \times 1 $ when $ \alpha > \beta $
and $ 1 \times 3 $ when $ \alpha < \beta $ respectively.

All the other phases separate into $ N \times 1 $ and $ 1 \times N $ co-planar spiral phase in the YZ plane.
As shown in Fig.\ref{diagstair}, there are also some tiny windows of C phases at $ \alpha=\frac{\pi}{N} n $ with $ n > 1 $
squeezed between the principle series. As shown below,
this kind of higher order co-planar phases with $ n > 1 $ are also common near the Abelian line $ ( 0 < \alpha < \pi/2,  \beta=0 ) $.
%and do contribute to the $ L_c $ at small $ \beta $.

Taking $ N \rightarrow \infty $ limit, one may approach the $ \alpha=\beta=0 $ Abelian point.
It suggests some IC phase near the point. To test this prediction, we first identify a spurious $ U(1) $ family of degenerate classical state which
is a FM state within XY plane. Then by performing a LSW on this degenerate manifold, the linear term indeed vanishes,
but the spin wave spectrum always become negative.
This fact indicates the FM is always unstable, the true ground state should be some IC phases corresponding
to $ N \rightarrow \infty $ limit in the FK model. The details are given in the following section.

\subsection{ The instability of FM state near $\alpha=\beta\sim0$}
To test instability of possible ferromagnetic (FM) order near $\alpha=\beta\sim0$,
we consider a $1\times1$ ansatz with all spins point in the same direction
$\mathbf{S}_i=S(\sin\theta_0\cos\phi_0,\sin\theta_0\sin\phi_0,\cos\theta_0)$.
The $1\times1$ ansatz lead to classic energy
\begin{align}
	E_c=-2JNS^2(\cos2\alpha+\sin^2\alpha\sin^2\theta_0)
\end{align}

If $ \alpha > 0 $, the minimization of $E_c$ leads to $\theta_0=\pi/2$, but gives no constraint on $\phi_0$.
This is a spurious $ U(1) $ family of degenerate classical state which is a FM state within XY plane.
In order to calculate the spin-wave spectrum,
one can introduce Holstein-Primakoff boson $a$ as
\begin{align}
	\begin{pmatrix}
	    S_i^x\\
	    S_i^y\\
	    S_i^z\\
	\end{pmatrix}
	=R_z(\phi_0)R_y(\pi/2)
	\begin{pmatrix}
	    \sqrt{2S}(a_i+a_i^\dagger)/2\\
	    \sqrt{2S}(a_i-a_i^\dagger)/(2i)\\
	    S-a_i^\dagger a_i\\
	\end{pmatrix}
\end{align}
and the Hamiltonian can be expand in the powers of $1/\sqrt{S}$,
\begin{align}
    H=-2NJS^2\cos^2\alpha+2JSH_2+\cdots
\end{align}
After a Fourier transformation, the quadratic order $H_2$ becomes
\begin{eqnarray}
	H_2 & = & \sum_k[
		2(\gamma_0-\gamma_{1,k}-\gamma_{2,k}) a_k^\dagger a_k    \nonumber   \\
		& + & \gamma_{3,k}a_ka_{-k}+\gamma_{3,k}^*a_k^\dagger a_{-k}^\dagger]
\end{eqnarray}
where we define
\begin{align}
	\gamma_0&=\cos^2\alpha    \nonumber   \\
	\gamma_1&=\sin2\alpha(\cos\phi_0\sin k_x+\sin\phi_0\sin k_y)/2  \nonumber \\
	\gamma_2&=[(\cos2\alpha+\sin^2\alpha\sin^2\phi_0)\cos k_x      \nonumber \\
            &+(\cos2\alpha+\sin^2\alpha\cos^2\phi_0)\cos k_y]/2   \nonumber \\
	\gamma_3&=\sin^2\alpha(\sin^2\phi_0 e^{ik_x}+\cos^2\phi_0 e^{ik_y})/2
\end{align}
A Bogliubov transformation lead to
\begin{align}
	H_2=-N\cos^2\alpha+\sum_k 2\omega_k(\alpha_k^\dagger \alpha_k+1/2)
\end{align}
where $\omega_k=\sqrt{(\gamma_0-\gamma_{2,k})^2-|\gamma_{3,k}|^2}-\gamma_{1,k}$.

One can also check that $\alpha=0$ lead to the well-known result of FM spin-wave dispersion $\omega_k=1-(\cos k_x+\cos k_y)/2$.
The FM state can be along any direction $ ( \theta_0, \phi_0 ) $ due to the spin $ SU(2) $ symmetry at $ \alpha=0 $.
However, when $\alpha=\beta\neq0$, it is easy to show $\omega_{k=0}=0$ and
long-wave length limit gives a linear dispersion
\begin{equation}
	\omega_k
	=\sqrt{v_x^2 k_x^2+v_y^2 k_y^2-2v_{xy}^2 k_x k_y}
	-c_x k_x-c_y k_y
\label{XYdis}
\end{equation}
where
\begin{eqnarray}
	v_x^2&= &\sin^2\alpha[\cos2\alpha+\sin^2\alpha\cos^2\phi_0(1+\sin^2\phi_0)]/4,  \nonumber \\
	v_y^2 & = &\sin^2\alpha[\cos2\alpha+\sin^2\alpha\sin^2\phi_0(1+\cos^2\phi_0)]/4,   \nonumber \\
	v_{xy}^2&= &(\sin^4\alpha\sin^2\phi\cos^2\phi_0)/4,    \nonumber \\
	c_x &= &(\sin2\alpha\cos\phi_0)/2,\quad
	c_y=(\sin2\alpha\sin\phi_0)/2
\label{phi0}
\end{eqnarray}
The fact $v_x v_y> 2v_{xy}^2>0$ suggest the first term of Eq.\eqref{XYdis} is always non-negative when $0<\alpha=\beta<\pi/4$,
but $v_x<c_x$ and $v_y<c_y$ suggest $\omega_k$ becomes negative near $k=(0,0)$ .
The negative spin-wave spectrum suggest an instability of FM state,
thus FM state is not be a ground-state.

Although FM state can not be a ground-state,
it is still interesting to study which $\phi_0$ have lowest energy in the $1\times1$ ansatz.
We evaluate $E_2(\phi_0)=\sum_k (\omega_k-\cos^2\alpha)$ and find it has two degenerate minima at $\phi_0=0$  and $\pi/2$
which indicate X-FM and Y-FM respectively.
Of course, as shown in Eq.\ref{XYdis} and \ref{phi0} which hold for any $ \phi_0 $, neither is stable.

\section{  Co-planar spiral phases and in-commensurate phases at a small $ 0< \beta \ll \alpha <\pi/2  $ near the Abelian line }

% Now we try to understand the global phase diagram Fig.1 near the whole Abelian line at the bottom $ 0 < \alpha < \pi/2, \beta =0 $
% pre-empty of the magnon condensation transitions and the meta-stable Y-x phase.

   Now we try to understand the global phase diagram Fig.1 near the whole Abelian line at the bottom $ 0 < \alpha < \pi/2, \beta =0 $.
  We will establish the classical phase diagram by mapping its lower half
  $ \beta < \alpha=\pi/N $ to the Frenkel-Kontorowa (FK) model with $ N \times 1 $ ( stripe ) ansatz.
  We consider a $ N \times 1 $ spin-orbital structure commensurate with a lattice with $ N \times M $ lattice sites.
  We will reach the incommensurate limit by taking $ N \rightarrow \infty $ limit.
  Within a general $ N \times 1 $ ansatz, applying the local spin rotation
  $ \tilde{\mathbf{S}}_n= R(\hat{x}, 2 \alpha n )\mathbf{S}_n $  in Eqn.\ref{rhgeneral}  to get rid of the $ R $ matrix along the $ x $ bonds,
  writing the spin as a classical unit vector in the rotated basis
  $ \tilde{\mathbf{S}}_n=(\cos\tilde{\eta}_n,\sin\tilde{\xi}_n\sin\tilde{\eta}_n,\cos\tilde{\xi}_n\sin\tilde{\eta}_n ) $,
  we find that any $ \beta > 0 $ picks up $\tilde{\eta}_n=\pi/2$ ( namely, a coplanar state in $ \tilde{Y}\tilde{Z} $ plane )
  and the trial energy per site is $
   E_{tri}( N \times 1 )=-\frac{J}{N}\sum_{n=1}^N [\cos(\tilde{\xi}_n-\tilde{\xi}_{n+1})-\sin^2\beta\cos(2\tilde{\xi}_n+ 4 \alpha n )+\cos^2\beta] $
   which can be transformed back to the original frame  using $ \xi_n=\tilde{\xi}_n+2n\alpha $
   ( so the spins remain in a coplanar state in the original $ YZ $ plane shown in Fig.\ref{allphases} ).
\begin{eqnarray}
	E_{YZ}& = & -\frac{J}{N}\sum_{n=1}^N[ \cos(\xi_{n+1}-\xi_{n}-2\alpha)- \sin^2\beta\cos2 \xi_n
                                                   \nonumber   \\
                    & + &  \cos^2\beta]
\label{fk}
\end{eqnarray}

     One can see that at a small $ \beta $, by using $ \cos(\xi_{n+1}-\xi_{n}-2\alpha) \sim 1- \frac{1}{2}(\xi_{n+1}-\xi_{n}-2\alpha)^2 $,
     $ E_{tri}( N \times 1 ) $ maps to the 1d Frenkel-Kontorova (FK) Model discussed in \cite{tom} at a finite size $ N $
     with the periodic boundary condition:
\begin{eqnarray}
	E_{FK}& = & \frac{J}{2N}\sum_{n=1}^N[ (\xi^{\prime}_{n+1}-\xi^{\prime}_{n}-2\alpha)^2 - \beta^2 \cos2 \xi^{\prime}_n ]
\label{fk1d}
\end{eqnarray}
    where $ \xi^{\prime}_n=\xi_{n} \pm \pi/2 $ and
    we dropped some irrelevant constants.  It also resembles the 2d Pokrovsky-Talapov (PT) which was used to discuss C-IC transition
    in 2d Bilayer quantum Hall systems \cite{blqh}.
    Note that the main difference between Eq.\ref{fk} and Eq.\ref{fk1d} is that the former works from
    the lower half $ \beta < \alpha $ upto the diagonal line $ \beta = \alpha^{-} $,
    while the FK model only works near the Abelian line $ 0< \beta \ll \alpha <\pi/2  $.

     In the appendix C, we perform the quantization of the mean field states in Eq.\ref{fk}.
     In principle, its quantized form Eq.\ref{generalS} can be used to study the quantum fluctuations in the  fractals in Fig.1
     at the lower half $ \beta < \alpha $ and along the diagonal line $ \beta = \alpha^{-} $.
     For example, as shown in appendices D,E,F, at the lower half $ \beta < \alpha $, it can be used to derive the excitation
     spectrum Eq.\ref{N1} and the quantum fluctuations Eq.\ref{Ntimes1eff} in the commensurate phases near $ \alpha=\pi/N $ and
     the effective quantum phason action Eq.\ref{phasonact} in the in-commensurate phases.
     From the Y-x state on the right, it was already used to re-derive the effective GL action Eq.\ref{Yx} along the diagonal line $ \beta = \alpha^{-} $ in the continuum limit in Sec. V-B.

\subsection{ The principle commensurate co-planar phases near $ \alpha=\pi/N $:  $ N $ Bragg  peaks }

%    The classical solutions on the FK model can be used to understand all the devil's staircases pinned at
%    the rational values $ \alpha=\pi/N, N=2,3,4,5,6... $ in Fig.\ref{square}a.

Some insights can be achieved from the FK model at a small $ \beta $. The kinetic term favors $ \xi_{n+1}=\xi_{n}+2\alpha $,
while the potential term favors $ \xi_n = \pm \pi/2 $. When $ \alpha=\pi/2 $, there is no frustration,
this leads to the Y-x state as the exact ground state.
However, when $ \alpha=\pi/N, N=3,4,5,\cdots $, frustrations comes in, the two terms compete against each other.
At a small $ \beta $, the kinetic term dominates over the potential term,
so $ \xi_{n+1} \sim \xi_{n}+2\alpha $ still holds approximately as shown for the $ 3 \times 1, 4 \times 1, 5 \times 1 $ spiral state in Fig.2.
In contrast to the Y-x state near  $ \alpha=\pi/2 $,
the commensurate phases near $ \alpha=\pi/3,\pi/4,\pi/5,\cdots $ are stripe co-planar ( in the YZ plane )
spiral phases shown in Fig.\ref{allphases} instead of a collinear phase.
As shown in Sec.V-C, we also find stable co-planar spiral phase at $ \alpha=\pi/N, N> 2 $ along the diagonal line.
So these phases found near the Abelian line $ \beta \ll 1 $  will expand all the way to the diagonal line $ \alpha=\beta $.

{\sl 1. Classical theory: Mean field state }

Obviously, the ordering wavevectors of any $ N \times 1 $ spiral state can only be $ Q_n= \frac{2 \pi}{N}n,n=0,1,\cdots,N-1 $.
 Because it is a spiral state at the $ YZ $ plane, so one can form  $ S^{\pm}= S^{z} \pm i S^{y} $.
 Then when $ \alpha= \pi/N, \beta \rightarrow 0^{+} $,  $ S^{+}(x)=e^{i ( \frac{2 \pi}{N} x + \xi_0) } $ which only contains one ordering wavevector $ Q_1 $.
 For $ N=q $ to be a prime number, $ \xi_0=\pm \pi/2 $.
 However, any small $ \beta >0 $ will introduce non-vanishing components at all the other ordering wavevectors
 which can be determined from the 1d FK model Eq.\ref{fk}. Because the spin is a unit vector in the YZ plane, equivalently, one can
 expand the angle $ \tilde{\xi}_n $ in the rotated basis in terms of $ Q_n= \frac{2 \pi}{N}n, n=0,1,\cdots,N-1 $:
\begin{equation}
  \tilde{\xi}(x)= \sum^{N-1}_{n=0} A_n e^{i \frac{2 \pi}{N}n x} +h.c
\label{xin1c}
\end{equation}
  where  $ \tilde{\xi}(x+N)=\tilde{\xi}(x) $  which leads to  $ S^{+}(x+N)= S^{+}(x) = e^{i( \frac{2 \pi}{N} x +  \tilde{\xi}(x) ) }  $.
  Of course, only the spin is experimentally measurable \cite{firstwinding}.

There are qualitative even-odd differences:
The net magnetization in the $ N \times 1 $ unit cell is small, but non-vanishing for odd $ N $, exactly vanishes for even $ N $.
%Odd $ N $ always contain $ \xi_0=\pm \pi/2 $, but even $ N $ do not.
Even $ N $ always has a larger  stable regime than its previous odd $ N-1 $.
There is a always cyclic degeneracy $ N $ for both even and odd $ N $.  The $ {\cal T } $ gives a different state for  $ N $ odd, but not for even $ N $.
So the degeneracy is $ 2 N $ for odd $ N $, just $ N $ for even $ N $. One can check other symmetries operations $ {\cal P}_x, {\cal P}_y, {\cal P}_z $ do not generate new states.

{\sl 2. Quantum effects: excitation spectrum }

The spin-coherent state path integral quantization of the classical 1d FK model is given in the appendix C.
Similar to that listed below Eq.\ref{Yx}, the $ \eta $ has been fixed to be at its classical value $ \eta_0=\pi/2 $.
Its quantum fluctuation $ S(\eta_i- \eta_0) $ plays the conjugate variable to $ \xi_i-\xi_0 $ ( see Eq.\ref{qpi} ).
It is used to re-derive the excitation spectrum of the Y-x state.
It is straightforward to extend the path integral calculation near $ \alpha=\pi/2 $ to  near $ \alpha= \frac{\pi}{N} $ to determine the
excitation spectra in these phases.
There are $ N $ branches, the lowest of which should take the similar form as
the $ 2 \times 1 $ Y-x state derived in Eq.\ref{relagap} which is also re-derived by the path integral method in appendix C
:
\begin{equation}
 E_{-}( \boldmath{q} )= \sqrt{ \Delta^2 + v^2_x q^2_x +  v^2_y q^2_y }
\label{N1}
\end{equation}
where $  -\frac{\pi}{N} < q_x < \frac{\pi}{N}, -\pi < q_y < \pi $ is confined in the RBZ.
It breaks the lattice symmetry to $ N $ sites per unit cell along the $ x $ direction, of course, also the time reversal symmetry.
%So it has a $ N $-fold degenerate for $ N $ even, $ 2N $-fold degenerate for $ N $ odd.

%rotated basis  $ \tilde{\mathbf{S}}_n $ near the Abelian line $ 0 < \alpha < \pi/2, \beta $.
%When $ \beta >0 $ is small, the ground state is very close to be a FM

Just like the Y-x state is pre-emptied by the first order transition into some IC phases through in-complete devil staircase discussed in the last subsection,
%we expect these principle commennsurate
%co-planar spiral phases will also be pre-emptied by first order transitions into some IC phases as shown in Fig.1.
it may also be interesting to determine the classical
first order transition boundaries between these robust gapped confined
C phases near  $ \alpha=\frac{\pi}{N} $ with some IC gapless phases ( see subsection D ) to determine
the corresponding plateaus of the in-complete devil staircase.
%and then to check if they will pre-empty the second order phase transitions due to the condensations of these magnons.

{\sl 3. Quantum effects: effective action and the SSCFs }

We already got the mean field ( classical state ) Eq.\ref{xin1c} and its low energy excitation spectrum Eq.\ref{N1}
of this commensurate co-planar phase. Now we will get the  low energy effective action to calculate the spin-spin correlation
functions (SSCFs).
Drawn the insight from the low energy effective action Eq.\ref{ppplusaction} for the C-magnons and
Eq.\ref{onlytildep} for the IC-magnons inside the Y-x phase and considering the main difference here is that
the critical mode will change from $ \delta \eta $ in the Y-x phase to the $ \delta \xi $ in the YZ coplanar phase,
which stands for the center of mass (COM) or uniform motion of the $ N \times 1 $ unit cell along the $ x $ direction,
is obviously independent of $ i=1,2,\cdots N  $ sublattice,
we can write down the low energy effective action describing the quantum fluctuations:
\begin{align}
 {\cal L}_{1/N} =\frac{N}{2\beta}( \omega_n^2 + \omega^2_{-}(q) ) \delta \xi (q,i\omega_n) \delta \xi (-q,-i\omega_n)
\label{Ntimes1eff}
\end{align}
  which leads to the low energy excitation spectrum Eq.\ref{N1}.
  Furthermore, from which one can compute the SSCFs
\begin{equation}
 \langle  S^{+}_i(x,y;\tau)  S^{-}_j (0,0;0)\rangle =F(i,j) e^{i [ \delta \xi (x,y;\tau)-\delta \xi (0,0;0) ] }
\label{Ntimes1spin}
\end{equation}
  where $ F(i,j)=  e^{ i [ \frac{2\pi}{N}(i-j) +\tilde{\xi}(i)- \tilde{\xi}(j) ]} $ is the form factor connecting sublattice
  $ i $ to $j $ in one unit cell,
  $ i,j=1,2,\cdots N $ label the $ N $ sublattices in a unit cell, it depends on $ i,j $ separately instead of just their differences
  $ i-j $. This is due to the translational symmetry breaking within the $ N \times 1 $ unit cell.
  It is completely determined by the classical configuration Eq.\ref{Ntimes1eff}. Setting $ i=j $ leads to the SSCF at the same sublattice.
  While $ (x,y,\tau ) $ label the 2d lattice sites corresponding to the RBZ listed below Eq.\ref{N1} and
  the imaginary time respectively. It is similar to the Debye-Waller (DW) factor $ e^{-2D} $  in the context of solid orders due to the phonons
  \cite{EPL,SS1,SS2,SSrev}:
\begin{equation}
 D =-\frac{1}{N \beta} \sum_{i \omega_n} \int \frac{d^2q}{(2\pi)^2}
 \frac{1- e^{i(q_x x + q_y y)-i \omega_n \tau}}{ \omega_n^2 + \omega^2_{-}(q) }
\label{DW}
\end{equation}
  which is completely determined by the quantum fluctuations.

  There is a separation between intra-cell represented by the static ( classical ) form factor $ F(i,j) $ and the inter-cell by the second
  dynamic (quantum DW ) factor in Eq.\ref{Ntimes1spin}. This is a quite appealing feature of the low energy effective description of the
  $ N \times 1 $ coplanar phase in YZ plane.
  One can similarly compute the anomalous SSCFs such as $ \langle  S^{+}_i(x,y;\tau)  S^{+}_j (0,0;0)\rangle $
  and $ \langle  S^{-}_i(x,y;\tau)  S^{-}_j (0,0;0)\rangle $.

\subsection{ Higher order Co-planar spiral states near $ \alpha=\frac{\pi}{N} n $ with $ n>1 $: $ N $ Bragg  peaks }

 As shown in Fig.5, following the principle series $ \alpha= \pi/N $ which stretches all the way down to the diagonal line,
 the Higher order co-planar phases for $ \alpha=\frac{\pi}{N} n,  n=2,3,\cdots, [N/2] $ will also extend to the diagonal line
 with much smaller measures.
 Similar to the constructions of the co-planar phases near  $ \alpha= \pi/N $,
 one can construct the spiral states near $ \alpha= \frac{\pi}{N}n, n=2,\cdots, [N/2] $.
 Just in terms of symmetry breaking patterns, they are  essentially the same states as those near $ \alpha= \pi/N $,
 therefore  have the same ground state degeneracies, the same form of excitations in Eq.\ref{N1} with smaller gaps
 and also the same set of  $ N $ Bragg  peaks in Eq.\ref{xin1c},
 but still can be distinguished by different topological winding numbers.

 Indeed, one can define a topological winding number for all the Co-planar spiral states:
 the homotopy group is $ S^1 \rightarrow S^1 $: the first $ S^{1} $ stands for the $ N $ lattice sites along the $ x $ direction,
 the second $ S^{1} $ stands for the spin orientation in the $ YZ $ plane (Fig.2) at a given lattice site along the $ x $ direction.
 So the winding number $ n $ describes the mapping due to:
\begin{equation}
  \Pi_1 ( S^{1} ) = n
\label{topon}
\end{equation}
 This may be similar to a 2d vortex which is also characterized by the winding number $ n $.
 Only the most fundamental vortex $ n =\pm 1 $ is stable, while higher ones $ |n | > 1 $ are unstable and decay into
 the $ |n|=1 $ one.

 Especially, we expect the gap $ \Delta_{W=n/N} $ is a monotonically decreasing function of the order $ n $ at a fixed $ N $.
 As $ N \rightarrow \infty $ at a fixed $ n $, then $ \alpha=\frac{\pi}{N}n \rightarrow 0 $.
 However, as $ N \rightarrow \infty, n \rightarrow \infty $, but the ratio $ \frac{n}{N} $
 approaching an irrational number $ \alpha $,
 the gap  $ \Delta_{W=n/N} \rightarrow \Delta_{W=\alpha} < \Delta_{W=1/N} \rightarrow 0 $ as $ N \rightarrow \infty $.
 So it gets smaller and smaller, eventually, leads to a gapless spiral co-planar IC-YZ-x phase to be discussed in the following.

\subsection{  The spiral co-planar IC-YZ-x phase with gapless phasons: its quantum melting to a quasi-1d Luttinger Liquid }

{\sl 1. Classical picture }

 Taking $ N \rightarrow \infty $ limit in Eqn.\ref{xin1c}, one can write the spins of the stripe coplanar ( spiral ) IC phase (
 denoted as IC-YZ-x ) $ S^{+}(x) = e^{i( 2 \alpha x +  \tilde{\xi}(x) ) }  $ at any $ ( 0 < \alpha < \pi/2, \beta ) $  in Fig.1:
\begin{equation}
    \tilde{\xi}(x)= \int^{\pi}_{-\pi} \frac{ d k_x}{2 \pi} A(k_x) e^{i k_x x} +h.c
\label{xinic}
\end{equation}
  which completely breaks the translational symmetry along the $ x $ axis.
  It is infinitely degenerate and has a gapless phason mode due to breaking a continuous $ U(1) $ symmetry.
So these co-planar IC phases have broad distributions of Bragg peaks, so very much resemble
the broad spectrum of spinons in the quantum spin liquid (QSL) phases in geometrically frustrated lattices.

{\sl 2. Quantum effects: excitation spectrum }

As $ N $ gets bigger,  but the ratio $ \frac{n}{N} $ approaching an irrational number $ \alpha $,
$ \Delta $ in Eq.\ref{N1} gets smaller, so the stability regimes
( or the widths of the devil staircases ) in Fig.\ref{phasedia} gets smaller.
As $ N \rightarrow \infty $, $ \Delta \rightarrow 0 $, the size of the RBZ along the $ q_x $ direction in  Eq.\ref{N1}  also shrinks to zero,
it becomes an IC phase,
which is a gapless state with the non-analytic anisotropic dispersion \cite{nonanalytic}:
\begin{equation}
E( q_y )=   v_y | q_y |
\label{phason}
\end{equation}
 where $ -\pi < q_y < \pi $. It may be called an anisotropic phason mode.

 It is responsible for its zero width in Fig.\ref{phasedia}.
Note that  although the spiral Co-planar  IC phases does not break the crystal translational along the $ y $ axis,
it completely breaks the crystal translational symmetry along the $ x $ axis.
So $ q_x $ is not even defined, the BZ can only be defined along $ q_y $.
The $ Z_N $ symmetry in the $ N \rightarrow \infty $ limit becomes a continuous $ U(1) $ symmetry,
its breaking  leads to a Goldstone mode which is nothing but the anisotropic gapless phason mode.

{\sl 3. Quantum melting: quasi-1d Luttinger liquid }

 In the $ N \rightarrow \infty $ limit, the $ i,j=1,2\cdots N $ labeling
 the $ N $ sublattices in a unit cell is promoted the $ x $ coordinate as the unit cell expands to cover the whole lattice along the $ x $ direction. In terms of $ \delta \xi $ which stands for the center of mass (COM) or uniform motion of the whole lattice along the $ x $ direction,
 is obviously independent of $ x $, the phason is described by the effective action:
\begin{equation}
   {\cal L}_{P}=\frac{K}{2}[ \frac{1}{v_y} ( \partial_{\tau}  \delta \xi (y,\tau))^2+  v_y ( \partial_{y}  \delta \xi  (y,\tau))^2 ]
\label{phasonact}
\end{equation}
%   where the mean field value of $ \tilde{\xi} (x,y,\tau) $ is given by Eq.\ref{xinic}.
   The two phenomenological Luttinger parameters $ K $ and $ v_y $, in principle, can be evaluated by the microscopic spin wave expansion.
   It only contains the spin sector, no charge sector.
   It is important to address that its gapless is protected by the IC- and the associated
   continuous $ U(1) $ symmetry breaking. The high oder terms in Eq.\ref{pathhigh} only modify $ K $ and $ v_y $
   without changing its gapless feature.

   Eq.\ref{phasonact} is essentially the same as the 1d Luttinger liquid (LQ) model. The gapless phason fluctuations
   lead to a infrared divergency even at $ T=0 $ which renders $ \langle  S^{+}(x) \rangle =0 $.
   Naively, this may suggest that both the translational symmetry and the time reversal breaking are restored.
   However, as shown immediately below, this is not the case.
   There is also an algebraic long-range order of the spin-spin correlation function at $ T=0 $:
\begin{equation}
 \langle  S^{+}(x_1,y;\tau)  S^{-}(x_2,0;0)\rangle =\frac{ F(x_1,x_2) }{ ( y^2/v_y + v_y\tau^2)^{\frac{1} {4 \pi K} }} ,
\label{phasonspin}
\end{equation}
    where the lattice constant $ a=1 $ and $ F(x_1,x_2)= e^{i [2 \alpha (x_1-x_2) + \tilde{\xi}(x_1)-\tilde{\xi}(x_2)] } $ is the form factor
    which  depends on $ x_1 $ and $ x_2 $ respectively instead of just their difference $ x_1-x_2 $.
    It is completely determined by the classical configuration Eq.\ref{xinic}.
    We also did the scaling $ ( y^{\prime}, \tau^{\prime} )= ( y/\sqrt{v_y}, \sqrt{v_y}  \tau)  $.
     Obviously, due to the gapless phason mode, all the anomalous SSCFs vanish !

    There is a clear separation between $ x $ and $ y $ coordinate.
    There is still a translational symmetry breaking along the x-direction, but an algebraic decay along the $ y $ direction.
    This is a salient feature due to the IC-, so justify its name as a quasi-1d Luttinger liquid ( LQx ).
    Away from the diagonal line $ \alpha > \beta $, as listed below Eq.\ref{rhgeneral}, the system has only translational and time reversal
    symmetry, both are broken by the static form factor $ F(x_1,x_2) $ in Eq.\ref{phasonspin} which depend on
    $ x_1 $ and $ x_2 $ separately and complex !
    As $ \alpha \rightarrow \beta^{-} $,
    this quasi-1d LQx also breaks the $ [C_4 \times C_4]_D $  symmetry. Then along the diagonal line $ \alpha = \beta $, it is a
    mixture of the LQx and LQy with any ratios.

%    Along the diagonal line $ \alpha = \beta^+ $, it breaks the $ [C_4 \times C_4]_D $ symmetry.
%    However, the state becomes a mixed state of $ LQx $ and $ LQy $ with any ratios.
%    Physically, the $ e ^{i 2\alpha x} $ plane-wave factor is due to

    Due to the strong quantum fluctuations from the gapless phasons, the classical
    spiral co-planar IC-YZ-x phase melts into a quasi-1d Luttinger liquid  even at $ T=0 $ which is described
    by the $ c=1 $ CFT. So at any finite temperature $ \beta = 1/k_B T $,
    the spin-spin correlation function can be obtained by performing
    a conformal transformation  $ f(\tau)= \tanh  \pi \tau/ \beta $ on Eq.\ref{phasonspin}.
    The form factor $ F(x_1, x_2) $ is static, so independent of $ T $.

    In the following, we use the name the spiral co-planar IC-YZ-x/LQx phase which stands for the physics of melting
    the classical phase IC-YZ-x into the quasi-1d Luttinger liquid LQx described by Eq.\ref{phasonspin}.

\subsection{ The pre-empty of the magnon condensations in the Y-x phase by the $ W=1/2 $ segement along the
in-complete devil staircase:
a complete picture on the instabilities of the Y-x phase  }

 Now we look at the possible transition driven by the C$_0 $ magnons in the lower part of the Y-x phase.
 As emphasized at the end of Sec.II, the mirror symmetry $ \beta \rightarrow \pi/2 - \beta $ relates the C$_0 $ and C$_\pi $ magnons only hold at the quadratic level.
 There is dramatic difference at the higher order. Indeed the OFQD phenomenon presented in Sec.III only happens
 near the  C$_\pi $ magnons in the upper part, but not near the
 C$_0 $ magnons in the lower part. The effective action along the diagonal line $ \alpha=\beta $ in the upper part
 was shown in Eq.\ref{Yxp}, in Eq.\ref{ppphi3} slightly away from it. Drawing the insights from Eq.\ref{ppphi3},
 one can directly write down the effective action driven by the condensation of the C$_{0}$ magnons in the lower part:
\begin{eqnarray}
	{\cal L}[  \eta ]_{Y-x,C_0} & = & \eta(-k,-i\omega_n)	[ \omega_n^2  + v^2_x q^2_x + v^2_y q^2_y       \nonumber   \\
      & + & \Delta^2 ] \eta(k,i\omega_n)  + \lambda \eta^3 +  \cdots
\label{eta33}
\end{eqnarray}
   where  $ 0 < k_x < \pi, -\pi < k_y < \pi/2 $ is in the BZ.
   The cubic term leads to a 1st order transition at $ \Delta^2_0= \lambda^2/2 \kappa > 0 $ which happens before
   the putative 2nd order transition $ \Delta^2=0 $. In contrast to near the diagonal line Eq.\ref{etaeta3}, the value $ \lambda $ could be very large  which spoils the $ \beta \rightarrow \pi/2 - \beta $ symmetry.

For any parameter $ \beta < \alpha=\pi/N $, Eq.\eqref{fk} gives the best estimation of the ground-state energy as
$\min\limits_{N\in[1,\infty)}E_{N\times1}$ which can be compared to that of the Y-x state
$E_{Y-x}=-2J\sin^2\alpha $. If one finds $\min E_{N\times1}<E_{Y-x}$ for some $ N $,
then it means Y-x becomes unstable against some spiral IC phase.
Note that even $\min E_{N\times1}$ may not give real ground-state energy,
but it does give a upper bound for the ground-state energy of the spiral IC phase whose precise nature is
difficult to determine using the $ N \times 1 $ ansatz in a finite size calculation.
The first order transition line \cite{firstorder} from the Y-x to some IC phases is drawn in Fig.\ref{phasedia}.
It also hits the $ \pm 2\pi/3 $ contour line inside the Y-x phase at one corner of the  $ 3 \times 3 $ SkX phase
which is a multi-critical ( M ) point at $ (\alpha_M,\beta_M)\approx(0.33952\pi,0.31284\pi) $ of several commensurate
and In-commensurate phases in Fig.1.
So all the C-$C_0$  regime and the IC- regime with $ 0 < k^{0}_y < 2\pi/3 $ in the Y-x phase are pre-emptied by
some spiral IC phases through the $ W=1/2 $ segment in the in-complete devil staircase.
%As shown above, these spiral IC phases are determined by the analytic arguments to be the IC-YZ-x/LQx phases.
%So the Y-x state becomes only a meta-stable state ( just a local minimum in energy landscapes ) between
%the first-order transition line ( solid line ) and the putative 2nd-order condensation boundary ( dashed line ) of the C-$C_0$  and the C-IC %magnons. So hysteresis behaviors are expected in this regime.
%So this physical picture provides a concrete example of a second order transition preempted by a 1st order \cite{firstorder} transition.

  Combining the results achieved in Sec.IV, Sec.V-B and here, we get the complete physical picture
  of the 3 piece-wise instabilities of the Y-x state in Fig.1 and shown in ( Fig.\ref{phasesarc} ):

  (1) The top segment: as established in Sec.IV, there is a first order transition from the Y-x state to the X-y state
      at $ \alpha=\beta $ near the Abelian point $ \alpha=\beta=\pi/2 $.
  One of the immediate consequence of the corrected spectrum Eq.\ref{gapspectrum} due to the order
  from quantum disorder is that the stability regime
  of the Y-x ( or X-y ) phase goes beyond the diagonal line and reaches the dashed line slightly above ( below ) the diagonal
  line in Fig.1. Between the dashed line and the diagonal line, the Y-x ( X-y ) phase becomes a meta-stale phase ( Fig.1).
  This 1st order transition pre-empties a putative 2nd order transition driven by the C$_\pi $ magnon condensation
  shown in Eq.\ref{ppphi3} and Eq.\ref{Yxp}.
  This dashed line in the top segment can be contrasted with that in the bottom segment
  drawn below the solid line connected from $ ( \pi/2, 0) $ to the M point in Fig.1 ( Fig.\ref{phasesarc} ).
%  The Y-x state also becomes meta-stable between the dashed line and the solid line to be discussed in Sec.VI-B.

  (2)  The middle segment:  as established in Sec.V-B,  there is a weakly first order quantum Lifshitz phase transition with the dynamic exponents
   $ z_x=z_y=1 $ described by the effective action Eq.\ref{etaeta3} from the
   Y-x to the gapped non-coplanar IC-SkX-y due to the condensations of IC- magnons with the ordering wavevectors
   $ \pi-\pi/3 < k^{0}_{y} < \pi- 0.18 \pi $. It reduces to the coplanar IC-XY-y phase along the diagonal
   line $ \alpha_{33} < \alpha < \alpha_{in} $ shown in Eq.\ref{Yxp} ( Fig.\ref{phasesarc} ).

  (3) The bottom segment: as established in here Sec.VI-D,
   numerically, there seems a 1st order \cite{firstorder} transition from the Y-x phase
   to the gapless IC-YZ-x/LQx phases along the counter lines of $ 0 < k^{0}_y < 2\pi/3 $
   which pre-empties a putative 2nd order transition driven by the  magnon condensation described by the effective action Eq.\ref{eta33}.
   In fact, it is the last segment $ W=1/2 $ of the  in-complete devil staircase.
   The top and middle segment meets at the counter line of $ k^{0}_y =\pi-q_{ic} $.
   The  middle and the bottom segment meets at the M point which is on counter line of $ k^{0}_y =2 \pi/3 $ ( Fig.\ref{phasesarc} ).

\section{  Rational and irrational topological winding numbers, Cantor function,
In-complete and complete devil's staircases. }

All the possible C- and IC- phases  along the diagonal line $ \alpha=\beta $ and near the Abelian line
$ \beta \ll \alpha $ were discussed in Sec.II-V and Sec.II/Sec.VI, but what is the organization pattern of all these phases ?
In this section, we will show that they are organized into the fractal structure  in Fig.1 and  Fig.\ref{trifeats}.
Especially, we introduce the Rational and irrational topological winding number $ W $
to characterize the in-complete devil staircase at $ \beta \ll \alpha $ and complete devil's staircase along $ \alpha=\beta^{-} $.
We conjecture that $ W $ becomes a Cantor function when $  0 < \alpha < \alpha^{-}_{33} $, but not defined anymore after $ \alpha >  \alpha^{-}_{33} $ which falls in the quantum phase transition regime in Fig.\ref{finiteT}.

\begin{figure}[!htb]
\centering
\includegraphics[width=0.7\linewidth]{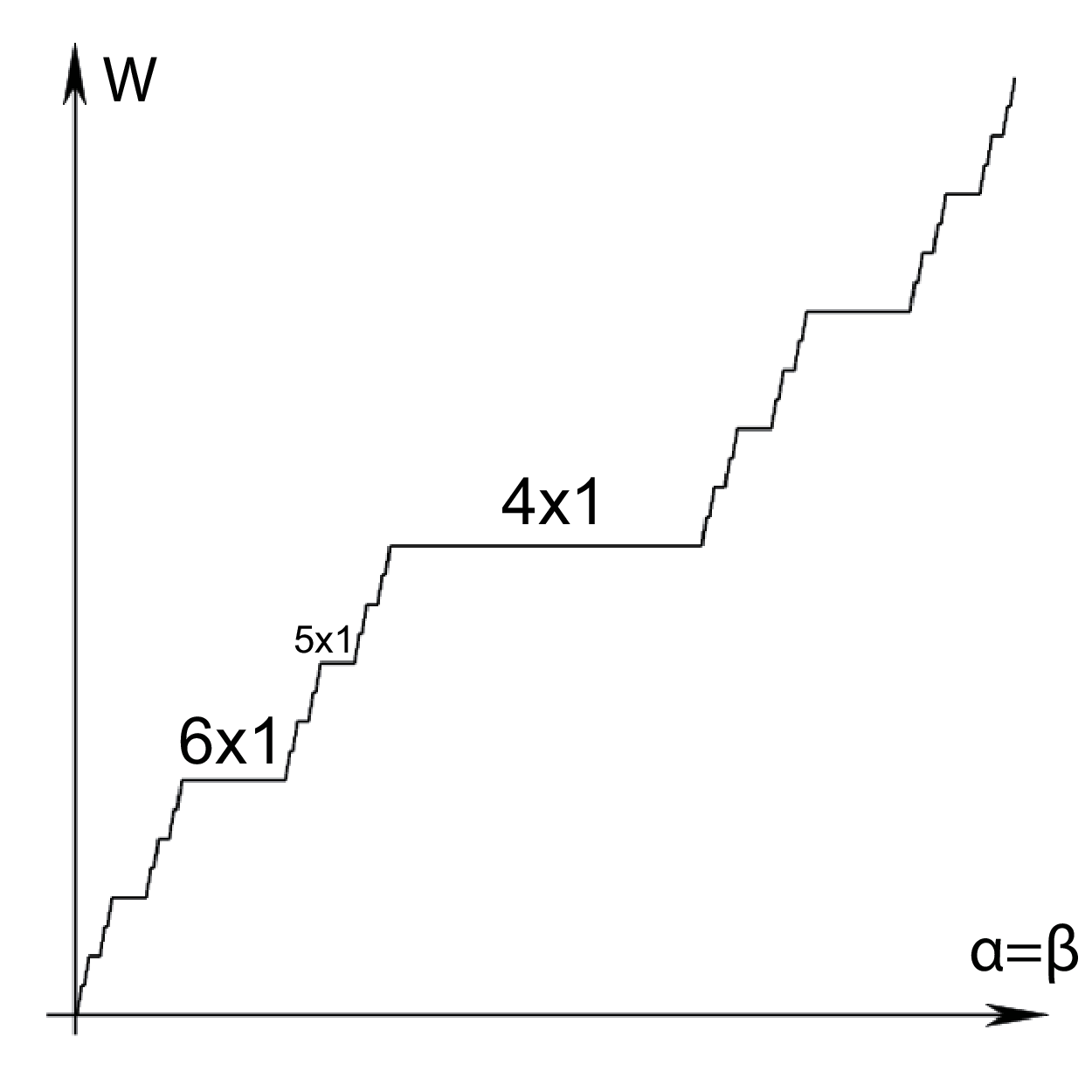}
%\quad
%\includegraphics[width=0.25\textwidth]{33SkX}
\caption{ The simplest ( the most original ) $1/3 $
 Cantor function taken from Wikipedia. It divides a line into 3 equal parts and remove the middle one, then repeat the procedure infinite number of times. It  is a monotonically increasing function
 and an odd function with respect to its half point $ x=1/2$ in a unit interval.
 It also has an oscillating width of plateaus.
 We expect the Cantor function $ W(\alpha) $ along the diagonal line is
 different, but hemeomorphic ( or topological equivalent ) to it.
So there is still a one to one correspondence between the principle series such as $ N \times 1 , N \geq 4 $ along the diagonal line
in Fig.1 and the major plateaus in the Cantor function. It remains a puzzle why the principle series take only half of the Cantor function and ends on the $ 3 \times 3 $ SkX ( hub )  phase.
Intervening between them are some small devil staircases with higher topological winding number $ W=n/N, n > 1 $.
The gapless IC-YZ-x/LQx phases form a Cantor set with a zero measure.
Because the fractal dimension is not topologically invariant, so we expect it is different than
$ d^F_{1/3}=\log 2/\log 3= 0.6309 $ associated to the $ 1/3 $ Cantor function.
It remains interesting to find this generalized Cantor function $ W(\alpha) $ and its associated fractal dimension.
Its crucial difference than the quantum Hall conductance $ R_H $ versus the filling factor $ \nu $ will be discussed in Sec.IX-7.  }
%It remains a puzzle to see why the complete devil staircase in Fig.1 only takes half of the Cantor function,
%then ends at the hub phase $ 3 \times 3 $ SkX. }
\label{cantor}
\end{figure}

\subsection{ Rational and irrational topological winding numbers W }
From all the co-planar spiral phases in Fig.\ref{allphases}, one can define the topological winding number
$ W= (\xi_N-\xi_0)/2\pi N $. For the C-phase at $ \alpha=\pi/N $, $ W=2 \alpha/2\pi=1/N $ is a rational winding number
which is independent of the intermediate values of $ \xi_{n}, n=0,1,.....,N-1 $.
For the other C phases at $ \alpha=\frac{\pi}{N} n $ with $ n > 1 $, the winding number is found to be $ W=2 \alpha/2\pi=\frac{n}{N} $
as shown in Eq.\ref{topon}.
The quantum fluctuations such as the DW factor in Eq.\ref{DW} will certainly reduce the
magnitude of spin at a given site $ \vec{M}_i= \langle \vec{S}_i \rangle $.
However, as long as the $ \vec{M}_i \neq 0 $, one can still define the fractional winding number $ W $ in terms of its phase.
So due to its topological features, the definition of the winding number $ W $ also hold in the quantum case.

For an In-commensurate phase, one can still define $  W= Lim_{N\rightarrow \infty } (\xi_N-\xi_0)/2 \pi N $ which becomes an irrational number.
%Drawing $ W/\pi $ versus $ \alpha/\pi $ leads to Fig.\ref{devil}.
Each C phase occupies a step with the length $ \Delta_c $, the total C length
$ L_{c}= \sum _{ \{C \} } \Delta_c $, its ratio over the total length $ L_0 $
gives the measure of all the C phases $ L_c/L_0 $.
For an in-complete devil staircases, $ L_c/L_0 < 1 $, the rest  $ 1-L_c/L_0 >0 $ goes to the measure of the IC phases.
For a complete devil staircases, $ L_c/L_0 =1 $, while the IC phases intervening all the C phases become a set of measure zero
forming a Cantor set with a fractal dimension.
For a harmless devil staircases, there is a direct first order transition between the two C phases with no intervening IC phases.

In the following, we will show that for $ \beta < \alpha $, it is in an in-complete devil-staircase in
all the regime $  \beta < \alpha < \pi/2 $.
However, when $ \beta = \alpha^{-} $ where the Hamiltonian has the $ [C_4 \times C_4]_D $ symmetry,
it becomes a complete devil-staircase when $ 0 < \alpha < \alpha^{-}_{33} $ where the topological winding number $ W $
become a singular continuous function ( Cantor function ) which has zero derivative everywhere except in the Cantor set with zero measure.

\subsection{ Complete devil's staircases along the diagonal line: Cantor set }

%Drawing $ W/\pi $ versus $ \alpha/\pi $ along the diagonal line leads to Fig.\ref{devil}b.
%\section{ Approach along the diagonal line $ \alpha =\beta $: the Complete devil staircases. }
Near the diagonal line $ \alpha=\beta $, the mapping to the FK model Eq.\ref{fk} may not be precise anymore.
  The classical ground state energies at $ \alpha=\pi/N $ devil staircases along the diagonal line $ \alpha=\beta $ are shown in Fig.\ref{diagstair}.
  In fact, every curve at $ N \times 1 $ in Fig.\ref{diagstair} contains $ N $ pieces: $ \alpha=\frac{\pi}{N}n, n=0,1,2,\cdots, N-1 $.
  Note that all $ N $ values contain the $ n=0 $ piece which stands for the FM state discussed in Sec.V-D.
  Because $ \alpha $ and  $ \alpha^{\prime}=\pi-\alpha $ ( its image about $ \pi/2 $ ) has the same set of Wilson loops, so they
  belong to the same equivalent class and have the same ground state energy.
  So one can just confine $ \alpha \leq \pi/2 $, so every curve in Fig.\ref{diagstair} has $ [N/2] $ pieces where $ [...] $ means the closest integer which is equal or larger than $ N/2 $.
  Starting from the Abelian line $ \beta \ll \alpha $,
  as one approaches the diagonal line $ \alpha=\beta^{-} $,  the principle $ \alpha=\pi/N $ staircase takes more and more measures,
  all the other higher order pieces at $ \alpha=\frac{\pi}{N} n, n \geq 2 $ take less and less measures.
  As shown in the last section, in fact, just from the symmetry breaking point of view, the C phases near $ \alpha=\frac{\pi}{N} n, n \geq 2 $
  are the same as those $ \alpha=\pi/N $.
  However, they can still be distinguished by different topological winding numbers $ W=n/N $.
  This could be a specific example where states share the same symmetry breaking patterns, but still
  can be distinguished by a topological winding number $ n $.

  Because all the $ N $ contain $ n=0 $ which is a FM along the $ Y $ direction, all the $ \alpha=\pi/N $ contains the FM piece near $ \alpha=0 $. As $ N \rightarrow \infty $, there are always in-commensurate phases below the FM phase.
  In fact, as shown in Sec.V-D, there is a degenerate family of FM state in the XY plane along the diagonal line near $ \alpha=0 $.
  In the spin wave expansion, although the linear term vanishes, the spin wave spectrum becomes negative indicating its instability against some
  IC phases. we expect that these IC phase are nothing but the IC-XY-x/LQx phases.

  In fact, there could always be a small regime of co-planar spiral C and IC phases sandwiched between two principle Commensurate
  phases $ N \times 1, N \geq 3 $ and $ (N+1) \times 1 $ along the diagonal line. For example,
  when following the $ 7 \times 1 $ ansatz in Fig.5, we find there is  a tiny regime between  $ 4 \times 1 $ and the $ 3 \times 3 $ SkX,
  $ \pi/4 < \alpha= 2\pi/7 < \pi/3 $ state has the lower energy than both, which indicates there could some C phases with small widths
  and IC phases with zero widths ( also total zero measures ) intervening between the $ 4 \times 1 $ and the $ 3 \times 3 $ SkX
  ( not shown in Fig.1, Fig.\ref{diagstair} and Fig.\ref{finiteT}).
  This suggests that the devil staircases is a complete one instead of a harmless one.

In short,  along the diagonal line, as shown in Fig.\ref{phasedia},
the principle series $ N \times 1 $ C phase occupies a step with the length $ \Delta_{ N \times 1 } $, the total principle C length
$ L_{C}= \sum^{\infty}_{N=4}  \Delta_{ N \times 1 } $ takes most of the total length
$ L_0= \alpha^{-}_{33} \sim 0.295 \pi $ ( Fig.\ref{finiteT} ), the other small part goes to higher order C-phases with $ W=n/N, n> 1 $,
while IC-XY-x/LQx phases take zero measure. So $ W $ becomes a Cantor function when $ 0 < \alpha < \alpha^{-}_{33} $.
In fact,  the enhanced $ [C_4 \times C_4 ]_D $ symmetry along the diagonal line $ \alpha=\beta $
makes it a special line which just becomes a complete devil staircase.
So all the IC phases along the diagonal line take zero measure, but form a Cantor set with an non-integer fractal dimension
which remains to be determined.

\subsection{ In-complete devil's staircases near the Abelian line $ ( 0< \alpha < \pi/2, \beta \ll 1 ) $. }

Near the Abelian line $ \beta \ll 1 $, Eq.\ref{fk} can be mapped to the FK model in the weak locking regime.
%Drawing $ W/\pi $ versus $ \alpha/\pi $ near the Abelian line leads to Fig.\ref{devil}a.
As shown in Fig.\ref{phasedia}, in addition to the $ N \times 1 $ C phase,
the C phases at $ \alpha=\frac{\pi}{N} n $ with $ n > 1 $  also contribute largely to  $ L_c= \sum _{ \{C \} } \Delta_c $.
The total length $  L_0=1/2- \beta/\pi $.
The C measure  $ L_c/L_0 < 1 $,
%( not counting the blue segments in Fig.\ref{devil}a ),
the IC measure $ 1-L_c/L_0 > 0 $. We expect that there are two following limiting cases in Fig.\ref{phasedia}:
As $ \beta \rightarrow 0 $ approaches the Abelian line, $ L_c/L_0 \rightarrow 0, 1- L_c/L_0 \rightarrow 1^{-} $,
so the IC phases takes almost all the measures.
As $ \beta $ approaches the diagonal line, $ L_c/L_0 \rightarrow 1^{-}, 1- L_c/L_0 \rightarrow 0^{+} $,
the C phases takes almost all the measures. At the transition point $ \alpha=\beta $
with the enhanced $ [C_4 \times C_4 ]_D $ symmetry, it just becomes the
complete devil staircases
%shown in Fig.\ref{devil}b
where the IC phases form a Cantor set with a fractal dimension presented in the last subsection.

  Near the Abelian line $ ( 0 < \alpha < \pi/2, \beta \ll 1 ) $, the FK model shows that the $ N \times 1 $ spiral state
  at $ \alpha=\pi/N $, $ \xi_{n+1} \sim \xi_n + n ( 2 \alpha ) $ which is a clockwise rotation ( positive winding number ).
  So at its image about $ \pi/2 $,
  $ \alpha^{\prime}=\pi-\alpha=\pi- \pi/N $, $ \xi_{n+1} \sim \xi_n - n ( 2 \alpha ) $ takes a counter-clockwise rotation
  ( negative winding number ).
  From the FK model \ref{fk}, one can also sketch some organization principle of these spiral phases ( Fig.\ref{trifeats} ).
  The staircase at $ \alpha=\frac{\pi}{6}= \frac{\pi}{2 \times 3 } $ can be considered as a composite of the one at $ \alpha= \frac{\pi}{2} $
  and $ \alpha= \frac{\pi}{3} $. In fact, any staircase at $ \alpha=\frac{\pi}{p q}= \frac{\pi}{p \times q } $ can be considered
  as a fusion of the one at $ \alpha= \frac{\pi}{p} $ and $ \alpha= \frac{\pi}{q} $.
  So one can first construct all the primary ( skeleton ) spiral states at $ \alpha=\pi/q $ with $ q $ a prime number.
  They always contain $ \xi_0=\pm \pi/2 $ ( namely, the Y axis ).
  Then one can construct all the principle staircases at $ \alpha=\pi/N $  where $ N= p \times q $, then
  the higher order ones at $ \alpha=\frac{\pi}{N} n, n= 2, \cdots, [N/2] $ which have the same
  symmetry breaking pattern as the principles ones at $ \alpha=\pi/N $,
  but can still be distinguished by the different winding number $ W=n/N $ defined in Sec.VII-A.
  So when $ \beta $ is small, one can construct all the $ \alpha=\frac{\pi}{N} n \leq \pi/2 $ staircases.

  As shown in the Fig.1, all the principle staircases at  $ \alpha=\frac{\pi}{N} $ reach the diagonal line and merge with
  those determined from the diagonal line shown in Fig.\ref{diagstair}.
  Some high order ones at $ \alpha=\frac{\pi}{N} n, n= 2, \cdots, [N/2] $ also reach the diagonal, but with very small measures
  compared to the principle ones.
  For example, the $ \alpha=2 \pi/5 $ staircase in Fig.1 is stable at a small $ \beta $, but may not reach the Y-x phase,
  there are some IC phases intervening between the two C phases. As argued in Sec.VI-D, these IC phases are nothing but the IC-YZ-x/LQx phases.
  Just in terms of symmetry breaking, the second order $ \alpha=2 \pi/5 $  is the same phase as the principle $ \alpha= \pi/5 $.
  However, they can still be distinguished by the different winding number $
  W=2/5 $ and $ W=1/5 $ respectively.

\section{ Quantum chaos and quantum information scramblings at a finite $ T $ in all phases in Fig.1. }

\begin{figure}[!htb]
	%\centering
\includegraphics[width=0.4\textwidth]{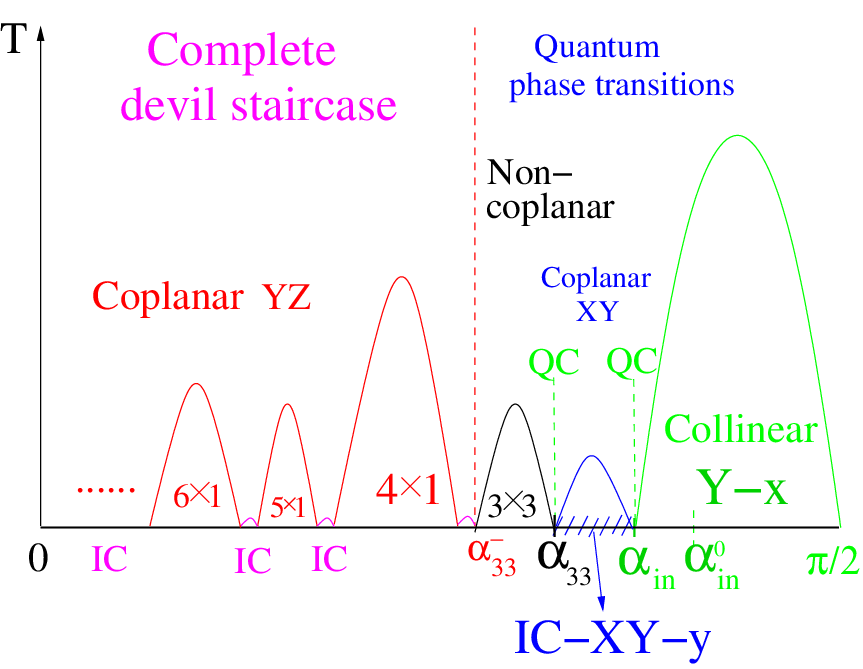}			
\caption{ (Color online) The finite temperature
 phase diagram  along the diagonal line $  \alpha= \beta^{+} $ in Fig.1.
 It consists of two regimes: the complete devil staircase
 $ 0 < \alpha < \alpha^{-}_{33} $ and the quantum phase transition (QPT) $  \alpha^{-}_{33} < \alpha < \pi/2 $.
 The  $3 \times 3 $ SkX  plays the hub (central node ) where the organization principle changes from the complete devil staircase
 to QPT ( Fig.\ref{trifeats} ). The spin orientation also twists from the YZ plane to the XY plane, then to collinear.
 The topological winding number $ W $ is only defined in the former, but not in the latter regime.
 In the QPT regime,
 there is an IC-XY-y phase with a finite measure intervening between the $3 \times 3 $ SkX and the Y-x phase
 through two quantum Lifshitz transitions.
 In the complete devil staircase regime, as shown in Fig.1, there is an oscillating of widths from odd to even
 in the principle series starting from $ 3 \times 3, 4 \times 1, 5 \times 1, 6 \times 1, \cdots $ which matches
 the salient feature of a Cantor function shown in Fig.\ref{cantor}.
 We also expect that the gaps decrease in an even-odd oscillating way as $ W=1/N, N \geq 3 $ decrease, the melting transition temperatures
 of the principle series also decrease in a similar even-odd oscillating way.
 The small domes stand for all the intervening phases between any two gapped principle $ N \times 1, N \geq 3$ phases:
 the C spiral phases at higher order $ W=n/N, n>1 $ with much smaller widths and also
  the  IC-YZ-x/LQx phases forming a Cantor set.
 The state at $ \alpha=\beta=\pi/2 $ is just an AFM state in $ \tilde{\tilde{SU}}(2) $ basis.
 The state near $ \alpha=\beta=0 $ is some IC-YZ-x/LQx phase instead of a FM  shown in Sec.V-D. }
\label{finiteT}
\end{figure}

   It was well known that at the classical level, there is an intimate connections between classical fractals and classical chaos.
   So it may interesting to look at if there is some intrinsic connections between fractals and quantum chaos at the quantum level.
   The quantum chaos have been extensively discussed in the context of SYK models in terms of both
   OTOC \cite{kittalk,syk1,syk2,syk3} and random matrix theory \cite{rmt1,rmt2,rmt3,rmt4,rmt5,sun1,sun2,sun3}. Here will look at quantum chaos
   in the fractals in Fig.1.

   It was shown in \cite{rh}, there is an Ising transition above the Y-x phase.
   Because the $ 3 \times 3 $ SkX has a $ 9 \times 2 $  fold degeneracy where the factor of $ 2 $ comes from the Time reversal symmetry,
   so it should be a $ Z_{18} $ Clock transition in Fig.6.
   As mentioned earlier, the ground state degeneracy of the coplanar $ N \times 1 $ state at $ \alpha= \frac{\pi}{N} n $ is $ N $ when $ N $ is even, $ 2 N $ when $ N $ is odd where the factor $ 2 $ comes from the Time reversal transformation
   ( Note the degeneracy here is determined only by the denominator, but independent of the numerator $ n $ ).
   So for $ N $ even or odd, it could be a $ Z_N $ or $ Z_{2N} $ clock transition.
%   However, for $ N $ odd, it is important to resolve the role of the factor $ 2 $ coming from the Time Reversal transformation.
   Taking $ N \rightarrow \infty $, the resulting spiral IC-YZ-x phase becomes infinitely degenerate,
   so there is an associated $ U(1) $ symmetry breaking leading to the gapless phason mode Eq.\ref{phason}.
%   At any finite temperature, due to the logarithmic IR divergence $ \int dq_y/q_y $, the broken $ U(1) $ symmetry
%   may become an algebraic order.
   As shown in Sec.VI-D, the strong quantum fluctuations due to the phason at $ T=0 $  melts this spiral IC-YZ-x phase
   into a quasi-1d Luttinger liquid (Fig.\ref{finiteT} ) which can be described by a 2d $ c=1 $ CFT at $ T=0 $.
   Its finite temperature behaviors is just putting the CFT at a finite temperature.

\subsection{ Quantum chaos near the 2nd order Quantum phase transitions }

The quantum Lifshitz transition from the collinear  Y-x state to the coplanar IC-XY-y state is described by the action
Eq.\ref{Yxp} and \ref{2times2order}.
The coplanar IC-XY-y state intervene between the collinear  Y-x state and the non-coplanar  $ 3 \times 3 $ SkX state with a finite measure
$ \alpha_{33} < \alpha < \alpha_{in} $. There should also be a melting transition above this IC-XY-y phase.
The nature of the transition should be determined by
the dilute gas of these repulsively interacting dis-commensurations and could be in the class of a 2d KT transition.
The transition from the non-coplanar  $ 3 \times 3 $ SkX to the coplanar IC-XY-y state  is expected to be also a  quantum Lifshitz transition.
Obviously, due to the non-linearities in Eq.\ref{Yxp}, the system shows quantum chaos \cite{SY,kittalk,syk1,syk2,syk3,on}.
From Eq.\ref{2times2order}, one can define the spin in XY plane $ S^{\pm}=S^{x}\pm i S^{y} $.
From Eq.\ref{Yxp}, we expect that the quantum information scrambling encoded in the transverse spin-spin out of time ordered correlation function (OTOC) at  a finite $ T $:
\begin{equation}
 \langle S^{+}(t, \vec{x} ) S^{+}(0, 0) S^{-}(t, \vec{x} ) S^{-}(0, 0) \rangle \sim e^{\lambda_L(t- x/v_B)}
\label{spinotoc}
\end{equation}
near the lightcone $  x=v_B t $ is greatly enhanced in the QC regime in Fig.\ref{finiteT}.
The Lyapunov exponent  $ \lambda_L$ reaches its maximal value $ \lambda_{L} \sim T $,
while the butterfly velocity $ v_{B} \sim T^{1-1/z} $. So for $ z_x=z_y=1 $, then $ v_{B} \sim const. $.
From the dimensional analysis, we conclude  $ v_{B} \sim  v_x $ in Eq.\ref{Yxp}.

\subsection{ Absence of the Quantum chaos above the complete devil staircase  }

In the complete staircase along the diagonal line $ \alpha=\beta^{+} $ away from the Abelian point $ \pi/2 $,
an IC-YZ-x/LQx phase  melts into a Luttinger liquid which
is described by a 2d CFT with the central change $ c=1 $.  Because it is integrable, so
it will not lead to any quantum chaos at a small $ T $, in sharp contrast to the
QPT from the Y-x state to the coplanar IC-XY-y state presented in Sec.VIII-A above.
In fact, for an integrable CFT, the Lyapunov exponent $ \lambda_L $ is not even defined,
the OTOC for an integrable CFT has been evaluated in \cite{RCFT}.

  It is instructive to compare the two in-commensurate co-planar phases:
  the gapped IC-XY-y and the gapless spiral IC-YZ-x/LQx phase.
  The former is ( countably or discrete ) infinitely degenerate due to the completely breaking of a discrete lattice
  symmetry along the $ y $ direction. It has only 4 Bragg peaks at   $ ( 0, \pm ( \pi-q^{0}_y ) ) $ and $ ( \pi, \pm ( \pi-q^{0}_y ) ) $
  and  gapped dis-commensurations. It leads to quantum chaos  near the two 2nd order transitions to its two neighbouring phases at a finite $ T $.
  While the latter, at the mean field level, is ( in-countably or continuously ) infinitely degenerate,
  has a dense set of  Bragg  peaks and a gapless phason mode due to the breaking of a continuous $ U(1) $ symmetry.
%  Because it has a broad distribution of Bragg peaks in the $ x $ direction, so just along x-direction, it may behave just like a
%  gapless QSL which does not show any Bragg peaks either.
   At the quantum level, it melts into a quasi-1d LQx which is described by $ c=1 $ CFT at $T=0 $ which shows no quantum chaos at a finite $ T $.
   They form a Cantor set with a fractal dimension along the complete devil staircase
on the diagonal line $ \alpha=\beta $ when $ 0 < \alpha < \alpha^{-}_{33} $.
As shown in Fig.\ref{finiteT}, despite being gapless ( conformally invariant ) and taking measure zeros, they can not be taken as QCPs.
So it makes no sense to talk about quantum critical scalings near the IC-YZ-x/LQx phases,
because there are so many other C- or IC- phases nearby.
In fact, in contrast to the effective GL action Eq.\ref{Yxp} in the continuum limit to describe the
quantum Lifshitz transition from the collinear  Y-x state to the coplanar IC-XY-y state, there is no way to
get to any continuum limit to describe the fractal structure in both the complete and in-complete devil stair cases.
One must stick to the original lattice to describe such a   fractal structure.
So the QPT/QCP  and the fractal structure in  Fig. are two completely different phenomena.

%The spiral co-planar C phases are confined phases with discrete Bragg peaks and gapped magnon excitations Eq.,
%while the IC-YZ-x/LQx phases  are similar to the deconfined QCP with broad distribution of Bragg peaks and
%gapless fractionized  spinon excitations.
%So the IC-YZ-x phase may also be called a Cantor deconfined phase.
%Similarly, there are deconfined excitations at the ( multiple )
%QCP between the columnar VBS and the staggered VBS in the quantum dimer model
%discussed in Sec. IV-A.

\subsection{ Contrast the gapped IC-SkX-y tuned by SOC with the gapless IC-SkX-$\phi$ tuned by the Zeeman field}

%  Because it keeps the translational symmetry along the $ y $ direction, it still
%  shows Bragg peaks along $ y $ direction.

  It is also instructive to compare the gapped IC-SkX-y Eq.\ref{SkXABlinear} found in Sec.V-B.
  with the gapless non-co-planar IC-SkX-$\phi$ phase Eq.\ref{icskxh} found in \cite{rhh}. The two states
  can be reached through quantum phase transitions from the Y-x phase driven by the SOC and a Zeeman field respectively. So
  the Y-x phase can be considered as the parent state of them.
  However, as shown in the appendix G, the two transitions take very different order parameters, the former is a real field,
  the latter is a complex field dictated by the $ U(1)_{soc} $ symmetry.

  As mentioned in the introduction, the Y-x phase is the exact ground state \cite{rh} along the solvable line  $ (\alpha=\pi/2, \beta) $ in Fig.1.
  There is an exact hidden spin-orbit coupled $ U(1)_{soc} $ symmetry generated by
  $ U_1(\phi)=e^{ i \phi \sum_{i} (-1)^x S^{y}_i } $ along this line. The Y-x state keeps the $ U(1)_{soc} $ symmetry.
  In the presence of a longitudinal Zeeman field  $ h_y $ along the $ Y $ direction which still keeps the $ U(1)_{soc} $ symmetry,
  in the high field case $ h> h_{c2} $, it becomes a Y-FM state. Intervening between the Y-x at low fields and Y-FM at a high field is
  the IC-SkX-$\phi$ phase. Namely, in the intermediate field strength $ h_{c1} < h < h_{c2} $,
  the IC-SkX-$\phi$ phase with the following spin-orbital configuration becomes the ground state:
\begin{eqnarray}
    S^y & = & A + B (-1)^x,   \nonumber  \\
    S^{+} &= & [C+D (-1)^x ]e^{i (-1)^x[k^{0}_y y + \phi ]}
\label{icskxh}
\end{eqnarray}
  where $   S^{\pm}= S^{x} \pm i S^{z}  $ stand for the transverse components. It is a non-coplanar phase with a non-vanishing Skyrmion density $ \vec{S}_i \cdot (\vec{S}_j \times \vec{S}_k ) \neq 0 $. Most importantly, it breaks the translation symmetry along the $ y $ axis and
   the $ U(1)_{soc} $ symmetry, therefore leads to a gapless Goldstone mode:
\begin{equation}
    \omega_{G}( \vec{k} )=\sqrt{v_{G,x}^2k_x^2+v_{G,y}^2k_y^2}-c_G k_y
\label{minusGold}
\end{equation}
  where $ c_{G} ( \beta, H )= -c_G( \pi/2-\beta, H ) $,  so $ c_G > 0 $ when $ \beta < \pi/4 $,
   $ c_G < 0 $ when $ \beta > \pi/4 $ and $ c_G = 0 $ when $ \beta = \pi/4 $
   However, it  still keeps their combination: $ y \rightarrow y+1, \phi \rightarrow \phi-k^{0}_y $
   denoted as $ [U(1)_{soc}]_{\phi \rightarrow \phi -k^{0}_y} \times ( y \rightarrow y+1 ) $.
   It dictates that the translational symmetry breaking along the $ y $ axis can be restored by making the corresponding
   rotation in the  $ U(1)_{soc} $ phase $ \phi \rightarrow \phi -k^{0}_y $. This fact
   is responsible for the gapless Goldstone mode excitation  above this IC-SkX-$\phi$ phase.
   So it can be written as the coset:
\begin{equation}
     U(1)_{soc} \times ( y \rightarrow y+1 )/[U(1)_{soc}]_{\phi \rightarrow \phi -k^{0}_y} \times ( y \rightarrow y+1 )
\label{coset}
\end{equation}
   This coset construction can be contrasted to gapless phason mode Eq.\ref{phason} in the IC-SkX-y/LQx phase
   which is dictated solely by the $ U(1) $  translational symmetry breaking along the $ x $ axis.

   As shown in \cite{rhh} and appendix G, at the mirror symmetric point $ \beta = \pi/4 $,
   the transition from the Y-x state to the IC-SkX phase
   at $ h=h_{c1} $ is in the same universality class as
   the zero density superfluid to Mott transition with $ z=2 $ with the order parameter $ \psi=e^{i \phi } $.
   As one can see from Eq.\ref{icskxh},
   $ \langle \psi \rangle =0 $ means the Y-x state, $ \langle \psi \rangle \neq 0 $ means the IC-SkX-$\phi$  state.
   The Lyapunov exponent due to the SF mode \cite{on} is $ \lambda_{L,sf} \sim T^3/\rho^2_s  $.
   So for $ z=2 $, $ v_{B} \sim \sqrt{T} $.
   From the dimensional analysis, we conclude  $ v_{B} \sim  v_{G,x} \sqrt{T} $.
%   This is different from the Y-x to the IC-XY-y transition described by the quantum Lifshitz action Eq.\ref{Yx}
%   and the associated order parameter Eq.\ref{2times2order}.

   However, as stressed in the introduction, appendix F and G, any deviation from this solvable line spoils this $ U(1)_{soc} $ symmetry.
   The Y-x state remains the classical state with quantum fluctuations.
   There is a 1st order  quantum Lifshitz transition  from the collinear Y-x phase from the right
   to the IC-SkX-y  listed in Eq.\ref{SkXABlinear} which is  a gapped phase
   and $ P_0, \phi_0 $ need to be fixed   by the cubic or 4th order term.
%   Both meet at the multi-critical point $ M $ with the $ 3 \times 3 $ non-coplanar SkX phase.
   The translational symmetry breaking along the $ y $ direction in the non-coplanar IC-SkX-y phase can not be restored
   by any rotation. This fact leads to the gapped dis-commensurations.
   Therefore, both $ \lambda_L $ and $ v_B $ are exponentially suppressed by the gap.

   For both phases' experimental relevanace to 4d/5d Kitaev materials and helical magnets in a Zeeman field see
   Sec.X-2.
%   While the $ U(1) $ continuous symmetry breaking along the $ x $ direction in the co-planar spiral IC-YZ-x/LQx
%   is restored by the quantum fluctuations of the anisotropic gapless phason mode Eq.\ref{phason}.
%   As stressed in Sec. VIII-B, one can construct an effective GL action in the continuum limit to
%   describe the quantum phase transition from the Y-x state to the IC-XY-y at $ h=0 $,
%   and the Y-x state to the IC-SkX phase at $ h=h_{c1} $ at the solvable line.
%   Both cases display quantum chaos at a finite $ T $.
%   But no such a the continuum limit can be taken to describe the fractal structures of the intertwining of  IC-YZ-x/LQx phase with its %neighbouring phases.

%   In short, due to the very different nature of the three in-commensurate phases,
%   the excitation spectra above the them  are quite different, therefore lead to quite different
%   behaviours at a finite temperature.
%   As shown in Sex.X, the DM term plays important roles leading to these IC phases.

\section{ Contrast with some previous works on different systems }

It is constructive to contrast Fig.1 with interacting SOC bosonic  system,
some other systems showing quantum Lifshitz transitions in a continuum, quantum or topological phases due to geometric or quenched disorders,  1d quasi-crystal,
2d deconfined quantum critical point, 2d fractional quantum Hall plateau-plateau transitions and
the 3d cubic code. The contrast with the 2d quantum dimer models was already made in Sec.IA.
We stress possible connections and also crucial differences between Fig.1 and these systems which could shed considerable lights
in other new systems.

{\sl 1. Contrast to the Quantum Heisenberg model with $ SU(2) $ symmetry in bipartite lattice:
        differences in the order parameter  }

  It was well known \cite{scaling,sachdev,aue} that the FM state is the exact eigen-state,
  in the non-linear sigma model effective action, the order parameter is simply a unit vector standing for the quantum spin  $ \vec{n}^2=1 $.
\begin{equation}
  {\cal L}_{FM}[\vec{n}]= i \vec{A}[\vec{n}] \cdot \partial_{\tau} \vec{n} + \frac{1}{2} \rho_s (\nabla \vec{n} )^2
\end{equation}
  which leads to one gapless mode with the dispersion $ \omega \sim k^2 $. The quantum spin is simply $ \vec{S}=\vec{n} $.

  For an AFM in a bipartite lattice, the AFM is just the classical ground state,
  the order parameter is the staggered magnetization $ \vec{n}=\vec{S}_1- \vec{S}_2 $
  of the unit cell,  while the uniform magnetization $ \vec{L}=\vec{S}_1 + \vec{S}_2 $ is subleading and can be integrated out.
\begin{equation}
  {\cal L}_{AFM}[\vec{n}]= \frac{1}{2} [ \chi_s  ( \partial_{\tau} \vec{n} )^2 + \rho_s (\nabla \vec{n} )^2]
\label{nafm}
\end{equation}
  which leads to the two transverse Goldstone modes with the dispersion $ \omega \sim k $. The quantum spin is
   $ \vec{S}= (-1)^i \vec{n} + \vec{L} $.

  Here, due to the SOC which breaks the $ SU(2) $ symmetry completely, the order parameter is much more complicated.
  For example, inside the Y-x phase, due to the two sublattice structure,  $ p^{(\pm)} $ and $ q^{(\pm)} $ set become conjugate variables.
  In the C- regime, the uniform pair $ ( p^{(+)}, q^{(+) }  ) $ become the relevant degree of freedoms, the other half
  ( the staggered pair ) are projected out.
  Then $ q^{(+) } $ becomes massive and can be integrated out to reach a final effective action in terms of critical mode
  $ p^{(+)} $. It leads to some commensurate phases in XY plane.
  In the IC-regime, $ \tilde{p}^{(\pm)} $ and $ \tilde{q}^{(\pm)} $ set become conjugate variables which are related to $ p^{(\pm)} $ and $ q^{(\pm)} $ just by a unitary transformation.
  Then the uniform pair $ ( \tilde{p}^{(+)}, \tilde{q}^{(+) }  ) $ become the relevant degree of freedoms, the other half ( the staggered pair ) are projected out.
  Then $ \tilde{q}^{(+) } $ becomes massive and can be integrated out to reach a final effective action in terms of critical mode
  $ \tilde{p}^{(+)} $ which is still equal to $ p^{(+)} $ at the low energy limit.

  In short, for both C- and IC-magnons, the order parameter for the Y-x phase is
\begin{equation}
  p^{(+)} =\sqrt{2} S \delta \eta
\end{equation}
  For the IC-magnon, there is an associated $  q^{(-)} $ listed in Eq.\ref{spinOP2}  which
  leads to the non-coplanar IC-SkX-y phase. So here it is always the uniform part of the unit cell which
  is the order parameter in both cases. The gapped Y-x phase may be contrasted with the FM state and the AFM state.
  While the gapless IC-YZ-x/LQx phase with the phason mode Eq.\ref{phason}
  in Fig.1 may be contrasted to the gapless FM mode with $ \omega \sim k $  and AFM state with the
  two transverse Goldstone modes $ \omega \sim k^2 $. The topological defects such as instantons
  have been discussed in the AFM. It is not know if they also play important roles in the SOC case.

  This crucial observation can be extended to the co-planar state in the YZ plane at $ W=1/N $ in Eq.\ref{Ntimes1spin} and
  the IC-YZ-x/LQx phase at $ W=\alpha $ in Eq.\ref{phasonact}. After considering the twist from the XY plane to the YZ plane,
  one can write down the order parameter $ q^{(+)} =\sqrt{\frac{N}{n}} S \delta \xi $ and
  $ q^{(+)} =\frac{1}{\sqrt{\alpha}} S \delta \xi $ for the two cases respectively.
  They were already used to derive Eq.\ref{Ntimes1eff} and Eq.\ref{phasonact}  (where
  $ \alpha $ was absorbed into the definition of $ K $ ) respectively.
  For the fermion case, we expect it may change to the staggered part \cite{rafhm}.

{\sl 2.  Contrast to the interacting SOC bosonic system: the connection between the effective actions and the OFQD }

   From the insights gained to derive the low energy effective actions Eq.\ref{ppplusaction} and Eq.\ref{onlytildep}  inside the Y-x phase
   corresponding to the C- and IC- magnons respectively, then from the general
   symmetry principle, we constructed the effective actions Eq.\ref{ppphi3} and Eq.\ref{etaeta3}  to describe the transitions from the
   Y-x phase to the X-y and IC-SkX-y phase respectively. It is difficult to evaluate the numerical values of
   the cubic $ \lambda $ and the quartic $ \kappa $  term by a microscopic calculation.
   However, when along the diagonal line near the Abelian point
   $ \alpha=\beta =\pi/2 $, the $ [C_4 \times C_4]_D $ symmetry dictates the cubic term must vanish $ \lambda $.
   One can also evaluate the numerical value of the quartic $ \kappa $  term by the OFQD as shown in Sec.III.
   This brings an intrinsic and deep  connection between the parameters in the effective action with a microscopic calculation by OFQD.
   This deep connection was established in the RFHM which is a quantum spin system in  Eq.\ref{rhgeneral}.
   It may be contrasted to its counterpart in the context of  interacting spinor SOC systems   \cite{pifluxgold,pifluxqsl,SFnon,NOFQD,SFQAH}.
   In the intermediate couplings, we developed a symmetry based effective action
 to classify all the possible phases, especially possible quantum spin liquids (QSL) phases with topological orders.
 We contrast the effective action with the microscopic calculations at both weak and strong
 coupling, therefore establish the mappings between the phenomenological parameters in the action and the bare parameters in the microscopic Hamiltonian, especially a deep and profound connection between the phenomenological action and
 the effective potentials generated by the OFQD mechanism.

{\sl 3. Contrast to other systems displaying quantum Lifshitz transitions in a continuum with a roton ring: The dual role of the Conjugate pairs  }

  It is very instructive to compare with the Bilayer quantum Hall (BLQH) systems \cite{blqh},
  which hosts the exciton superfluids (ESF)
  at the total filling factor $ n_T=1 $. It may also be called QH ferromagnet (QHFM) systems.
  The phase and density of excitons are conjugate variables.
  At short distance,  the phase mode becomes critical, it is convenient to integrate out the density mode to study the
  gapless quantum phase fluctuations inside the ESF.
  However, as the distance increases, the roton ring drops at a finite momentum,
  the density fluctuation become critical at a finite momentum, so it is convenient to
  integrate out the phase fluctuations to study the exciton superfluid to a possible pseudo-spin density wave (PSDW) transition.
  One can also construct a quantum Lifshitz action to study such a transition.
  However, because the BLQH systems are on a continuum, so there is a translational and rotational symmetry,
  the roton minima is in a roton ring.
  Due to the exchange $ Z_2 $ symmetry between the two layers in the balanced case, the cubic terms is absent which is crucial to determine the underlying PSDW to be a square lattice
  instead of a triangular lattice. However, for the quantum Lifshitz
  transition from a quantum Hall state to a Wigner crystal state in a single layer discussed in the appendix B of  \cite{blqh},
  the cubic term exists which indeed leads to a triangular lattice.
  Similar quantum Lifshitz transitions driven by the roton dropping in a continuum system such as a superfluid Helium,
  exciton bilayer systems, Larkin-Ovchinnikov-Fulde - Ferrell (LOFF) state  were addressed
  in \cite{SS1,SS2,SSrev,loff}. The general lattice structure of the resulting phases was classified in \cite{loff}.
  Here, due to the underlying square lattice and the SOC,
  the "roton" minima of the IC-magnons are located at two discreet points $ ( 0, \pm k^0_y ) $.

  Notably, it is instructive to compare with the quantum Lifshitz transition in the repulsively interacting spin-orbit coupled Fermi gas \cite{fm} with the 3d Weyl coupling. A putative ferromagnetic state (FM) is always unstable against a stripe spiral spin density wave (S-SDW) or a stripe longitudinal SDW (LSDW) at small or large SOC strengths, respectively.
  The stripe-ordering wave vector is given by the nesting momentum of the two SOC-split Fermi surfaces with the same or opposite helicities at small or large SOC strengths, respectively.
  The LSDW is accompanied by a charge density wave (CDW) with half of its pitch.  The transition from the paramagnet to the SSDW or LSDW+CDW is described by quantum Lifshitz-type actions, in sharp contrast to the Hertz-Millis types for itinerant electron systems without SOC.
   The collective excitations and Fermi surface reconstructions inside the SSDW and LSDW+CDW are also studied.
   When comparing with Fig.\ref{finiteT}, we can see some interesting analogy:
  at small SOC,  the putative FM state is always unstable to any small SOC,
  they are  stripe spiral co-planar phases characterized by  $ W $.
  At large SOC, there is twist from YZ plane to the XY plane,
  then to the collinear. There is a non-co-planar $ 3\times 3 $ SkX acting as the hub phase. Unfortunately,
  the original paper \cite{fm} is in-conclusive at intermediate SOC couplings $ 0.5 < \gamma < 1.5 $.
  This analogy suggests there could be some non-coplanar itinerant magnetic phases at this intermediate SOC
  which act as the hub phase bridging the SSDW and LSDW+CDW.

{\sl 4. Contrast to geometric frustrations: possible Quantum spin liquids (QSL) }

   The order from quantum disorder (OFQD) phenomena at the isotropic Rashba limit $ \alpha=\beta $
   in the quantum phase transition regime $ \alpha_{33} \sim 0.34 \pi < \alpha < 0.5 \pi $ in Fig.\ref{finiteT}
   is due to the Rashba SOC which is a completely different mechanism than the geometric frustrations \cite{sachdev,aue,wen,SLrev1}.
   All the magnetic phases
   in Fig.1 can be contrasted to the co-planar phases in geometrically frustrated magnets \cite{scaling,sachdev,aue,wen}.
  For example, for an AFM in a frustrated lattice such as a triangular lattice,
  the classical ground state has the  $ 120^{\circ} $ co-planar structure,
  the order parameter is the $ SO(3) $ matrix characterized by the 3 orthogonal unit vectors
  $ \vec{n}_1, \vec{n}_2, \vec{n}_3= \vec{n}_1 \times \vec{n}_2$. The effective action is given by
\begin{equation}
  {\cal L}_{120^{\circ}}[\vec{n}]= p_1  [ (\partial_{\mu} \vec{n}_1 )^2 + (\partial_{\mu} \vec{n}_2 )^2] + p_3  (\partial_{\mu} \vec{n}_3 )^2
\label{n123}
\end{equation}
  which leads to 3 Goldstone modes \cite{sachdev}. Setting $ p_1=0 $ leads back to Eq.\ref{nafm},
  then the 3 Goldstone modes reduce to two.
  The quantum spin of the co-planar state is $
  \vec{S}= \vec{n}_1 \cos \vec{Q} \cdot \vec{x}  + \vec{n}_2 \sin \vec{Q} \cdot \vec{x} =
  (\vec{n}_1 -i \vec{n}_2) e^{i \vec{Q} \cdot \vec{x} }+ (\vec{n}_1 +i \vec{n}_2) e^{-i \vec{Q} \cdot \vec{x} } $
  where $ \vec{Q}= 4\pi(1/3, 1/\sqrt{3} ) $ is the ordering wavevector.

   It was also speculated that strong quantum fluctuations in a Kagome lattice may melt
   any magnetic ordered state to a possible quantum spin liquids (QSLs) \cite{SLrev1}.
   In fact, the SOC could also be a new mechanism leading to new classes of quantum spin liquids even in a bipartite lattice
%   In the presence of SOC,  the spin rotation symmetry is completely broken, so LSMOH theorem \cite{lsm,Os,has} may not apply.
%   Then the possible new mechanisms \cite{filling} only assuming  $ {\cal T } $ may apply to search for possible spin liquid states
%   in the presence of SOC in some lattices. The RFHM with a generic SOC parameter $ ( \alpha,\beta) $ does keep the  $ {\cal T } $
%   symmetry and also holds spin $ 1/2 $ per unit cell.
%  So in principle, it may support some spin liquid
   which may have a good chance to be sandwiched between two commensurate magnetic phases.
   However, in Fig.\ref{phasedia}, sandwiched between
   any two commensurate magnetic phases are some IC- phases:
   for example, along the diagonal line, the IC-XY-y phase intervening between the $3\times 3 $ and the Y-x phase,
   away from the diagonal line, the IC-SkX-y phase intervening between the $3\times 3 $ and the Y-x phase from the right.
%   IC-SKX or IC phase. However,

   Along the diagonal line, the gapless spiral co-planar IC-YZ-x/LQx phases form a Cantor set with a fractal dimension.
   Away from the diagonal line, it take a finite measure. Near the Abelian line,   all the gapped C phases embedded in the sea
of the IC phases have very small gaps and small widths, so all the interleaved C and IC phases near the Abelian line
show fractal structures. It remains challenging to resolve the fractal structures by the resonant magnetic X-ray diffractions \cite{kitaevlattice,kitaevlattice1,kitaevlattice2}. If not, one may just see similar experimental signature as a 2d gapless QSL.
   There should be intimate connections between In-commensurability and QSLs.
   For possible QSLs in a honeycomb or a cubic lattice due to the SOC,
   see the conclusion Sec.XI-2 and -5.

From a different perspective, it maybe also inspiring to contrast with the possible 2nd order deconfined quantum critical point (DQCP)
between a VBS and a Neel state \cite{senthil} in a square lattice.
The two states on the two sides of the DCQP break completely different symmetries of the Hamiltonian.
There are gapless deconfined spinons right at the DQCP. Its gapless is due to its QCP.
Here, the co-planar C-phases at $ W=n/N   $ along the complete devil staircase $ \alpha= \beta^{-} $ break
different lattice translational symmetries. The IC-YZ-x/LQx phases
immerse inside these C phases and form a Cantor set with a fractal dimension.
As stressed in the previous paragraph, they do not act as QCPs with the dynamic exponent $ (z_x=0, z_y=1 ) $.
So there is no associated QC scalings. Its gapless is protected by the IC- and associated $ U(1) $ lattice symmetry breaking
instead of a QCP.
Similarly, along the complete devil staircase $ \alpha= \beta^{+} $, the IC-XZ-y/LQy phases do not act as
QCPs with the dynamic exponent $ (z_x=1, z_y=0 ) $.

%   In fact, both IC phases are particularly vulnerable to some small parameter changes.
%   So it would be important to study if a spin liquid phase can creep in a honeycomb lattice with
%   three SOC parameters $ ( \alpha, \beta, \gamma ) $ putting on its three bonds.
%   For example, it is tempting to see if the two kinds of IC phases will be fractionized into chiral
%   spin liquids or spin  liquids under the third SOC parameter $ \gamma $ in a honeycomb lattice.
%   Indeed, there remains some possible still unknown deep connections among the in-commensurability,
%   the hierarchy structures  of fractals, spin liquids and their fractionalized excitations.

{\sl 5. Contrast to quenched disorders:  Quantum chaos versus fractals}

    The strong correlations and quenched disorders lead to a new class of state of matter: quantum spin glass or gapless quantum spin liquids \cite{SY,kittalk,glass1,glass2}.
    The multiple local ( meta-stable ) states, hysteresis and fractals in Fig.\ref{phasedia} may resemble the complex multiple local
    minimum landscape in quantum spin glass (QSG).  In fact, we suspect this landscape in the QSG may be just multi-fractals !
    However, the former is SOC induced, the latter is due to quenched disorders.
    So the SOC may induce some similar complex quantum glassy phenomena as the disorders do.
    For example, all the spiral commensurate co-planar phases near $ \alpha =\frac{\pi}{N}n $ embedded in the soups of
    the IC-YZ-x/LQx phases near the Abelian line $ \beta \ll 1 $ can be contrasted to the QSG phase.
%    Indeed, as mentioned above, the coplanar IC phase in Fig.\ref{phasedia} completely breaks the translational symmetry along
%    the $ x $ axis. Because it has a broad distribution of Bragg peaks, so it may also behave like a glassy phase.
%    The excitations above all the C phases in Fig.1 are just magnons.
%    But it is not still known what are the excitation spectra above the non-coplanar IC-SkX and coplanar ( spiral ) IC in Fig.1.
    As to be stressed in the conclusion section XI-4, the possible connections between the fractals in Fig.1 due to the SOC
    and the quantum  chaos in the $ 0+1 $d gapless quantum spin liquids or  quantum spin glass in the SYK models \cite{SY,kittalk,syk1,syk2,syk3} or quantum rotor models \cite{glass1,glass2}  due to the quenched disorders  deserve to be explored further.
%    \cite{syk0,mess3,kittalk,syk1,syk2}.

{\sl 6. Contrast to topological states in ( non-interacting )
        1d Aubry-Andre (AA) model and its associated Hofstadter problem: the role of In-commensurability  }

   There are previous theoretical \cite{QC1} and experimental  works \cite{QC2} on topological states in 1d Aubry-Andre (AA) model
   ( also known as the Harper model \cite{AA} with a
   quasi-periodic ( or in-commensurate ) potential  $ V_H= \lambda \cos( 2 \pi \alpha n + \phi ) $.
   When $ \alpha $ becomes irrational, the bulk energy spectrum becomes independent of $ \phi $.
   The Chern number defined for the whole family $ -\pi < \phi < \pi $ also becomes independent of $ \phi $.
   So a Chern number can also be defined for each $ \phi $. Then as sweeping $ \phi $ from $ - \pi $ to $ \pi $, there
   are boundary states transverse across the bulk gaps. However, when $ \alpha =p/q $ is  a rational number, namely,
   $ V_{AA} $ is an commensurate  potential, the bulk spectrum is invariant only by translating the lattice for
   $ m=1,2,\cdots q $ sites which correspond to $ q $ shifts of $ \phi=2\pi/q \times (1,2,\cdots,q) $.
   So the bulk spectrum depends on $ \phi $ with the periodicity $ 2\pi/q $.
   The Chern density also depends on $ \phi $ with the periodicity $ 2\pi/q $.
   Because the Chern number is the integration of $ \int^{2\pi/q}_0 d\phi $, which is associated  with the whole family of
   1d systems, a single 1d system belongs to a trivial phase.
   This is the main difference between $ \alpha $ is rational or irrational.
   It can be contrasted with the spiral C phases when  $  W = \frac{n}{N} $ is rational or spiral IC-YZ-x/LQx phases
   when $ W $ becomes irrational,
   the former has a gap and discrete Bragg peaks, while the latter has a gapless phason mode protected by
   the irrationality and associated $ U(1) $ symmetry breaking. It shows no Bragg peaks, so resemble a gapless QSL
   to some extent.

   In fact, these 1d topological properties of AA model inherit from those of the 2d electrons hopping in an Abelian flux $ \alpha=p/q $,
   namely, the  Hofstadter problem \cite{hh}, with $ \phi =ka $ identifying as the extra momentum in the 2d lattice,
   so playing the role of an "synthetic" dimension. When the flux $ \alpha=p/q $ is rational, the spectrum has $ q $ bands
   with $ q $ energy gaps. When the Fermi surface is within one of the $ q $ gaps, there is an integer
   quantized Hall conductance $ \sigma_H= \nu e^2/h $
   with $ \nu $ as the number of filled bands \cite{TKNN}.
   As $ q \rightarrow \infty, p \rightarrow \infty $, but the ratio $ \alpha=\frac{p}{q} $ approaching an irrational number,
   the spectrum becomes a Cantor set, the wavefunctions and energy gaps exhibit
   a fractal structure displaying self-similar behaviours. Assuming the first time a gap within this fractal structure emerges
   at some $ p $ and $ q $, then the gap remains as $ q \rightarrow \infty, p \rightarrow \infty $ and approaches $ \alpha=\frac{p}{q} $.
   When the Fermi surface is within this gap, there is a quantized Hall conductance $ \sigma_H= \nu e^2/h $
   with $ \nu $ satisfying the diophantine equation $ tp+rq=1 $.
   This so called fractal integer QH has been experimentally realized in Moire superlattice due to a
   slight twisted angle between the two layers
   of graphene sheet \cite{BLgraphene}.
   Of course, these are 1d or 2d free electron systems,
   ours are 2d quantum spin models subject to SOC. There are always dramatic differences between interacting and non-interacting systems.

 {\sl 7. Contrast to 2d fractional QH plateau-plateau transitions: its chiral edge state protected by
 the bulk-edge correspondence }

This picture may also be contrasted to fractional QH plateau-plateau transitions \cite{hlr,zwang1,zwang2,QH1,QH2}.
The precise nature of the fractional  QH plateau-plateau transitions
is still not known yet, but it maybe a 2nd order transition with the dynamic exponent $ z=1 $. So there is a gapless state intervening
between the two plateaus. There should be universal quantum Hall conductances at the QCP.
The two plateaus are two different topological phases with different quantum Hall conductances,
the gapless quantum critical state between the two topological states may be described by a $2+1 $ dimensional CFT.
If  replacing the spin operator in Eq.\ref{spinotoc} by the electron operators, it clearly leads to quantum chaos at a finite $ T $.
Here the topological winding numbers $ W $ defined in Sec.VII-A is a Cantor function which has zero derivative everywhere except
in the Cantor set with zero measure. The Cantor set consists of the IC-YZ-x/LQx phases which do not act as QCPs !
So there is no associated QC scalings. The gapless of the IC-YZ-x/LQx phase is protected by the IC- and associtaed $ U(1) $ symmetry breaking
instead of being a QC point.
%the quantum Hall edge states to some extent.
%The former is the critical state between the two topological states in the parameter space such as the filling factor, the latter
%the critical state between the two topological states ( or one topological, the other is an insulator ) in the real space.

%This analogy may suggest that the quantum Hall conductances versus the filling factors in the fractional quantum Hall
%effects may also form a complete devil staircase, the
%QCPs of the fractional QH plateau-plateau transitions may form a Cantor with a fractal dimension.
As stressed in Sec.VIII-B,  the quantum Hall conductances versus the filling factors in the fractional quantum Hall
effects consist of QH plateaus and QCPs between plateaus. This organization pattern is nothing but the QPT, so
quite different than a complete devil staircase
of the topological winding numbers $ W=n/N $ versus the SOC parameter $ \alpha=\beta $ in Fig.\ref{cantor}.
Even so, there could be also some deep connections between the fractals and
the hierarchy of fractional quantum Hall Abelian or non-Abelian states.

It maybe instructive to  contrast spiral IC-YZ-x/LQx phase with the edge mode of a quantum Hall state which also has the same dispersion
$ E( q_y )=   v_y q_y  $ for a edge along $ y $ direction, so $ q_y $ remains a good quantum number.
The former is a LQ respecting the Time-reversal symmetry, the latter is a chiral Luttinger liquid ( CLQ )
 due to the explicit Time-reversal symmetry breaking.
Due to the two edges at $ x=0, x= L $, $ q_x $ stops to be a good quantum number.
However, the physical origins of the gapless mode is completely different, here it is a Goldstone mode
due to the translational $ U(1) $ symmetry breaking, while that in the QH is due to topological bulk-edge correspondence
instead of a symmetry breaking.

{\sl 8. Contrast to 3d cubic model: low dimensional excitations embedded in a higher dimension }

   There are several salient features of the cubic model \cite{cubic} and its extension called fracton model which maybe contrasted with
   the global phase diagram Fig.1:
(1) because of its hierarchy structure of fractals, it can never be effectively described by a smoothed-out ( or coarse-graining ) description of the underlaying cubic lattice, so it is beyond a continuum quantum field theory. In this regard,
 it is similar to the fractal structure
shown in Fig.1 and  the gluing rule 2 in the right box of Fig.2 which also defies a global continuum quantum field theory description.
 However, starting from
a general action Eq.\ref{generalS}, we are still able to come up with
a continuum quantum field theory description to describe all the QPTs in the gluing rule 1 in the left box of Fig.2.
Eq.\ref{Ntimes1eff} to describe all the C- co-planar phases with $ W=n/N $.
Eq.\ref{phasonact} to describe the melting of the IC-YZ-x into a quasi-1d Luttinger liquid.
Of course, Eq.\ref{generalS} is well defined on a square lattice.
%As stressed in Sec.VIII-B, there is no way to construct a continuum quantum field theory from it to describe the fractal structure in Fig.1.
(2) It contains localized excitations, or confined to move along a line or in a plane in real space.
    This is in sharp contrast to the non-interacting topological materials with point like excitations such as Dirac or Weyl points,
    line nodes or plane nodes in momentum space.
%    Of course, the localized excitations in real space is dramatically different than those in the momentum space.
    In the present problem, the  spiral co-planar IC-YZ-x/LQx phase in Fig.1
%host a ground state with extensive ground state degeneracy.
    supports the 1d gapless  phasons Eq.\ref{phason} and
    represents a quasi-1d Luttinger liquid embedded in a 2d square lattice.
    It spin-spin correlation function Eq.\ref{phasonspin} is static and has no decay along x-axis, but only algebraic decay along y-axis
    with an exponent depending on the SOC parameter $ (\alpha, \beta) $.
    It maybe interesting to explore if all the foliated topological phases in the cubic model are organized in terms of in-complete/complete devil staircase.
% (5) Due to the above features, the cubic code could be useful for quantum cryptography or topological quantum computing.

\section{  Implications on cold atom experiments and 4d or 5d Kitaev materials with SOC. }

In this section, we discuss the applications of our results in cold atoms, 4d/5d materials, high $ T_c $ cuprates and lattice QCD
with a parity violation respectively.

{\sl 1. Implications in cold atom experiments in the strong coupling limit: possible heating issue }

 There have been some remarkable experimental advances to  generate various kinds of 2D SOC for cold atoms in both continuum and optical lattices.
A 2d Rashba SOC was implemented by Raman scheme in the fermion $ ^{40} K $ gas \cite{expk40,expk40zeeman}.
Using an optical Raman lattice scheme, Wu {\sl et al} \cite{2dsocbec}  realized the tunable quantum Anomalous Hall (QAH) SOC
of spinor bosons $ ^{87} Rb $ in a square lattice.
% The heating issue still hinders the observation of true many body phenomena, especially for alkali fermions.
More recently, the fermionic optical lattice clock \cite{clock} scheme was successfully implemented to generate a strong SOC
for $ ^{87} Sr $ clock \cite{clock1}, $ ^{173} Yb $ clock \cite{clock2} and also $ ^{87} Rb $ \cite{SDRb},
where  the heating and atom loss from spontaneous emissions are eliminated, the exceptionally long lifetime
$ \sim 100 s $ of the excited clock state have been achieved.
%The " synthetic dimensions " idea has also been used \cite{SDRb} to generate a strong SOC  in an effective
%two-dimensional manifold of discrete atomic momentum states of $ ^{87} Rb $.
%They open up a new frontier of combining clock precision measurement, metrology and many body phenomena unique to SOC.
In parallel, by using the most magnetic fermionic element dysprosium to eliminate the heating due to the spontaneous emission,
the authors in \cite{ben} created a long-lived SOC gas of quantum degenerate atoms. The long lifetime of this weakly interacting SOC degenerate
Fermi gas will also facilitate the experimental study of quantum many-body phenomena manifest at longer time scales.
These ground-breaking experiments set-up a very promising platform to observe novel many-body phenomena shown in Fig.\ref{phasedia} due to
interplay between SOC and interaction in optical lattices.
%An optical lattice clock scheme  was proposed to generate a 2d SOC in an optical lattice, it has the advantages to suppress the possible heatings issue.

  The thermodynamic quantities such as magnetization, uniform and staggered susceptibilities, specific heat
  at the low temperatures for Y-x phase have been done in Eq.\ref{mc}.
  One can similarly work out these thermodynamic quantities  in all the phases in Fig.\ref{phasedia}.
  However, the SSCFs are much more involved.
  The orbital ordering wavevectors and spin-orbital structures  of all the phases in Fig.1 have been listed in the previous sections,
  and appendix F.
  As said below Eq.\ref{rhgeneral}, there is no spin-orbital coupled $ U(1)_{soc} $ symmetry anymore away from
  the solvable line $ (\alpha=\pi/2, \beta) $,
  so one need to calculate the $ 3 \times 3 $ tensor $ \langle S_{i}(\vec{k},\omega) S_{j}(-\vec{k},-\omega) \rangle $ spin-spin correlation functions (SSCFs)
  which, in principle, can be achieved from the general path integral approach Eq.\ref{Yxpathk} in the appendix C
  or the canonical quantization approach Eq.\ref{H2Yx} in the appedix A.
  However, as explicitly pointed out in the appendix C-3, even for the simplest Y-x state, such a calculation can not lead to any physically
  transparent results. As summarized in Sec.IX-1, it is highly non-trivial to identify the low energy modes even  in the simplest Y-x phase.
  So in the following appendix D,E,F, we derived low energy effective actions corresponding to
  C- and IC- magnons respectively which lead to quite different behaviours in the leading SSCFs in the two cases.
  One can extend these procedures to all the other phases in Fig.1,
  so one can work out various kinds of SSCFs  in all the phases in Fig.\ref{phasedia}
  such as Eq.\ref{Ntimes1spin} and Eq.\ref{phasonspin} at the low and high temperatures.
  In the cold atom contexts, all these physical quantities can be detected by atom or light Bragg spectroscopies \cite{lightatom1,lightatom2},
  specific heat measurements \cite{heat1,heat2} and  the {\sl In-Situ } measurements \cite{dosexp}.
  In materials, they can be easily measured by magnetic resonant X-ray diffraction or neutron scattering techniques \cite{kitaevlattice,kitaevlattice1,kitaevlattice2}.

  Unfortunately, the RFHM Eq.\ref{rhgeneral} can only be reached in the strong coupling limit where the heating issues remain
  serious in the current cold atom experiments. Now, we turn its more promising applications in the strongly correlated 4d
  or 5d materials with strong SOC.

%Most recently, using the optical Raman lattice scheme, the authors in the experiment \cite{2dsocbec} realized the 2d SOC for $ ^{87}Rb $
%with tunable $ (\alpha, \beta) $ in a square lattice.
%However, so far, the experiments are still in very weakly interacting regime where the systems are in various spin-orbital correlated superfluid states
%which will be presented in a separate publication.
%As estimated in Sec.VII of Ref.\cite{rh}, the strong coupling regime can be easily realized in a relatively deep optical lattice.
%It seems the possible heating issues in such a strong coupling regime are well under control.
%So the novel phases and phase transitions in Fig.\ref{phasedia} are ready to be
%explored in the experiment setup in \cite{2dsocbec} and also in \cite{clock} using the atomic clock scheme.

{\sl 2. Implications on 4d or 5d materials: the effects of the DM term leading to the IC-SkX-y phase }

 In the so called  5d-orbital Kitaev materials such $ A_2 Ir O_3 $ with $ A=Na_2, Li_2 $ or more recent 4d-orbital materials $ \alpha-Ru Cl_3 $, so far,
   only Zig-Zag phase or an IC-SkX phase were observed experimentally \cite{kitaevlattice,kitaevlattice1,kitaevlattice2},
   no quantum spin liquids ( QSLs ) have been found. For example, an IC-SkX phase with the ordering wavevector $ \vec{q}=(0,0,q), q= \pi + \delta, \delta \sim 0.14 \pi $ lying along the orthorhombic $ \vec{a} $ axis
was also detected on 3d hyperhoneycomb iridates $\alpha,\beta,\gamma$-Li$_2$IrO$_3$
by resonant magnetic X-ray diffractions \cite{kitaevlattice,kitaevlattice1,kitaevlattice2}.

The IC-SkX-y phase of the 2d RFHM Eq.1 are quite similiar to the IC-SkX phase detected in these Kitaev materials.
Naively, due to its microscopic bosonic nature, the RFHM Eq.\ref{rhgeneral}  may not be useful to describe the magnetism in
these so called Kitaev materials such as Iridates or Osmates \cite{SLrev1}. However,
as shown in \cite{rh}, the RFHM along the diagonal line $ \alpha=\beta $ can be written as the Heisenberg- ferromagnetic Kitaev  \cite{kit}-Dzyaloshinskii-Moriya (DM) form.
\begin{equation}
 H_{R}  = -J [  \sum_{\langle i j \rangle  } J_{H} \vec{S}_{i} \cdot \vec{S}_{j} +
     \sum_{\langle i j \rangle a } J_{K} S^{a}_{i} S^{a}_{j}  + \sum_{\langle i j \rangle a } J_{D} \hat{a} \cdot \vec{S}_{i} \times \vec{S}_{j}]
\label{expand}
\end{equation}
 where $ \hat{a}=\hat{x}, \hat{y} $ and $  J_{H}= \cos 2 \alpha, J_{K}= 2 \sin^{2} \alpha, J_{D}=  \sin 2\alpha $.
One can estimate their separate numerical values near $ \alpha= \alpha^{0}_{in}=\arccos \frac{1}{\sqrt{6}} $
in the IC-SkX-y phase with the ordering wavevectors $ ( 0, \pm ( \pi-q^{0}_y ) ) $ and
 $ ( \pi, \pm ( \pi-q^{0}_y ) ) $ where  $ 0.18 \pi <  q^{0}_y <  \pi/3 $ in the inset of Fig.\ref{phasedia}:
the Heisenberg interaction $ J^{x}_{H}=  J^{y}_{H} =\cos 2 \alpha \sim -2/3 $, so it is an AFM coupling,
the Ferromagnetic Kitaev interaction $ J^{x}_{K}=  J^{y}_{K}= 2 \sin^{2} \alpha \sim 5/3 $,
the DM term $  J^{x}_{D}=  J^{y}_{D}= \sin 2\alpha \sim \sqrt{5}/3 $.
So the model becomes a dominant FM Kitaev term plus a small AFM Heisenberg term and a small DM term.
In fact, setting $ \alpha=\pi/4 $, the Heisenberg term drops out, the Ferromagnetic Kitaev term
$ J^{x}_{K}=  J^{y}_{K}= 1$ and the DM term $  J^{x}_{D}=  J^{y}_{D}= 2  $.

So the RFHM Eq.\ref{rhgeneral} could be an alternative minimal model to the
Heisenberg-Kitaev-Ising  $(J,K,I)$  model used in \cite{kitaevlattice1,kitaevlattice2}
or Heisenberg-Kitaev-Exchange ($J,K,\Gamma$) model used in \cite{kim,kim1}
to fit the experimental data phenomenologically.
One common thing among all the three models is a dominant FM Kitaev term plus a small AFM Heisenberg term.
Of course, the FM Kitaev sign in these materials originates from
the Hunds rules instead of the bosonic spinor  nature of the underlying microscopic models.
It was known that there is no such IC-SkX phase in the Heisenberg-Kitaev model with only $ ( J, K ) $ term.
So the DM term in our RFHM plays quite important roles. In fact, there are always appreciable DM terms
in these materials, especially in Herbertsmithite \cite{SLrev1,SLrev2,SLrev3}.

%  Indeed, as demonstrated in Sec.X,
%  the IC-SkX-y phase display very similar properties as those observed in the 4d or 5d Kitaev materials.

Various IC-SkX phase have also been observed in some helical magnets with a strong Dzyaloshinskii-Moriya (DM) interaction \cite{dm1}.
Indeed, a 2D Skyrmion lattice has been observed between $ h_{c1}=50$ mT  and $ h_{c2}=70$ mT
in some chiral magnets MnSi or a thin film of Fe$_{0.5}$Co$_{0.5}$Si \cite{sky4}.
These 2d Skyrmion lattices are induced by a magnetic field which breaks the Time reversal symmetry explicitly.
The IC-SkX-$\phi$ phase discovered in \cite{rhh} and revisited in Sec.VIII-C can be used to explain this phenomena induced by a Zeeman field.
Of course, there is no $ U(1)_{soc} $ symmetry, so should be gapped magnons in MnSi.
%The IC-XY-y phase in Fig.1 happens at zero field and breaks the Time reversal symmetry spontaneously.
%In the presence of a Zeeman field along the $ z $ direction, we expect it to become a non-coplanar IC-SkX phase.

  There are many previous works \cite{SLrev1} on the more conventional SOC   $ \vec{L} \cdot \vec{S} $ which is even
  under both parity and time reversal, so a scalar. It is important in many 4d or 5d transition metal oxides.
  As shown here,  one can simply take the RFHM Eq.1 near the IC-SkX-y regime in Fig.1 as a suitable effective quantum spin model to
  describe these so called Kitaev materials. Of course, here the DM term which breaks the parity plays some important roles.

{\sl 3. Compare the  IC-XY-y  to in-commensurate co-planar magnetic phases in high $ T_c $ cuprates }

There are previous theoretical works on in-commensurate spin density waves (IC-SDW) in the $ J_1-J_2-J_3 $ frustrated
    quantum Heisenberg model  with the spin $ SU(2) $ symmetry on a square lattice \cite{j1j2j3}.
    The in-elastic neutron scattering experiments \cite{Csdw} on the high $ T_c $ cuprate  La$_{2-x}$Sr$_{x}$CuO$_4 $
    indeed found that the magnetic peak at momentum $ ( \pi,\pi) $ in the AFM state near half filling splits into four incommensurate peaks at
    $ ( \pi \pm \delta, \pi \pm \delta ) $ in the underdoped and superconducting regime. The
    incommensurability $ \delta $ scales as the doping concentration $ x $.
    It was known that this IC-SDW is co-planar and is due to the geometric frustrations in the quantum Heisenberg model.
    Our theoretical work here discovered that along the diagonal line $ \alpha_{33} < \alpha < \alpha_{in} $ in Fig.1,
    the state is a mixed state of the two In-commensurate states with any ratio: the IC-XY-y with  4 incommensurate peaks at
    $ (0, \pm(\pi-q^{0}_{y} ) )$ and $ (\pi, \pm(\pi-q^{0}_{y} )) $ and the IC-XY-x with  4 incommensurate peaks at
    $ ( \pm(\pi-q^{0}_{x}), 0 ) $ and $ ( \pm(\pi-q^{0}_{x}), \pi ) $, $ \pi-\pi/3< q^{0}_{x}=q^{0}_{y}<\pi-q_{ic} $ is determined by the SOC
    parameter $ \alpha=\beta $,   the two phases are related by $ [C_4 \times C_4]_D $ symmetry.
    So the geometric frustrations and the SOC are two completely different mechanisms leading to the
    two in-commensurate co-planar phases respectively
    which also display some similar properties.

{\sl 4. Implications on lattice QCD calculations with a parity violation. }

  Note that the SOC studied here is the Weyl type of SOC $ \vec{k} \cdot \vec{S} $ in 3d which
  keeps the time reversal, but breaks the inversion symmetry, so it is
  a pseudo-scalar which is  odd under the parity and even under the Time reversal.
  It was well known that it is this type of SOC which exists in the electro-weak interaction and is responsible for the parity violation in  the weak interaction \cite{Lee}. In the strong coupling limit, it is the DM term encoded in the RFHM Eq.\ref{expand}
  which breaks the inversion symmetry.
%   It earns T. D. Lee  and C. N. Yang's 1957 Nobel prize in physics.
  In 2d system, it is nothing but the well known Rashba SOC.  Surprisingly, its effects in a lattice system
  have not been studied in any depth until this work. So results achieved here, especially its counterpart in
  the fermionic case \cite{rafhm}  could have some impacts in lattice QCD calculations with a parity violation.
  Of course, the non-abelian gauge fields may have their own dynamics in lattice QCD systems.

%   It has been demonstrated the dominant Kitaev term \cite{kit} is extremely fragile against
%   various kinds of small perturbations. It is extremely unlikely to
%   observe the Kitaev spin liquid in these so called Kitaev materials. In view of recent ground breaking experiments generating
%   2d SOC which can tune various SOC parameters, lattice geometries and weak to strong interactions \cite{expk40,expk40zeeman,clock,2dsocbec},
%   there may be some promising platforms   to quantum simulate spin liquid phase in cold atom systems.

%We can say $\min E_{N\times1}<E_{Y\textmd{-}x}$ is enough
%to show $Y$-$x$ state is not a ground-state any more.

\section{ Conclusions and  Discussions }

In this work, we used exact symmetry analysis,  microscopic calculations such the canonical quantization method such as spin wave expansion,
non-perturbation OFQD analysis, augmented by some exact diagonization (ED ) study.
 Then by using the combinations of canonical quantization and coherent spin path integral,
we identify the correct critical modes, we construct symmetry based phenomenological effective actions
to study all the phases and phase transitions.
We also contrast the effective action with the microscopic calculations, especially the OFQD analysis, therefore establish the mappings between the phenomenological parameters in the effective action and the bare parameters in the microscopic Hamiltonian, especially a deep and profound connection between the phenomenological action and the effective potentials generated by the OFQD mechanism.

Our results demonstrate that the interplay among strong correlations, Rashba SOC and lattice geometries
opens a new avenue to explore whole new classes of quantum or topological phenomena, especially new organization principle which
may have wide implications in cold atoms, various strongly correlated materials with SOC and lattice QCD with parity violation.

{\sl 1. What are so special about the diagonal line in Fig.1 and Fig. 2 }

 The Hamiltonian has the $ [C_4 \times C_4]_D $ symmetry along the diagonal line which has many important implications.

  (1) There is an OFQD phenomena near the Abelian point $ \alpha=\beta=\pi/2 $.
      It hosts the mixture of the two sole collinear phases Y-x and X-y phase which are related by the $ [C_4 \times C_4]_D $ symmetry.

  (2) The non-coplanar IC-SkX-y phase away from it reduces to the co-planar IC-XY-y phase. The first order transition line
  with a cubic term $ \lambda \neq 0 $
  in Fig.\ref{phasesarc} ends at a second order transition point $ \alpha_{in} $ with $ \lambda=0 $ which is a bi-critical point.

  (3) The $ 3 \times 3 $ SkX phase is the only phase respecting the $ [C_4 \times C_4]_D $ symmetry.
      It is the hub phase which connects the two different organization principles in Fig.2

  (4) Away from the diagonal line, the gapless IC-YZ-x/LQx phase breaks the translational and the time reversal of the Hamiltonian.
  But it still shows no Bragg peaks, so behaves just like a gapless 2d QSL. Along the diagonal line,
    this quasi-1d LQ also breaks the $ [C_4 \times C_4]_D $  symmetry, it is a
    mixture of the LQx and LQy with any ratios. Its gapless is protected by the In-commensurate and the associated
   continuous $ U(1) $ symmetry breaking.

  (5) The in-complete devil staircase away from it becomes a complete one along it shown in Fig.\ref{cantor}.
      In fact, it only takes half of the  Cantor function, ending at the hub phase, then turns into
      a different organization principle: quantum phase transition.

%   We have computed the LSW spectrum in the Y-x phase. As said in the text, it can be easily extended to the
%   robust commensurate non-coplanar $ 3 \times 3 $ SkX phase and the spiral $ N \times 1 $ phases.
%   However, it still not know how to evaluate the excitation spectra above the non-coplanar IC-XY-y phase and the coplanar ( chiral )
%   IC phase.
%   One can show that after some modifications,
%   the qualitative features of Fig.1 also applies to fermions in a bi-partite lattice \cite{rafhm}.
%   The Y-x ( X-y ) state is replaced by the Y-y ( X-x ) state, the C-$ C_{\pi} $
%   magnons become the same as the C-$ C_0 $ magnons, all the orbital orders in other phases also need to be  multiplied by $ (-1)^{x+y} $.
%   Of course, all the numbers listed at the end of the caption in Fig.1 need to be recalculated for the RAFHM.

{\sl 2. The classification of all the phases in Fig.1 and Fig.\ref{trifeats}: compare to an exact theorem without assuming a spin
       $ SU(2) $ symmetry }

    We may classify the zoo of phases in Fig.1 as the following 5 classes and also outline
    their important roles played in the global phase diagram Fig.1:
   (1) The Y-x phase with $ N=2 $ is the only collinear phase. It is the exact ground state  along the anisotropic line
   $ (\alpha=\pi/2, \beta) $, but becomes just a classical state away from it.  It also takes most measures in Fig.1.
       The analytic calculations done in this phase in the appendices have shaded considerable lights
       in all the other phases and phase transitions in Fig.1.
   (2) The $ 3 \times 3 $ SkX phase is the only commensurate non-coplanar phase. It is also the only phase
       respecting the $ [C_4 \times C_4]_D $ symmetry along the diagonal line. It has gapped excitations.
       It is the hub which connects quantum phase transition (QPT) and the complete devil staircase along the diagonal line.
   (3) The in-commensurate non-coplanar IC-SkX-y phase which reduces to the co-planar IC-XY-y phase along the diagonal line.
       The quantum fluctuations can be represented  by  2d gapped dis-commensurations.
       They appear around the $ 3 \times 3 $ SkX phase
   (4) At mean field level,  the $ N \times 1, N \geq 3 $ commensurate co-planar spiral phases
       characterized by the topological winding numbers $ W=n/N $
       are 1d stripe phases embedded in a 2d lattice. The primary $ n=1 $ and the higher orders with $ n > 1 $
       break the same symmetry, but still can be described by $ W $ due to the different homotopy properties in Eq.\ref{topon}.
       The quantum fluctuations are still described by  2d gapped excitations.
       The gaps go to zero as $ N \rightarrow \infty $.
   (5) At mean field level, the IC-YZ-x/LQx are 1d stripe phases embedded in a 2d lattice which completely break
       the lattice translational symmetry along the x-axis. However, the quantum fluctuations  due to the 1d gapless phason modes
       melt them into quasi-1d Luttinger liquid phases. The translational symmetry breaking along the x- axis and the Time reversal symmetry break
       can still be seen in the form factor of the SSCFs Eq.\ref{phasonspin}

    A recent work \cite{lacksoc} concluded that for a system with only translational and Time reversal
    symmetry, and spin $ s=1/2 $ in a unit cell,  but without assuming any spin-rotational symmetry,
    if a ground state  breaks no symmetries, then it can only be  a non-trivial state with a topological order.
    Our system  belongs to this category without SOC symmetry.
    However, all the 5 classes of phases break both the translational and the Time reversal symmetry,
    the IC-YZ-x/LQx is the closest state to a 2d gapless QSL. For its fate in a honeycomb or cubic lattice, see {\sl 4} below.

{\sl 3. The in-complete and complete staircase in Fig.1 and Fig.\ref{trifeats} is a new class of phenomena  beyond any symmetry or topological classification }

    All the 5 classes of phases are glued together by QPT or in-complete/complete devil stair cases with a fractal structure
    shown in Fig.1,2,\ref{trifeats}, \ref{cantor} and Fig.\ref{finiteT}. It is not known if there exists a 3rd class of organization principle.

   There are classifications of quantum phases according to their symmetry breaking patterns \cite{scaling,sachdev,aue,wen0}.
   The discovery of fractional quantum Hall effects inspires the classifications of various topological phases
   with long-range entanglements, recently dubbed symmetry enriched topological (SET) phases \cite{frusrev,SLrev1,SLrev2,SLrev3,tenfold,wen}.
   The discovery of non-interacting topological insulators\cite{kane,zhang} leads to
   the classifications of various interacting topological phases
   with short-range entanglements, recently dubbed symmetry protected topological (SPT) phases\cite{tenfold,wen}.
   There could be also some intimate connections between SET and SPT \cite{tenfold,wen}.
   There are also corresponding quantum or topological phase transitions  between different quantum  or topological phases.
   One may construct various symmetry based  GL action  or topological quantum field theories (TQFT)  to
   describe such phase transitions. If the transition is a continuous one (2nd order or above), various physical quantities
   also satisfy quantum critical scaling functions with various critical exponents \cite{scaling,sachdev,tqpt,tqpt2}.
   However, the fractal structure in Fig.1 and Fig.\ref{finiteT} is beyond all these classification schemes:
   any two quantum phases may also be connected by a segment in a complete or in-complete devil staircase
   instead of any 1st or 2nd order transitions. There is no associated QC scalings either.
   This fractal structure existing at any finest segment defies descriptions by any effective GL or
   topological action in the continuum limit. It can be easily smeared out in any continuum limit.
   Obviously, it presents a new class of problems beyond any symmetry or topological classifications.

   For example, along the diagonal line $ \alpha^{-}_{33} < \alpha=\beta <\pi/2 $ shown in Fig.1, Fig.5 and Fig.6,
   one can construct GL effective action Eq.\ref{Yxp} to describe
   the quantum Lifshitz transition from the Y-x phase to the IC-XY-y phase. However, after $ 0 < \alpha < \alpha^{-}_{33} $,
   although one can introduce topological rational or irrational winding number (the Cantor function)
   $ W $ to characterize  all the principle spiral C-phase with $ W=1/N $, higher order spiral C-phase with $ W=n/N $, then the IC-YZ-x/LQx
   phases form a Cantor set,
   there is no such  effective actions available anymore to describe the complete devil staircase.
   Similarly, there is no  such  effective action available either to describe the in-complete devil staircase at $ \beta < \alpha $.
   In fact, as stressed in Sec.VI-D, the solid line connecting the $ M $ point to $ (\pi/2,0) $ of the boundary of the Y-x phase in Fig.1 is just $ W=1/2 $ segment of the whole in-complete staircase instead of a 1st order transition line. Of course, in
   any numerical calculation on a finite size, it appears as a 1st order transition line.

{\sl 4. The gluing pattern is also beyond any quantum chaos classification scheme }

   As stressed in Ref.\cite{rmt1,rmt2,rmt3,rmt4,rmt5,sun1,sun2,sun3},
   there is another classification scheme from a different perspective: the possible organization patterns of strongly interacting  matter
   can also be classified by the 10 fold-way of quantum chaos in terms of random matrix theories.
   It was known that there could be some relations between  fractal/non-integer fractal dimension and classical chaos/classical Lyapunov exponent.
   However, when going to quantum cases, it is not known if such relations still exist between
   the fractals and quantum chaos. In Sec.VIII, we tried to address such a possible relation.
   Because quantum chaos can only be defined at a finite temperature, so they are suppressed in any gapped phases.
   It was greatly enhanced in the QC regime from the Y-x to the IC-XY-y phase, but still absent in the IC-YZ-x phases due to its integrability at $ T=0 $.  It remains interesting to explore possible hidden relations  between quantum chaos and the complete devil stair-case from a different angle such as wavefunction landscape or entanglement. As presented in \cite{sun1,sun2,sun3}, quantum phase transitions and quantum chaos
   are controlled by low energy excitations and bulk excitations respectively.

{\sl 5. A few possible future perspectives }

   In this paper, we only discussed the spinor bosons in the strong coupling limit which leads to the RFHM Eq.1.
   In the weak coupling limit, the spinor bosons are in various spin-bond correlated superfluid phases \cite{SFnon}
   along the solvable line $ (\alpha=\pi/2, \beta) $. In a future publication, starting from this solvable line
   in the weak coupling limit, we will work out the SF phases in the whole SOC parameter space which is the weak coupling
   analog of Fig.1. We will also explore the transitions from the spin-bond correlated SFs in
   weak couplings  to all the spin-bond correlated magnetic phases in Fig.1 in the strong couplings.
   Gaining the insights achieved in our recent work on $ \pi $ flux in a square lattice \cite{pifluxqsl},
   we expect that possible new phases such as topological quantum spin liquids (QSL) with
   fractionized excitations emerge in the intermediate couplings.

   From both exact symmetry analysis, analytical calculations from the three lines and also augmented by some numerical
   calculations, we conjecture that there is a complete devil staircase along the diagonal line shown in Fig.1, Fig.\ref{cantor} and Fig.\ref{finiteT}.
   Recall that the original ( simplest ) Cantor set carries the fractal dimension $ d^F_{1/3}=log2/log3=0.6309 $.
   As mentioned in the caption of Fig.\ref{cantor}, the Cantor function $ W(\alpha) $ should be different from
   the $1/3 $ one, so the Cantor set formed by the IC-YZ-x/LQx phases in the
   $ \alpha=\beta^{+} $ complete devil staircase $ 0 < \alpha < \alpha^{-}_{33} $ should carry a different
   fractal dimension than $ 0.638 $.
   There are many generalized Cantor functions which are homeomorphic (topologically equivalent) to the the original ( simplest ) Cantor one.
   They also carry different fractal dimensions.
   It remains outstanding to find one which describes the $ W(\alpha) $ in Fig.1.
   The deep mathematical and topological structure of the fractals in Fig.1 deserve more investigations.

   It is also very interesting to extend Fig.1 to 2d honeycomb lattice and 3d cubic lattice, both of which have 3
   SOC parameters $ (\alpha, \beta, \gamma ) $, so they
   should contain even richer fractal or even multi-fractal structures.
   At the classical level, it is interesting to see if the Cantor set in a square lattice becomes a Cantor dust or Sierpinski carpet.
   When incorporating quantum fluctuations, it is important to see if complete or-incomplete devil staircases melt into
   $ Z_2 $ or $ U(1) $ QSLs in the honeycomb or cubic lattice respectively, if the IC-SkX-y or IC-XY-y melt into QSL or chiral QSL
   receptively. It may also be interesting to study the fate of the analog of the IC-YZ-x/LQx phase in both cases.
   Indeed, the SOC in the 3d cubic lattice is a more suitable place to compare with the 3d cubic model \cite{cubic} mentioned in Sec.IX-4.

   Obviously,  in view of the direct relevances of the fermionic case
   to materials and lattice QCD with parity violation in particle physics, it is important to investigate
   the fermions which should show quite different behaviours.
   In the fermionic case,
   starting from the results achieved in \cite{rafhm} on both weak and strong coupling (RAFHM)
   along the anisotropic line $ (\alpha=\pi/2, \beta) $, we will map out the global quantum phase diagram for spin 1/2 fermion case in the generic
   $ (\alpha, \beta ) $ SOC parameter space in both weak and strong coupling   limits in a
   separate publication. We may also explore the quantum or topological transitions between
   the two limits.

%   Although classical RFHM and RAFHM makes no differences \cite{classdm1,classdm2} in a bipartite lattice,
%   the quantum RFHM and RAFHM are dramatically different.
%   Indeed, as shown in the recent work \cite{rafhm}, the RAFHM along the line $ (\alpha=\pi/2, \beta) $ show dramatically different phenomena
%   than those in RFHM in Fig.\ref{phasedia}. So we expect the RAFHM in the generic case will also be dramatically different and will be
%   investigated in a future publication.

%     The $ Z-x $ state gets into the IC-SkX state directly when above $ h_{c1} $, but the FM state
%    gets to the two canted states near small and large $ \beta $ when below $ h_{c2} $.
%It was shown in \cite{rh} that $ \beta=\pi/4 $ falls into the most frustrated regime
%where the Wilson loop $ W_R=-1 $,
%the Dzyaloshinskii-Moriya (DM) term \cite{dmterm1,dmterm2} dominates.

%     In short, quantum spin systems with SOC subject to a Zeeman field opens a new platform to
%     display rich and novel class of quantum In-commensurate phases, excitations and quantum C-IC phase transitions, which can be observed in
%     both cold atoms and materials with SOC.
%     The results achieved in this paper just reveals a tip of an iceberg.

{\bf Acknowledgements }
  J. Y thank Iva Martin for the hospitality during his visit in 2016 at Argon national lab and also helpful discussions,
   Niu Qian for the discussions on the integer QH fractal gaps in the  Hofstadter problem  at an irrational flux $ \alpha $.
  We thank  AFOSR FA9550-16-1-0412 for supports.
  The work at KITP at UCSB was supported by NSF PHY11-25915.

\vspace{0.25cm}
\appendix

{\bf Appendix }

\begin{widetext}

In the main text, we mainly demonstrate the new and important concepts. In this appendix, we support these important claims by
specific calculations by both canonical quantization, path integral, especially the shifts between the two approaches.

(1) The spin wave calculations ( canonical quantization ) to the order of $ 1/S $ and extract the parameters of
   the 3 kinds of relativistic magnons in the Y-x state away from the solvable line.

(2) A systematic order from quantum disorder analysis ( canonical quantization ) to evaluate
not only the mass gap, but also the spectrum in the Y-x state
along the diagonal line near the Abelian point in Fig.1.

 (3) Path integral Quantization of the classical 1d
 FK model discussed in Sec.VI. The main goal is to construct a general 2+1 d effective action which,
 in principle, it can be used to derive  all the quantum phases and quantum phase transitions in Fig.1 and
 describe the fractal structures in Fig.1. Then we apply it to study the excitation spectrum in the Y-x phase successfully.
 Unfortunately, we are not able to derive a low energy effective action to perform any practical calculations to evaluate
 the spin-spin correlation function (SSCFs) in the Y-x phase from the path integral approach.
 We point out the orgins of the difficulty and motivate the following sections.

 (4)  We combine the canonical quantization method used in (1) and (2) with the path integral approach used in (3).
  By identifying the correct low energy critical mode,
  Then we derive the low energy effective action corresponding to C- and IC- magnons respectively.
  We spell out the appealing physical picture in both C- and IC-magnons low energy effective Hamiltonian where  only half pair of degree freedoms appear, the other half are projected out. Then we will push the effective action to the magnon condensation boundary
  corresponding to the C- and IC-magnons respectively.

  (5) We use the effective actions for the C- and IC-magnons derived in (4) to derive their SSCFs respectively.
  We find the actions take different forms in C-regime and IC-regime which can be  experimentally distinguished.

  (6) We push the effective actions for the C- and IC-magnons to higher orders to study
  the quantum phase transitions from the Y-x state
  to the commensurate collinear X-y state and non-co-planar IC-SkX-y state with non-vanishing Skyrmion density respectively.
  We re-derive the continuum effective action  Eq.\ref{Yxp},\ref{2times2order} which are reached along the diagonal line in the main text.
%  which was used to study the finite temperature behaviours such as quantum chaos and quantum information scramblings in  Fig.\ref{finiteT}.
  We also write down the spin-orbital structure of the non-coplanar IC-SkX-y
  phase which reduces to the co-planar IC-XY -y phase along the diagonal line $ \alpha=\beta $.
  As alerted in Sec.IX-1, these effective actions can be considered as a non-linear Sigma model in the presence of SOC
  which extends the NLSM with $ SU(2) $ symmetry to the SOC case without any spin-rotation symmetry.

  (7) For a comparison, we also derive an effective low energy action inside
      the Y-x phase in the presence of a longitudinal Zeeman field in \cite{rhh}.
      We point out its dramatic different than the IC-magnons in (5) and (6) and also
      stress the important roles played by the $ U(1)_{soc} $ symmetry. See Sec.VIII-C.

  Although, we mainly focus on the Y-x which is the $ 2 \times 1 $ state. The method developed here
  can be transformed to study the $ N \times 1 $ co-planar states in Sec.VI-A,B, also the
  effective phason action Eq.\ref{phasonact} in Sec.VI-C. It may also be used to study the SSCF of the Y-x state in the presence of
  a longitudinal $ h_y $ Zeeman field \cite{rhh} which still keeps the $ U(1)_{soc} $ symmetry ( see appendix G ) and also
  in the two transverse fields $ h_x, h_z $ \cite{rhtran} which still breaks the $ U(1)_{soc} $ symmetry.

%(2) The quantum C-IC Lifshitz transition from the Y-x state to the non-coplanar IC-XY-y state along the diagonal line $ \alpha=\beta $ in Fig.M1
%(3) Determination of some robust commensurate phases at $ \alpha=\pi/N, N=2,3,4,5,6 \cdots $ on
%the complete devil staircases along the diagonal line  $ \alpha=\beta $
%and compare with those achieved  from the FK model on the in-complete devil staircases near the Abelian line at a small $ \beta $.
%(4) The detailed spin structures  and ordering wavevectors of the only non-coplanar commensurate $ 3 \times 3 $ SkX phase in both real and  momentum space,
%(5) The nature of the co-planar spiral confined gapped phases at $ \alpha=\frac{\pi}{N} n $ and the  gapless co-planar IC phase.

\section{ The gap and velocities of  the 3 kinds of magnons in the Y-x state away from the solvable line:
Canonical quantization approach }.

 Away from the solvable line $ (\alpha=\pi/2, \beta=0 ) $, there is no $ U(1)_{soc} $ anymore, the Y-x state remains the classical ground state.
 We will first calculate its excitation spectrum by  the spin wave calculations to the leading order $ 1/S $, then we will
 extract its low energy branch and three kinds of magnons which dominate the physical measurable quantities at the low temperatures.

\subsection{ The spin-wave calculations to evaluate the excitation spectrum  to the order $1/S $.}

For $Y$-$x$ state,
we can easily work out the classic energy
\begin{align}
	\langle\textmd{Y-x}|H|\textmd{Y-x}\rangle
	=-2NJS^2\sin^2\alpha
\end{align}
When $\alpha=\pi/2$ (at exact solvable line),
we recover fully saturated result $-2NJS^2$.

Before applying Holstein-Primakoff (HP) transformations,
we need a globe rotation $R_x(\pi/2)$ to align spin to new Z-axis.
Now we can rewrite Hamiltonian Eq.\ref{rhgeneral} as
\begin{align}
	H=-J\sum_i
	[\bar{\mathbf{S}}_iR_x(2\alpha)\bar{\mathbf{S}}_j
	+\bar{\mathbf{S}}_iR_z(2\beta)\bar{\mathbf{S}}_j]
\label{eq:RH_r}
\end{align}
We can rewrite Eq.\eqref{eq:RH_r} in an explicit form
\begin{align}
	H=-J\sum_i[
	S_i^xS_{i+x}^x+\cos(2\alpha) S_i^yS_{i+x}^y+\cos(2\alpha) S_i^zS_{i+x}^z
	+\sin(2\alpha) (S_i^y S_{i+x}^z-S_i^z S_{i+x}^y)]\nonumber\\
	-J\sum_i[
	\cos(2\beta) S_i^xS_{i+y}^x+\cos(2\beta) S_i^yS_{i+y}^y+S_i^zS_{i+y}^z
	+\sin(2\beta) (S_i^x S_{i+y}^y-S_i^y S_{i+y}^x)]
\end{align}

We need the following HP transformation:
\begin{align}
	\bar{S}_i^+=\sqrt{2s-a_i^\dagger a_i}a_i,\quad
	\bar{S}_i^-=a_i^\dagger\sqrt{2s-a_i^\dagger a_i},\quad
	\bar{S}_i^z=s-a_i^\dagger a_i,\quad \forall i\in A;      \\ \nonumber
	\bar{S}_i^+=b_i^\dagger\sqrt{2s-b_i^\dagger b_i},\quad
	\bar{S}_i^-=\sqrt{2s-b_i^\dagger b_i}b_i,\quad
	\bar{S}_i^z=-s+b_i^\dagger b_i,\quad \forall i\in B;
\label{HPab}
\end{align}

When we expand the Hamiltonian with respect to the powers of
$1/\sqrt{2S}$,
we obtain the following series ( Eq.\ref{swcubic} in the main text ):
\begin{align}
	H=H_0+2JS\Big[H_2
	+\Big(\frac{1}{\sqrt{2S}}\Big)H_3
	+\Big(\frac{1}{\sqrt{2S}}\Big)^2H_4+\cdots\Big]
\end{align}
where the symbol $H_n$ denotes the $n$-th polynomial of the boson operators.
For example $H_0=-2NJS^2\sin^2\alpha$ and
\begin{align}
	H_2 & =2\sin^2\alpha\sum_k(a_k^\dagger a_k+b_k^\dagger b_k)
	-\sum_k[
	\sin^2\alpha\cos k_x(a_k^\dagger b_k+b_k^\dagger a_k)
	+\cos^2\alpha\cos k_x(a_k b_{-k}+a_k^\dagger b_{-k}^\dagger)   \nonumber   \\
	& +\cos(k_y\!-\!2\beta) a_k^\dagger a_k
	+\cos(k_y\!+\!2\beta) b_k^\dagger b_k]
\end{align}
  where $ k $ is confined in the reduced BZ $  0 < k_x < \pi, -\pi < k_y  < \pi $.

 in a matrix form $ H_2=-N\sin^2\!\alpha\! + H^{\prime}_2 $ where ( we drop $ \prime $ in the following )
%{\small
\begin{align*}
	H_2\!=
	\sum_k\!
	\begin{pmatrix}
	  a_k^\dagger\\ b_k^\dagger\\ a_{-k}\\ b_{-k}\\
	\end{pmatrix}^\intercal\!\!
	\begin{pmatrix}
	  \sin^2\!\alpha\!-\!\frac{1}{2}\cos(k_y\!-\!2\beta)
	    &-\frac{1}{2}\sin^2\alpha\cos k_x
	      &0
	        &-\frac{1}{2}\cos^2\!\alpha\cos k_x\\
	  -\frac{1}{2}\sin^2\alpha\cos k_x
	    &\sin^2\!\alpha\!-\!\frac{1}{2}\cos(k_y\!+\!2\beta)
	      &-\frac{1}{2}\cos^2\!\alpha\cos k_x
	        &0\\
	  0
	    &-\frac{1}{2}\cos^2\!\alpha\cos k_x
	      &\sin^2\!\alpha\!-\!\frac{1}{2}\cos(k_y\!+\!2\beta)
	        &-\frac{1}{2}\sin^2\alpha\cos k_x\\
	  -\frac{1}{2}\cos^2\!\alpha\cos k_x
	    &0
	      &-\frac{1}{2}\sin^2\alpha\cos k_x
	        &\sin^2\!\alpha\!-\!\frac{1}{2}\cos(k_y\!-\!2\beta)\\
	\end{pmatrix}\!\!
	\begin{pmatrix}
	  a_k\\ b_k\\ a_{-k}^\dagger\\ b_{-k}^\dagger\\
	\end{pmatrix}
\end{align*}
%}
 The first step is to  perform a unitary transformation:
\begin{align}
	U_k=
	\begin{pmatrix}
	     \sin\frac{\theta_k}{2} &\cos\frac{\theta_k}{2}\\
	    -\cos\frac{\theta_k}{2} &\sin\frac{\theta_k}{2}\\
	\end{pmatrix}
\end{align}
  where the  auxiliary angle $\theta_k$ is defined by:
\begin{align}
	\sin\theta_k=
	\frac{\sin^2\alpha\cos k_x}
	     {\sqrt{\sin^4\alpha\cos^2 k_x+\sin^22\beta\sin^2 k_y}},
	~~~~~~
	\cos\theta_k=
	\frac{\sin2\beta\sin k_y}
	     {\sqrt{\sin^4\alpha\cos^2 k_x+\sin^22\beta\sin^2 k_y}},
\label{evenodd}
\end{align}
then we need work on
\begin{align}
	\begin{pmatrix}
	    U_k^\dagger &0\\
	    0 &U_{-k}^\intercal
	\end{pmatrix}
	\begin{pmatrix}
	    D_k &A_k\\
	    A_k &D_{-k}\\
	\end{pmatrix}
	\begin{pmatrix}
	    U_k &0\\
	    0 &U_{-k}^*
	\end{pmatrix}
	=
	\begin{pmatrix}
	    U_k^\dagger D_k U_k & U_k^\dagger A_k U_{-k}^*\\
	    U_{-k}^\intercal A_k U_k &U_{-k}^\intercal D_{-k} U_{-k}^*\\
	\end{pmatrix}
\end{align}
Further calculations show
\begin{align}
	U_k^\dagger D_k U_k
	=
	\begin{pmatrix}
	    \lambda_k^+ &0\\
	    0 &\lambda_k^-\\
	\end{pmatrix},
	\quad
	U_k^\dagger A_k U_{-k}^*
	=
	\begin{pmatrix}
	    \chi_k &0\\
	    0 &-\chi_k\\	
	\end{pmatrix}
\end{align}
where we define $\chi_k=\frac{1}{2}\cos^2\alpha\cos k_x$
and $\lambda_k^{\pm}
=\sin^2\alpha-\frac{1}{2}\cos2\beta\cos k_y
\pm\frac{1}{2}\sqrt{\sin^4\alpha\cos^2k_x+\sin^22\beta\sin^2 k_y}$.

Notice $\lambda_k^{\pm}=\lambda_{-k}^{\pm}$, $ H_2 $ takes the form:
\begin{align}
	H_2=&
        \sum_k
	\begin{pmatrix}
	  \bar{a}_k^\dagger\\
	  \bar{b}_k^\dagger\\
	  \bar{a}_{-k}\\
	  \bar{b}_{-k}\\
	\end{pmatrix}^\intercal\!\!
	\begin{pmatrix}
	    \lambda_k^+ &0 &\chi_k &0\\
	    0 &\lambda_k^- &0 &-\chi_k\\
	    \chi_k &0 &\lambda_k^+ &0\\
	    0 &-\chi_k &0 &\lambda_k^-\\
	\end{pmatrix}
	\begin{pmatrix}
	  \bar{a}_k\\
	  \bar{b}_k\\
	  \bar{a}_{-k}^\dagger\\
	  \bar{b}_{-k}^\dagger\\
	\end{pmatrix}    \\  \nonumber
	=&
	\sum_k
	\begin{pmatrix}
	  \bar{a}_k^\dagger\\
	  \bar{a}_{-k}\\
	\end{pmatrix}^\intercal\!\!
	\begin{pmatrix}
	  \lambda_k^+ &\chi_k\\
	  \chi_k &\lambda_k^+\\
	\end{pmatrix}
	\begin{pmatrix}
	  \bar{a}_k\\
	  \bar{a}_{-k}^\dagger\\
	\end{pmatrix}
	+
	\sum_k
	\begin{pmatrix}
	  \bar{b}_k^\dagger\\
	  \bar{b}_{-k}\\
	\end{pmatrix}^\intercal\!\!
	\begin{pmatrix}
	  \lambda_k^- &-\chi_k\\
	  -\chi_k &\lambda_k^-\\
	\end{pmatrix}
	\begin{pmatrix}
	  \bar{b}_k\\
	  \bar{b}_{-k}^\dagger\\
	\end{pmatrix}
\end{align}
 The second step is to perform a single-mode Bogoliubov transformation:
\begin{align}
	\begin{pmatrix}
		\bar{a}_k\\
		\bar{a}_{-k}^\dagger\\
	\end{pmatrix}
	=
	\begin{pmatrix}
		u_k^a &v_k^a\\
		v_k^a &u_k^a\\
	\end{pmatrix}
	\begin{pmatrix}
		\alpha_k\\
		\alpha_{-k}^\dagger\\
	\end{pmatrix},
	\quad
	\begin{pmatrix}
		\bar{b}_k\\
		\bar{b}_{-k}^\dagger\\
	\end{pmatrix}
	=
	\begin{pmatrix}
		u_k^b &v_k^b\\
		v_k^b &u_k^b\\
	\end{pmatrix}
	\begin{pmatrix}
		\beta_k\\
		\beta_{-k}^\dagger\\
	\end{pmatrix}
\label{eq:Symplectic}
\end{align}
where
\begin{align}
	(u_k^a)^2=\frac{1}{2}(\frac{\lambda_k^+}{\omega_k^+}+1),
	\quad
	(v_k^a)^2=\frac{1}{2}(\frac{\lambda_k^+}{\omega_k^+}-1),
	\quad
	u_k^a v_k^a=-\frac{\chi_k}{2\lambda_k^+\omega_k^+}  \\  \nonumber
	(u_k^b)^2=\frac{1}{2}(\frac{\lambda_k^-}{\omega_k^-}+1),
	\quad
	(v_k^b)^2=\frac{1}{2}(\frac{\lambda_k^-}{\omega_k^-}-1),
	\quad
	u_k^b v_k^b=+\frac{\chi_k}{2\lambda_k^-\omega_k^-}
\end{align}
Finally, we obtain the Hamiltonian in a diagonal form
\begin{align}
	H_2=\sum_k(\omega_k^++\omega_k^--2\sin^2\alpha)
	+2\sum_k(\omega_k^+\alpha_k^\dagger\alpha_k
		+\omega_k^-\beta_k^\dagger \beta_k)
\label{H2Yx}
\end{align}
where the spin-wave dispersion $\omega_k^{\pm}=\sqrt{(\lambda_k^{\pm})^2-\chi_k^2}$. One can also see
at the $ H_2 $ order, it still has the mirror symmetry $ \beta \rightarrow \pi/2 - \beta $ symmetry.

\begin{figure}[!htb]
\includegraphics[width=4.0cm]{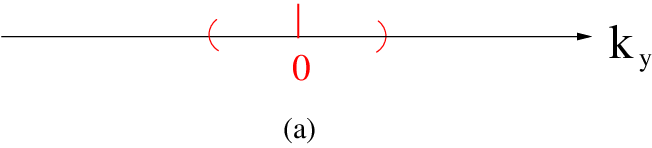}
\hspace{0.2cm}
\includegraphics[width=4.0cm]{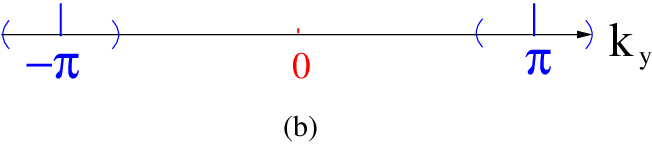}
\hspace{0.2cm}
\includegraphics[width=4.0cm]{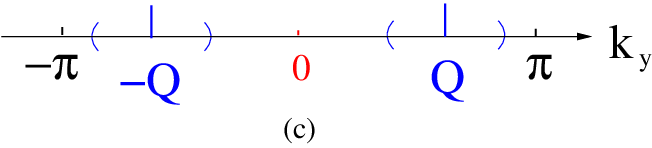}
%\includegraphics[width=4.25cm]{Fig2-3.eps}
%\hspace{-0.5cm}
%\includegraphics[width=4.0cm]{globalphaseM.eps}
%\hspace{0.2cm}
%\includegraphics[width=4.5cm]{contourline.eps}
%\hspace{0.2cm}
%\includegraphics[width=6.5cm]{contours2.eps}
%\hspace{0.2cm}
%\includegraphics[width=4.0cm]{Min0.eps}
\caption{ The dispersion minima of the three kinds of magnons in the Y-x phase along the $ k_y $ axis.
The braket stands for the cutoff around the minima.
(a) C$_0$ with one minimum at $ k^{0}_y=0 $, (b) C$_\pi$ with one minimum at $ k^{0}_y=\pi $
(c) IC-magnons with two minima at $ k^{0}_y=\pm Q  $ which are related by the time reversal symmetry. }
\label{threemag}
\end{figure}

\subsection{ The low energy mode and the parameters of the three kinds of magnons }

As shown in the main text, away from the solvable line $ \alpha=\pi/2 $, all the $ C_0 $, IC- and $ C_{\pi} $
magnons take the relativistic form Eq.\ref{relagap}:
\begin{equation}
 \omega_{-}(q) = \sqrt{ \Delta^2 + v^{2}_x q^{2}_x + v^{2}_y q^{2}_y }
\label{rela}
\end{equation}

  In the following, we list $\Delta$, $v_x, v_y$ for $ C_{\pi} $,  $ C_0 $ and IC- respectively ( Fig.\ref{threemag} ).
  Setting $  \lambda=\frac{\cos(2\beta)}{\sin(2\beta)}	\sqrt{\sin^4(\alpha)+\sin^2(2\beta)} $,
  then when $ \lambda \in (-\infty,-1),[-1,1],(1, \infty) $, $ k_y^0=\pi,\arccos\left[\lambda \right ],0 $ corresponds
  to the magnons C-$C_{\pi} $, IC-, C-$C_0$ respectively.

  For C$_{\pi}$, $ k=( 0, \pi ) + q $, the parameters   are:
\begin{align}
    \Delta=\sqrt{\frac{1}{2}\cos^2\beta(\cos2\beta-\cos2\alpha)},  \nonumber  \\
    v_x^2=\frac{1}{4}(\cos^4\alpha+\cos2\beta\sin^2\alpha+\sin^4\alpha),  \nonumber  \\
    v_y^2=\frac{1}{4}[1+\cos2\beta(\sin^2\alpha+\frac{\sin^22\beta}{\sin^22\alpha})]
\label{mass0}
\end{align}

 As shown in the main text, along the diagonal line $ \arccos(1/\sqrt{6}) = \alpha^{0}_{in} < \alpha < \pi/2 $,
 $ \Delta=0 $. It stands for the spurious Goldstone mode Eq.\ref{gapless}.
 The order from disorder mechanism generates a mass $ \Delta_B $ at the order of $ \sqrt{S} $ shown in Eq.\ref{diagap}, so transfers it
 into the pseudo-Goldstone mode Eq.\ref{gapspectrum}.

  For C$_{0}$, $ k= q $, the parameters are:
\begin{align}
    \Delta=\sqrt{-\frac{1}{2}\sin^2\beta(\cos2\beta+\cos2\alpha)},   \nonumber \\
    v_x^2=\frac{1}{4}(\cos^4\alpha-\cos2\beta\sin^2\alpha+\sin^4\alpha), \nonumber \\
    v_y^2=\frac{1}{4}[1-\cos2\beta(\sin^2\alpha+\frac{\sin^22\beta}{\sin^22\alpha})]
\end{align}

  For IC-magnons,  $ k=( 0, k^{0}_{y} ) + q $, the parameters in terms of $ \alpha, \beta, k^{0}_{y} $ are:

\begin{eqnarray}
    \Delta&=&\omega_{k=(0,k_y^0)}^-
    =\sqrt{\Big(\sin^2\alpha-\frac{1}{2}\cos2\beta\cos k_y^0
		    -\frac{1}{2}\sqrt{\sin^4\alpha+\sin^22\beta\sin^2 k_y^0}\Big)^2
	    -\frac{1}{4}\cos^4\alpha},  \nonumber \\
    v_x^2 &= &\sqrt{(\Delta^2+\frac{1}{4}\cos^4\alpha)}
	\frac{\sin^4\alpha}{2\sqrt{\sin^4\alpha+\sin^22\beta\sin^2 k_y^0}}
	+\frac{1}{4}\cos^4\alpha, \nonumber \\
    v_y^2&= &\sqrt{(\Delta^2+\frac{1}{4}\cos^4\alpha)}
	\Big[\frac{1}{2}\cos2\beta\cos k_y^0
	-\frac{\sin^22\beta\cos2 k_y^0}{2\sqrt{\sin^4\alpha+\sin^22\beta\sin^2 k_y^0}}
	+\frac{\sin^42\beta\cos^2 2k_y^0}{(2\sqrt{\sin^4\alpha+\sin^22\beta\sin^2 k_y^0})^3}
	\Big]    \nonumber
\end{eqnarray}
   Plugging in $ \cos k_y^0= \frac{\cos(2\beta)}{\sin(2\beta)}
	\sqrt{\sin^4(\alpha)+\sin^2(2\beta)} $ leads  to
\begin{eqnarray}
    \Delta &= & \sqrt{\Big( \sin^2\alpha-\frac{\sqrt{\sin^4\alpha+\sin^22\beta}}{2\sin2\beta}
		 \big)^2-\frac{1}{4}\cos^4\alpha}   \nonumber \\
    v_x^2 &= &\frac{\sin^6\alpha}{2\sin2\beta\sqrt{\sin^4\alpha+\sin^22\beta}}
	  -\frac{\sin^4\alpha}{4\sin^22\beta}-\frac{1}{4}\cos^4\alpha    \nonumber  \\
    v_y^2 &= & \Big( \frac{\sin^2\alpha}{2\sin2\beta\sqrt{\sin^4\alpha+\sin^22\beta}}
	    -\frac{1}{4\sin^22\beta}\Big)
	    \Big( \sin^22\beta -\frac{\cos^22\beta}{\sin^22\beta}\sin^4\alpha
	    \Big)
\end{eqnarray}

    As shown in the main text, these parameters can be extracted from all the physical measurable quantities such as
    magnetization, specific heat, various susceptibilities discussed in Sec.II and the
    spin-spin correlation functions to be evaluated in the appendix E.

\section{ The gap and spectrum generated from the OFQD phenomena along the diagonal line $ \alpha=\beta^{+} $:
 a canonical quantization approach to study the pseudo-Goldstone mode }

 In Sec.III, we develop a systematic spin coherent state path integral to evaluate not only the mass gap, but also the
 the spectrum generated by the order from quantum disorder in the $ \tilde{\tilde{SU}}(2) $ basis.
 Here, by using the canonical HP boson quantization in the original basis, we achieve the same goal, therefore confirm the results achieved in the main text. The canonical quantization approach is complementary to
 the path integral approach developed in the main text.
 Both methods have its own advantages and dis-advantages.  The canonical quantization approach is more physically transparent and intuitive than
 the path integral approach.
 However, the main advantage of the path-integral method using the polar coordinates automatically incorporate the
 non-linear interactions between spin waves, so can be  directly applied to study the finite temperature behaviours near the
 Y-x to IC-XY-y transition in  Fig.\ref{finiteT}.

  The degenerate family of the $2\times2$ vortex state is given in Eq.\ref{2times2order}:
\begin{align}
	\mathbf{S}_i=S[\sin\phi ~e^{iQ_yr_i}(1,0,0)+\cos\phi ~e^{iQ_xr_i}(0,1,0)]
\end{align}
   where $ Q_x=(\pi,0), Q_y=(0,\pi) $. $ \phi=0,\pi/2 $ leads to the Y-x and X-y state  \cite{rh} respectively.

   For a small $\phi$ deviation from the Y-x state:
\begin{align}
	\delta \mathbf{S}_i=(\delta S_i^x,\delta S_i^y,\delta S_i^z)
	=S \phi e^{iQ_yr_i}(1,0,0)
\end{align}
thus
\begin{align}
	\phi=S^{-1}\delta S_i^x e^{iQ_yr_i}= (-1)^{i_y} S^{-1}\delta S_i^x
\label{phiofqd}
\end{align}
  which stands for the quantum fluctuations near  $ Q_y=(0,\pi) $ signifying a transition to the X-y state.

  The quantum spin fluctuations around the Y-x state can also be written in terms of the  HP boson \cite{rh}:
\begin{align}
	\delta S_i^x=\frac{S_i^+ + S_i^-}{2}=\sqrt{\frac{S}{2}}(a_i^\dagger+a_i) \text{, if }i\in A;~~~~~~
	\sqrt{\frac{S}{2}}(b_i^\dagger+b_i) \text{, if }i\in B
\end{align}

  In the unit of $ 2JS $ in Eq.\ref{swcubic}, the quantum correction to the Hamiltonian
  due to the order from quantum disorder in Eq.\ref{quantumphi} can be re-written as:
\begin{align}
	\delta \mathcal{H}_2=\frac{B}{2}\sum_i\phi_i^2
	=\frac{B}{4S}\left[\sum_{i\in A}(a_i^\dagger+a_i)^2+\sum_{j\in B}(b_j^\dagger+b_i)^2\right]
\end{align}

  Combining with the  $ H_2 $ in Eq.\ref{h2e2} leads to:
\begin{align}
	\mathcal{H}_\text{OFD}=\mathcal{H}_2+\delta\mathcal{H}_2
\end{align}

  The re-diagonalization of $ \mathcal{H}_\text{OFD} $ leads to the pseudo-Goldstone mode Eq.\ref{gapspectrum}:
\begin{align}
	\omega_-=\sqrt{ \Delta^2_B + v_x^2q_x^2+v_y^2q_y^2}
\end{align}
where
\begin{align}
    \Delta^2_B  =  \frac{B}{S}\cos^2\alpha,~~~	v_x^2=\frac{1}{4}\cos^2\alpha-\frac{B}{4S}\cos 2\alpha,~~~~
	v_y^2  =\frac{1}{4}\cos^2\alpha(1-6\cos^2\alpha) - (\frac{1}{2}\cos2\alpha+2\cot^2\alpha)\frac{B}{2S}
\label{vxvy}
\end{align}

    Putting back the unit $ 2 JS $ leads to the same mass gap $ \Delta_B $  as Eq.\ref{gapspectrum}.
    The correction to $ v_y^2 $ leads a very small shift on $ \alpha_{in} $.
    As argued in the main text, we also evaluated the contributions from $ H_3 $ and $ H_4 $ in Eq.\ref{swcubic}
    and found they are subleading to  $ \delta \mathcal{H}_2 $ by $1/\sqrt{S} $.

    These parameters can be extracted from all the physical measurable quantities such as
    magnetization, specific heat, various susceptibilities and spin correlation functions \cite{rh}.

\section{Quantization of the 1d Frenkel-Kontorowa (FK)  model: spin-coherent state path integral approach  }
   In Sec.VI, we presented the classical 1d FK model, here we provide the Quantization of the 1d Frenkel-Kontorowa (FK)  model by the spin-coherent state path integral which leads to a 2d quantized model.
   In principle, it can be applied to  investigate the fractal structure and calculate the excitation spectrum in any states in the global phase diagram Fig.1.
   Then we use it to re-derive the excitation spectrum of the Y-x state.
   Then we try to derive a low energy effective theory and find it is difficult to achieve this goal from
   the path integral approach. We shift back to the canonical quantization Eq.\ref{H2Yx} in the next section, then use the combination of
   both to achieve the goal.

\subsection{ The quantization of the 1d FK model at any  $ (\alpha, \beta) $. }
  In the spin coherent state, it is convenient to use the parametrization in the polar coordinate along the $ X $ direction:
\begin{align}
    \mathbf{S}_i=S(\cos\eta_i,\sin\xi_i\sin\eta_i,\cos\xi_i\sin\eta_i),
\label{spinYZ}
\end{align}

For the $N\times 1$ state at $ \beta < \alpha $, we have the classical state $(\xi_i,\eta_i)=(\xi_i^0,\eta_i^0)$,
where $\eta_i^0=\pi/2$ and
$\xi_i^0$ can obtained from the saddle point equation:
\begin{align}
    2\sin^2\beta\sin2\xi_i^0
    -\sin(\xi_i^0-\xi_{i-x}^0+2\alpha)
    +\sin(\xi_{i+x}^0-\xi_i^0+2\alpha)=0\>.
\end{align}
    So the classical state $ \mathbf{S}_i=S(0,\sin\xi^0_i,\cos\xi^0_i) $ is in YZ-plane, in contrast to
    Eq.\ref{2times2order} which is in the XY plane. The main difference is that the former is completely
    classical, while the latter involves OFQD which is a non-perturbative quantum effect.
    Of course, both coordinate contains Y-axis, so can be used to study the collinear Y-x phase.
    Indeed, the Y-x state was studied in Sec.III-IV in the $ (\theta, \phi) $ polar coordinate  along the $ Z- $ quantization axis,
    here it will be investigated in the  $ (\eta, \xi) $ polar coordinate Eq.\ref{spinYZ} along the $ X- $ quantization axis.

 To compute the quantum fluctuations, we also work out the second order derivatives
\begin{align*}
    &\left.\frac{\partial^2 H}{\partial\xi_i\partial\xi_j}
    \right|_{\substack{\eta=\eta_0\\\xi=\xi_0}}
    =-JS^2\big\{
     \cos[(\xi_i^0\!+\!2i_x\alpha)-(\xi_j^0\!+\!2j_x\alpha)](\delta_{i+x,j}+\delta_{i-x,j})
    +(\cos^2\xi_i^0+\cos2\beta\sin^2\xi_i^0)(\delta_{i+y,j}+\delta_{i-y,j})
    -2\epsilon_i^0\delta_{ij}\big\}\\
    &\left.\frac{\partial^2 H}{\partial\eta_i\partial\eta_j}
    \right|_{\substack{\eta=\eta_0\\\xi=\xi_0}}
    =-JS^2\big[
     \delta_{i+x,j}+\delta_{i-x,j}+\cos2\beta(\delta_{i+y,j}+\delta_{i-y,j})
    -2\epsilon_i^0\delta_{ij}
    \big]\\
    &\left.\frac{\partial^2 H}{\partial\xi_i\partial\eta_j}
    \right|_{\substack{\eta=\eta_0\\\xi=\xi_0}}
    =-JS^2[\sin2\beta\sin\xi_i^0 (\delta_{i+y,j}-\delta_{i-y,j})]\\
    &\left.\frac{\partial^2 H}{\partial\eta_i\partial\xi_j}
    \right|_{\substack{\eta=\eta_0\\\xi=\xi_0}}
    =-JS^2[-\sin2\beta\sin\xi_i^0 (\delta_{i+y,j}-\delta_{i-y,j})]
\label{secondd}
\end{align*}
where we have defined:
\begin{align}
    \epsilon_i^0=\big[\cos(\xi_{i+x}^0-\xi_i^0+2\alpha)
		+\cos(\xi_{i}^0-\xi_{i-x}^0+2\alpha)\big]/2
	-\sin^2\beta\cos2\xi_i^0+\cos^2\beta
\end{align}

Note that the ground state energy of the 1d FK model Eq.\ref{fk} can be written as
$\min_{\xi}E_{\rm FK}=-JS^2\sum_i\epsilon_i^0$.

The  expansion of the Hamiltonian to the second-order yields
\begin{align}
    H=H[\xi^0,\eta^0]+\frac{1}{2}\Bigg[
	\left.\frac{\partial^2 H}{\partial\xi_i\partial\xi_j}
	\right|_{\substack{\eta=\eta_0\\\xi=\xi_0}}\delta\xi_i\delta\xi_j
	+
	\left.\frac{\partial^2 H}{\partial\eta_i\partial\eta_j}
	\right|_{\substack{\eta=\eta_0\\\xi=\xi_0}}\delta\eta_i\delta\eta_j
	+
	\left.\frac{\partial^2 H}{\partial\xi_i\partial\eta_j}
	\right|_{\substack{\eta=\eta_0\\\xi=\xi_0}}\delta\xi_i\delta\eta_j
	+
	\left.\frac{\partial^2 H}{\partial\eta_i\partial\xi_j}
	\right|_{\substack{\eta=\eta_0\\\xi=\xi_0}}\delta\eta_i\delta\xi_j\Bigg]
	+\cdots
\label{pathhigh}
\end{align}
   where $ \cdots $ means the higher order terms which will be revisited in appendix F.

   One can define the conjugate variable  $ q_i=\delta\xi_i, \quad p_i=S\delta\eta_i $ satisfying the commutation relation:
\begin{align}
      [ q_i, p_j] = i \hbar \delta_{ij}
\label{qpi}
\end{align}
    which leads to the quantization of the classical FK model.

   Plugging in Eq.\ref{secondd} leads to:
\begin{align}
    H &=-JS^2\sum_i\epsilon_i^0+\frac{1}{2}JS^2\sum_i\Big[
	2\epsilon_i^0(\delta\xi_i\delta\xi_i+\delta\eta_i\delta\eta_i)
	-\cos(\xi_{i+x}^0-\xi_i^0+2\alpha)\delta\xi_i\delta\xi_{i+x}
	-\cos(\xi_{i}^0-\xi_{i-1}^0+2\alpha)\delta\xi_i\delta\xi_{i-x}	\nonumber\\
	&-(\cos^2\xi_i^0+\cos2\beta\sin^2\xi_i^0)
	 (\delta\xi_i\delta\xi_{i+y}+\delta\xi_i\delta\xi_{i-y})
	-(\delta\eta_i\delta\eta_{i+x}+\delta\eta_i\delta\eta_{i-x})
	-\cos2\beta(\delta\eta_i\eta_{i+y}+\delta\eta_{i}\delta\eta_{i-y})  \nonumber\\
	&-\sin2\beta\sin\xi_i^0(\delta\xi_i\delta\eta_{i+y}-\delta\xi_i\delta\eta_{i-y})
	+\sin2\beta\sin\xi_i^0(\delta\eta_i\delta\xi_{i+y}-\delta\eta_i\delta\xi_{i-y})	
	\Big]
\label{eq:H2d}
\end{align}
which can be simplified to:
\begin{align}
    H&=-JS^2\sum_i\epsilon_i^0+JS^2\sum_i\Big[
	\epsilon_i^0\delta\xi_i\delta\xi_i
	-\cos(\xi_{i+x}^0-\xi_i^0+2\alpha)\delta\xi_i\delta\xi_{i+x}
	-(\cos^2\xi_i^0+\cos2\beta\sin^2\xi_i^0)\delta\xi_i\delta\xi_{i+y} \nonumber\\
	&+\epsilon_i^0\delta\eta_i\delta\eta_i
	-\delta\eta_i\delta\eta_{i+x}-\cos2\beta\delta\eta_i\delta\eta_{i+y}
	-\sin2\beta\sin\xi_i^0(\delta\xi_i\delta\eta_{i+y}-\delta\eta_i\delta\xi_{i+y})
	\Big]
\end{align}

   The corresponding action on the square lattice is:
\begin{equation}
  {\cal L}= S \delta\xi_i \partial_{\tau} \delta\eta_i + H[ \delta\xi_i, \delta\eta_i]
\label{generalS}
\end{equation}
  which, in principle, can be used to study the fractal structure in Fig.1.
  For example, it can be used to compute the excitation spectrum in any $ N \times 1 $ state in Fig.1.
  In the following, we will use it to re-derive the $ N=2 $ case which is nothing but the Y-x state
  which, in contrast to $ N \geq 3 $ cases, is a collinear state instead of a co-planar state.

%\clearpage
\subsection{ The excitation spectrum in the Y-x state }

Near $\alpha\sim\pi/2$ and $0\le\beta\le\pi/2$, the ground state is the Y-x state in Fig.1.
It's saddle point is $(\xi^0_i,\eta^0_i)=((-1)^{i_x}\pi/2,\pi/2)$.
Plugging into Eq.\ref{spinYZ} leads to $ \mathbf{S}_i=S(0, (-1)^{i_x},0 ) $.
Note that we reach this state completely by classical calculations instead of involving any OFQD.
We can confirm this by checking Eq.(C2). Indeed the linear term vanishes.
%Without loss of generality,
\begin{align}
    &2\sin^2\beta\sin(2\times(-1)^{i_x}\pi/2)
     -\sin[(-1)^{i_x}\pi/2-(-1)^{i_x-1}\pi/2+2\alpha]
      +\sin[(-1)^{i_x+1}\pi/2-(-1)^{i_x}\pi/2+2\alpha]\nonumber\\
    =&0-\sin(\pi/2+2\alpha)+\sin(\pi/2+2\alpha)=0
\end{align}
   We can also evaluate
\begin{align}
    \epsilon_i^0=-[\cos(2\alpha)+\cos(2\alpha)]/2-\sin^2\beta\cos(2\times(-1)^{i_x}\pi/2)+\cos^2\beta
	=-\cos2\alpha+1=2\sin^2\alpha
\end{align}

 The general quantum fluctuations in the classical Y-x state becomes:
\begin{align}
    \mathbf{S}_i=-S(\sin \delta \eta_i,  -(-1)^{i_x}\cos \delta \xi_i\cos \delta\eta_i, (-1)^{i_x} \sin \delta\xi_i\cos \delta\eta_i ),
\label{spinYx}
\end{align}
Now, Eq.\eqref{eq:H2d} takes the form
\begin{align}
    H & =-2JS^2N+JS^2\sum_i\Big[
	2\sin^2\alpha\delta\xi_i\delta\xi_i
	+\cos2\alpha\delta\xi_i\delta\xi_{i+x}
	-\cos2\beta\delta\xi_i\delta\xi_{i+y}  \nonumber \\
	&+2\sin^2\alpha\delta\eta_i\delta\eta_i
	-\delta\eta_i\delta\eta_{i+x}
	-\cos2\beta\delta\eta_i\delta\eta_{i+y}
	-(-1)^{i_x}\sin2\beta(\delta\xi_i\delta\eta_{i+y}-\delta\eta_i\delta\xi_{i+y})
	\Big]
\label{Yxhamil}
\end{align}
   In terms of the conjugate variable in Eq.\ref{qpi}, it becomes:
\begin{align}
    H & =-2JS^2N+J\sum_i\Big[S^2(
	2\sin^2(\alpha) q_i^2+\cos2\alpha q_iq_{i+x}-\cos2\beta q_iq_{i+y}) \nonumber \\
	&+2\sin^2(\alpha) p_i^2-p_ip_{i+x}-\cos2\beta p_i p_{i+y}
	-(-1)^{i_x}S\sin2\beta(q_i p_{i+y}-q_ip_{i+y})
	\Big]
\label{HYx}
\end{align}
where one can extract the mass, force constant and the coupling matrices  as:
\begin{align}
    M_{ij}^{-1} & =JS^2[4\sin^2\alpha\delta_{ij}
	+\cos2\alpha(\delta_{i+x,j}+\delta_{i-x,j})
	-\cos2\beta(\delta_{i+y,j}+\delta_{i-y,j})], \nonumber \\
    K_{ij} & =J[4\sin^2\alpha\delta_{ij}-\delta_{i+x,j}-\delta_{i-x,j}
	-\cos2\beta(\delta_{i+y,j}+\delta_{i-y,j})], \nonumber \\
    P_{ij} & =-(-1)^{i_x}JS\sin2\beta(\delta_{i-y,j}-\delta_{i+y,j})
\end{align}

After introducing A/B sub-lattice structure and incorporating the quantum commutation relations $ [ q_i, p_i] = i \hbar $,
we can express the characteristic equation in the $(q_k^A,q_k^B,p_k^A,p_k^B)$ basis
\begin{align}
    \begin{pmatrix}
	JS^2[4\sin^2\alpha-2\cos2\beta\cos k_y] &JS^2[2\cos2\alpha\cos k_x] &i\omega_k+i2JS\sin2\beta\sin k_y &0\\
	JS^2[2\cos2\alpha\cos k_x] &JS^2[4\sin^2\alpha-2\cos2\beta\cos k_y] &0 &i\omega_k-i2JS\sin2\beta\sin k_y\\
	-i\omega_k-i2JS\sin2\beta\sin k_y &0 &J[4\sin^2\alpha-2\cos2\beta\cos k_y] &-J[2\cos k_x]\\
	0 &-i\omega_k+i2JS\sin2\beta\sin k_y &-J[2\cos k_x] &J[4\sin^2\alpha-2\cos2\beta\cos k_y]\\
    \end{pmatrix}
\end{align}
  Setting its determinant vanishing leads to the excitation spectrum in the Y-x state:
\begin{align}
    \omega_k&=JS
    \sqrt{(4\sin^2\alpha-2\cos2\beta\cos k_y)^2
	  +4\sin^4\alpha\cos^2 k_x-4\cos^4\alpha\cos^2 k_x
	  +4\sin^22\beta\sin^2k_y
	\pm 2X
    }   \nonumber \\
    X&=\sqrt{(4\sin^2\alpha-2\cos2\beta\cos k_y)^2
	    (4\sin^4\alpha\cos^2 k_x+4\sin^22\beta\sin^2k_y)}
\end{align}
  which can be simplified as
\begin{align}
    \omega_k =JS\sqrt{
    \big(4\sin^2\alpha-2\cos2\beta\cos k_y
	\pm\sqrt{4\sin^4\alpha\cos^2 k_x+4\sin^22\beta\sin^2k_y}\big)^2
    -4\cos^4\alpha\cos^2 k_x}
\label{Yxex}
\end{align}

Comparing with results achieved from the HP boson calculation in the main text and appendix A,
we find they are identical.

\subsection{ The difficulty with the path-integral approach to find  the spin-spin correlation function inside the Y-x state and
  to construct the quantum Lifshitz transition from the Y-x state on the right to its neighbouring phases }

  In the last section, we derived the excitation spectrum in the Y-x state within the first ( the largest lob in Fig.1 ).
  In Ref.\cite{rh}, the Y-x state is the exact ground state alone the solvable line $ ( \alpha=\pi/2, \beta ) $.
  it has the exact $ U(1)_{soc} $ symmetry which put exact constraints on the SSCF at any finite temperature.
  So one can use the canonical quantization approach to evaluate the SSCFs at a small finite $ T $ below the
  finite melting transition $ T_m $. Here away from the solvable line, the exact $ U(1)_{soc} $ symmetry
  is absent,  the Y-x is just a classical ground state with strong quantum fluctuations,
  so the SSCF is non-trivial even at $ T=0 $. So it is much more involved to evaluate the SSCF  with strong quantum
  fluctuations in the absence of the $ U(1)_{soc} $.

 After splitting into sublattice A and B, in $k$-space, the spin Hamiltonian Eq.\ref{Yxhamil} takes the form
\begin{align}
	H_2[\delta\xi,\delta\eta]
		=-2JS^2N+\frac{1}{2}JS^2\sum_k[
		&(4\sin^2\alpha-2\cos2\beta\cos k_y)
		 (\delta\xi_k^A\delta\xi_{-k}^A+\delta\xi_k^B\delta\xi_{-k}^B)
		 +4\cos2\alpha\cos k_x \delta\xi_k^A\delta\xi_{-k}^B  	\nonumber\\
		&+(4\sin^2\alpha-2\cos2\beta\cos k_y)
		  (\delta\eta_k^A\delta\eta_{-k}^A+\delta\eta_k^B\delta\eta_{-k}^B)
		 -4\cos k_x \delta\eta_k^A\delta\eta_{-k}^B		\nonumber\\
		&+i4\sin2\beta\sin k_y(\delta\xi_k^A\delta\eta_{-k}^A-\delta\xi_k^B\delta\eta_{-k}^B)]
\label{Yxhamilk}
\end{align}
then the Lagrangian is
\begin{align}
	L[\delta\xi,\delta\eta]
	=\sum_k S\omega (\delta\xi_k^A\delta\eta_{-k}^A+\delta\xi_k^B\delta\eta_{-k}^B)
	+H_2[\delta\xi,\delta\eta]
\label{Yxpathk}
\end{align}
% For simplicity, we take $J=1$ and $S=1$ in below.
 In principle, one can evaluate the SSCF
from Eq.\ref{spinYx} and Eq.\ref{Yxhamilk}. Unfortunately,
it becomes essentially impossible to extract any physics from such  general and complicated expressions.

It becomes important to get a compact and physical expression for the SSCFs corresponding to the low energy mode $ \omega_{k}^{-} $ in Eq.\ref{Yxex}.
Unfortunately, despite we can recover the whole dispersion Eq.\ref{Yxex} easily from the path integral Eq.\ref{Yxpathk},
we find it is very difficult to derive the effective low action corresponding to the low energy mode in Eq.\ref{rela}
from the path-integral Eq.\ref{Yxpathk}. It is very difficult to even isolate the low
energy mode $ \omega_{k}^{-} $ from the high energy one $ \omega_{k}^{+} $ in the path integral language.
So one has to turn back to the canonical quantization approach in appendix A where this crucial isolation can be  done easily
and intuitively.
To linear order, one can transfer back and forth between the canonical quantization and the path integral.
This transfer back and forth strategy  was also used in dealing with the infra-red divergence due to the photon condensation in the super-radiant phase in the $ U(1) $ Dicke model \cite{u1dicke}.

 The goal of the following sections is to find an effective action in terms of critical fields ( or order parameters ), then the relations
 between the quantum spin in terms of the critical fields,  evaluate the leading
 spin-spin correlations functions inside the Y-x phase and also find the spin-orbital structure of the IC-SkX it may get into.
 For notational convenience, we list the variables ( or order parameters ) and physical quantities frequently used in this appendix.

 (a)  The order parameters appearing in the effective actions:  $ q= \delta \xi, p= S \delta \eta $ which satisfy the commutation relation
 $ [q_i, p_j]=i \delta_{ij} $,
 For the A/B sublattice in the Y-x phase, one forms $ q^{\pm}, p^{\pm} $ which satisfy the commutation relations $[q_{k}^{(s)},p_{-k}^{(s)}]=i\delta_{ss'}$ where $s,s'=\pm$.

 (b)  The relations between the spin and the order parameters:
 To then linear order $ S^x_i= S \delta \eta_i, S^z_i= (-1)^{i_x} S \delta \xi_i $, namely,
 $ S^x_A= S \delta \eta_A, S^x_A= S \delta \eta_B  $   and $ S^z_A=  S \delta \xi_A,   S^z_B=  -S \delta \xi_B $.
 To higher order, one need use the expression in Eq.\ref{spinYx}.

\section{ The low energy mode and low energy effective action inside the Y-x state and close to the transitions: the combination of
the canonical quantization and the path integral   }

   Now we shift from the path integral to the canonical quantization.
   As argued below Eq.\ref{Yxpathk}, in principle, from the relations between the original bosons
   HP bosons $ a,a^{\dagger}, b, b^{\dagger} $ and the quasi-particle operators
   $ \alpha_k, \alpha_k^\dagger, \beta_k, \beta_k^\dagger $ in Eq.\ref{H2Yx}, one can
   one can evaluate the SSCF. Unfortunately,
   it becomes essentially impossible to extract any physics from such  general and complicated expressions either.
%   In fact, this is even worse than the path integral method mentioned below Eq.\ref{Yxpathk},
%   because the canonical quantization method to the order $ 1/S $ did not take into account the magnon/magnon interactions,
%   so can not be extended to a finite $ T $, but the path-integral method can.

   Obviously, one can extract the following low energy effective action  directly from Eq.\ref{H2Yx}:
\begin{align}
    H_\text{eff}=4JS\sum_k \omega_{k}^{-}\beta_k^\dagger\beta_k
\label{effYx}
\end{align}
where $\omega_k^{-}=\sqrt{(\lambda_{k}^-)^2-\chi_k^2}$. It suggests that the low energy field can be taken as $\beta_k$:
\begin{align}
	\beta_k
	=u_k^b \bar{b}_k-v_k^b \bar{b}_{-k}^\dagger
	=u_k^b [\cos(\theta_k/2) a_k+\sin(\theta_k/2) b_k]
	-v_k^b [\cos(\theta_{-k}/2) a_{-k}^\dagger+\sin(\theta_{-k}/2) b_{-k}^\dagger]
\label{betak}
\end{align}

Unfortunately, this naive suggestion does not work as the magnon condensation boundary is approached.

\subsection{ $ \beta_k $ is not the suitable critical field }

  This is because at the critical SOC parameters $(\alpha,\beta)$ and the magnon condensation momentum $(0,k_0)$,
  $\omega_{k}^{-}\to0$, $u_k$ and $v_k$ diverges. Then  the magnon field  $ \beta_k $ in  Eq.\ref{betak} also diverges, so it can not be used
  as the magnon critical field near the phase boundary. This is also closely related to the
  key observation as stressed in the appendix F-3 that $ \omega_{k}^{-} $ in Eq.\ref{rela} is non analytic !
  However, the fact that the effective Hamiltonian Eq.\ref{effYx} remain finite suggests the following
  finite combination as the critical magnon field \cite{dvm}:
\begin{eqnarray}
	& \lim_{k\to k_0}\sqrt{\omega_{k}^{-}}\beta_k  \\ \nonumber
    & =\lim_{k\to k_0}
	\frac{1}{2}\{
	[(\lambda_k^-)^{1/2}\cos(\theta_k/2)a_k-\chi_k^-(\lambda_k^-)^{-3/2}\sin(\theta_k/2)a_k^\dagger]
	+[(\lambda_k^-)^{1/2}\sin(\theta_k/2)b_k-\chi_k^-(\lambda_k^-)^{-3/2}\cos(\theta_k/2)b_k^\dagger]\}  \\ \nonumber
	& = \lim_{k\to k_0}
	\frac{1}{2}(\lambda_k^-)^{1/2}\{
	[\cos(\theta_k/2)a_k-\sin(\theta_k/2)a_k^\dagger]
	+[\sin(\theta_k/2)b_k-\cos(\theta_k/2)b_k^\dagger]\}
\label{fakebeta}
\end{eqnarray}
 which is indeed finite at the boundary. Thus we only need to evaluate $\theta_k$ at $(0,k_0)$.
Now apply it to three types of magnon condensation.

\begin{itemize}
	\item For C$_0$,  $k_0=0$ and $\theta_{k_0}=\pi/2$,
		then $\cos(\theta_k/2)a_k-\sin(\theta_k/2)a_k^\dagger\propto \bar{S}_A^+-\bar{S}_A^-=\bar{S}_A^y=S_A^z$
		and $\sin(\theta_k/2)b_k-\cos(\theta_k/2)b_k^\dagger\propto \bar{S}_B^--\bar{S}_B^+=\bar{S}_B^y=S_B^z$.
	\item For C$_\pi$,  $k_0=\pi$ and $\theta_{k_0}=\pi/2$,
		then $\cos(\theta_k/2)a_k-\sin(\theta_k/2)a_k^\dagger\propto \bar{S}_A^+-\bar{S}_A^-=\bar{S}_A^y=S_A^z$
		and $\sin(\theta_k/2)b_k-\cos(\theta_k/2)b_k^\dagger\propto \bar{S}_B^--\bar{S}_B^+=\bar{S}_B^y=S_B^z$.

	\item For IC-,  $k_0$ is incommensurate and
		$\theta_{k_0}=\arcsin(\frac{6}{1+9\sin^22\beta})\in[0.205\pi,0.5\pi]$,
		then $\cos(\theta_k/2)a_k-\sin(\theta_k/2)a_k^\dagger
		\propto\sin(\frac{\pi}{4}-\frac{\theta}{2})\bar{S}_A^x
			+\cos(\frac{\pi}{4}-\frac{\theta}{2})\bar{S}_A^y$
		and $\sin(\theta_k/2)b_k-\cos(\theta_k/2)b_k^\dagger
		\propto \cos(\frac{\pi}{4}+\frac{\theta_k}{2})\bar{S}_B^x
			+\sin(\frac{\pi}{4}+\frac{\theta_k}{2})\bar{S}_B^y$,
\end{itemize}
  which is a linear combination of $ S^x $ and $ S^{z} $ tuned by the angle $ \theta_k $.
  Because we already set the field at the phase boundary and the condensation point, so we can not see
  the critical behaviour, only the massive modes survive. So for the C-magnons, setting $ S^z=0 $ implies
  the condensation of C-magnons may lead to a co-planar state in XY plane.
  For the IC-magnons, setting $ \cos(\frac{\pi}{4}+\frac{\theta_k}{2})\bar{S}_B^x
			+\sin(\frac{\pi}{4}+\frac{\theta_k}{2}) S_B^z=0 $  implies
  the condensation of IC-magnons may lead to a non coplanar state.

  In a sharp contrast, despite the effective Hamiltonian Eq.\ref{effYx} takes the identical form as Eq.\ref{rhhlow}
  which describes the Y-x state subject to a longitudinal Zeeman field. Here, $ \beta_k $  can be taken as the critical field ( or order parameter ), so one must search for a suitable critical field,  but it is in the latter case dictated by the
  $ U(1)_{soc} $  as demonstrated in appendix.G. This is also directly related to the fact that here $ \omega_{k}^{-} $ is non-analytic,
  while it is in  Eq.\ref{rhhlow} with the dynamic exponent $ z=2 $.

\subsection{ Construct the effective action in terms of the  critical field ( or order parameter ). }

   In order to extract the clear physical meanings of the above equations, it is important to move away
   from the condensation points $ (0, \pm k_0) $ and also the  condensation boundary.
  Substituting  Eq.\ref{betak} into Eq.\ref{effYx} leads to its  form in terms of the original HP boson:
\begin{equation}
    H_\text{low}=2JS
    \sum_k
	\begin{pmatrix}
	  a_k^\dagger\\
	  b_k^\dagger\\
	  a_{-k}\\
	  b_{-k}\\
	\end{pmatrix}^\intercal
	\begin{pmatrix}
	 \lambda_k^-(1+\cos\theta_k) &\lambda_k^-  \sin\theta_k & -\chi_k  \sin\theta_k & -\chi_k(\cos\theta_k+1) \\
	 \lambda_k^-\sin\theta_k & \lambda_k^-(1-\cos\theta_k) & \chi_k(\cos\theta_k-1) & -\chi_k\sin\theta_k \\
	 -\chi_k  \sin\theta_k & \chi_k(\cos\theta_k-1) & \lambda_k^- (1-\cos\theta_k) & \lambda_k^-\sin\theta_k\\
	 -\chi_k  (\cos\theta_k+1) & -\chi_k\sin\theta_k &\lambda_k^-\sin\theta_k &\lambda_k^-(1+\cos\theta_k) \\
	\end{pmatrix}
    	\begin{pmatrix}
	  a_k\\
	  b_k\\
	  a_{-k}^\dagger\\
	  b_{-k}^\dagger\\
	\end{pmatrix}
\label{HabYx}
\end{equation}
which is still $4\times4$ matrix. Diagonalising it  lead to the eigen mode $\omega_k^-$ and one extra zero mode.
The zero mode has no physical meaning, because it is due to the fact that we project out the high energy mode $ \omega_k^+ $
in Eq.\ref{H2Yx} \cite{takelimit}.

 In the polar coordinate Eq.\ref{spinYx}, the quantum spin is expressed  in terms of $ ( \delta\eta_i, \delta\xi_i ) $ so to compute the
 SSCF, it would be most convenient to
 write the HP boson in the above equation in terms of $ ( \delta\eta_i, \delta\xi_i ) $.
 For small quantum fluctuations around the classical Y-x state in Eq.\ref{spinYx},  one can write
 \begin{align}
   \delta\mathbf{S}_i
	=(\delta S_i^x,\delta S_i^y,\delta S_i^z)
	=-S(\delta\eta_i,0,(-1)^{i_x}\delta\xi_i)
	=-(p_i,0,(-1)^{i_x}Sq_i)
\end{align}
  which leads to the following relations between HP bosons $ (a, b) $ in the canonical quantization and
  the $ ( p, q ) $ coordinates in the spin-coherent state path-integral representation  \cite{dvm}:
\begin{align}
    a_i=(S_i^x+iS_i^z)/\sqrt{2S}=-(p_i+iSq_i)/\sqrt{2S},\quad
    a_i^\dagger=(S_i^x-iS_i^z)/\sqrt{2S}=-(p_i-iSq_i)/\sqrt{2S},\quad
    \forall i\in A     \\ \nonumber
    b_i^\dagger=(S_i^x+iS_i^z)/\sqrt{2S}=-(p_i-iSq_i)/\sqrt{2S},\quad
    b_i=(S_i^x-iS_i^z)/\sqrt{2S}=-(p_i+iSq_i)/\sqrt{2S},\quad \forall i\in B
\end{align}
    whose Fourier transformation lead to:
\begin{align}
    a_k=-(p_{-k}^A+iSq_{-k}^A)/\sqrt{2S}, \quad b_k=-(p_{-k}^B+iSq_{-k}^B)/\sqrt{2S}
\label{ab-pq}
\end{align}

  Introducing the uniform  ( or the center of mass ) and staggered ( or relative ) combinations
   $q_{k}^{(\pm)}$ and $p_{k}^{(\pm)}$  of the two sublattices defined by \cite{differpm}
\begin{align}
	q_{k}^{(\pm)}=(q_k^A\pm q_k^B)/\sqrt{2},\quad
	p_{k}^{(\pm)}=(p_k^A\pm p_k^B)/\sqrt{2}
\end{align}
 Then the commutation relations become $[q_{k}^{(s)},p_{-k}^{(s)}]=i\delta_{ss'}$ where $s,s'=\pm$.

 Eq.\ref{HabYx} leads can be expressed in terms of two decoupled $ 2 \times 2 $ matrix form as
\begin{align}
    H_\text{low}
    =J\sum_k
	\begin{pmatrix}
	    q_{k}^{(+)}\\
	    p_{k}^{(-)}\\
	\end{pmatrix}^\intercal
	\begin{pmatrix}
	    S^2(\lambda_k^-+\chi_k)(1+\sin\theta_k) &-iS(\lambda_k^-+\chi_k)\cos\theta_k\\
	    iS(\lambda_k^-+\chi_k)\cos\theta_k &S^2(\lambda_k^-+\chi_k)(1+\sin\theta_k)\\
	\end{pmatrix}
	\begin{pmatrix}
	    q_{-k}^{(+)}\\
	    p_{-k}^{(-)}\\
	\end{pmatrix}           \\  \nonumber
	+
	J\sum_k
	\begin{pmatrix}
	    q_{k}^{(-)}\\
	    p_{k}^{(+)}\\
	\end{pmatrix}^\intercal
	\begin{pmatrix}
	    S^2(\lambda_k^--\chi_k)(1-\sin\theta_k) &-iS(\lambda_k^--\chi_k)\cos\theta_k \\
	    iS(\lambda_k^--\chi_k)\cos\theta_k &S^2(\lambda_k^--\chi_k)(1-\sin\theta_k)  \\
	\end{pmatrix}
    	\begin{pmatrix}
	    q_{-k}^{(-)}\\
	    p_{-k}^{(+)}\\
	\end{pmatrix}
\end{align}

 Finally, we reach the low energy effective Hamiltonian in the  $ ( p^{\pm}, q^{\pm} ) $  representation:
\begin{eqnarray}
    H_\text{low} & = & J\sum_k
	(\lambda_k^--\chi_k)[S^2(1-\sin\theta_k)q_{k}^{(-)}q_{-k}^{(-)}
			+(1+\sin\theta_k)p_{k}^{(+)}p_{-k}^{(+)}
			-iS\cos\theta_k(q_{k}^{(-)}p_{-k}^{(+)}-p_{k}^{(+)}q_{-k}^{(-)})]    \\  \nonumber
	& + & (\lambda_k^-+\chi_k)[S^2(1+\sin\theta_k)q_{k}^{(+)}q_{-k}^{(+)}
			+(1-\sin\theta_k)p_{k}^{(-)}p_{-k}^{(-)}
			-iS\cos\theta_k(q_{k}^{(+)}p_{-k}^{(-)}-p_{k}^{(-)}q_{-k}^{(+)})]
\label{Hlowall}
\end{eqnarray}
   which contains the two sets of variables $ ( p^{-}, q^{-} ) $ and $ ( p^{+}, q^{+} ) $ and their mutual couplings.

   In the following, we apply this Hamiltonian to the three magnon regimes.
   Due to the mirror symmetry $ \beta \rightarrow \pi/2 - \beta $ at the quadratic order, the two commensurate magnons
   C$_0$ or C$_\pi$  can be discussed together with the caution that C$_\pi$ magnons's minimum is at $ k^0_y=\pi $
   ( Fig.\ref{threemag} ).

\subsection{ The commensurate magnons }

In C$_0$ or C$_\pi$ regimes, $\theta_k$ at the minima takes value $\pi/2$,
then the coupling between the two set of variables
are small,  only one half of the degree of freedoms $ ( p^{+}, q^{+} ) $ set survives in the effective Hamiltonian.
They stand for the center of mass (COM) ( or uniform )  fluctuations between the two sublattices $ A $ and $ B $  which
 are controlled by the lower energy  mode $ \omega_k^- $.
The other half of the degree of freedoms $ ( p^{-}, q^{-} ) $ set only make sub-leading contributions, so can be dropped.
They stand for the relative ( or staggered )  fluctuations between the two sublattices $ A $ and $ B $ which are controlled
by the higher energy  mode $ \omega_k^+ $.

Thus Eq.\ref{Hlowall} can be expressed in terms of only half degree of freedoms $ ( p^{+}, q^{+} ) $ set
which stand for the translational ( or uniform )  fluctuations between the two sublattices $ A $ and $ B $:
\begin{equation}
	H_\text{low} =2J\sum_k [(\lambda_k^--\chi_k)p_{k}^{(+)}p_{-k}^{(+)}+S^2(\lambda_k^-+\chi_k)q_{k}^{(+)}q_{-k}^{(+)}]
\label{hpqplus}
\end{equation}

Due to the commutation relation $ [q^{+}, p^{+}]= i $, the low-energy effective action corresponding to
the Hamiltonian Eq.\ref{hpqplus} is
\begin{equation}
    S_\text{low}=\int d\tau\sum_k\{
	p_{k,\tau}^{(+)}\partial_\tau q_{-k,\tau}^{(+)}
	+2J[(\lambda_k^--\chi_k)p_{k,\tau}^{(+)}p_{-k,\tau}^{(+)}
	+S^2(\lambda_k^-+\chi_k)q_{k,\tau}^{(+)}q_{-k,\tau}^{(+)}]\}
\label{pqplusaction}
\end{equation}

  As shown in Eq.\ref{c0crit}, $q_{k}^{(+)}$ remains massive, integrating it out leads to the low energy effective action:
\begin{align}
	S_\text{eff}[p^{(+)}]=\frac{1}{2}\sum_{k,\omega_n}
	\frac{(4JS\omega_k^-)^2+\omega_n^2}{4JS^2(\lambda_k^-+\chi_k)}
	p^{(+)}(k,i\omega_n)p^{(+)}(-k,-i\omega_n)
\label{ppplusaction}
\end{align}

   The process of deriving this low energy effective action is just transfer the non-analyticity in Eq.\ref{effYx}
   to the analytic one here which will be used to evaluate the spin-spin correlation functions (SSCFs) in the next section  \cite{nonanalytic}.

\subsection{ The In-commensurate magnons }

In IC- regimes, $\theta_k$ at $(k_x,k_y)=(0,\pm k_0)$ will be
in the range $(0.205\pi,0.5\pi)$ for $+$ sign
or $(0.5\pi, 0.795\pi)$ for $-$ sign.  The $H_\text{low}$ in Eq.(77-78) can be \textit{re}-written as
\begin{align}
    H_\text{low}=J\sum_k
	\Big\{\frac{\lambda_k^--\chi_k}{1+\sin\theta_k}
	[-iS\cos\theta_k q_k^{(-)}+(1+\sin\theta_k)p_k^{(+)}]
	[-iS\cos\theta_{-k} q_{-k}^{(-)}+(1+\sin\theta_{-k})p_{-k}^{(+)}]    \\  \nonumber
	+\frac{\lambda_k^-+\chi_k}{1-\sin\theta_k}
	[-iS\cos\theta_k q_k^{(+)}+(1-\sin\theta_k)p_k^{(-)}]
	[-iS\cos\theta_{-k} q_{-k}^{(+)}+(1-\sin\theta_{-k})p_{-k}^{(-)}]\Big\}
\end{align}

   This form  suggests that if one introduces the complex conjugate variables
\begin{align}
	\tilde{p}_k=[-i S\cos\theta_kq_k^{(-)}+ (1+\sin\theta_k )p_k^{(+)}]/2,\quad
	\tilde{q}_k=q_k^{(+)}+\frac{i(1-\sin\theta_k)}{S\cos\theta_k}p_k^{(-)}
\end{align}
  which satisfy the commutation relation $ [ \tilde{p}_k,  \tilde{q}^{*}_k ]=1 $.
  It also automatically ensures its
  complex conjugate $ [ \tilde{p}^{*}_k,  \tilde{q}_k ]=1 $.
  One can also check that as $ \theta_k \rightarrow \pi/2 $, only $ ( p^{+}, q^{+} ) $ set survives, so
  they also smoothly connected to the C- magnons.
  Then  $H_\text{low}$ can be written in terms of the complex conjugate pairs:
\begin{align}
    H_\text{low}=J\sum_k
	\Big[\frac{4 ( \lambda_k^--\chi_k) }{1+\sin\theta_k}\tilde{p}_k\tilde{p}_{k}^*
	+S^2  ( 1+\sin\theta_k )( \lambda_k^-+\chi_k )   \tilde{q}_k\tilde{q}_{k}^*\Big]
\label{hpqcomplex}
\end{align}

  However, one can push this further by observing that $\tilde{p}_{k}^{(+)*}=\tilde{p}_{-k}^{(+)}$.
  As shown in Eq.\ref{evenodd},  $ \cos\theta_k $ ( $ \sin \theta_k $ )  is an odd ( even ) function under
  $ \vec{k} \rightarrow  - \vec{k} $. One can show that
\begin{align}
	\tilde{p}_k^{(+)*}
	=[+iS\cos\theta_k q_{-k}^{(-)}+(1+\sin\theta_k)p_{-k}^{(+)}]/2
	=[-iS\cos\theta_{-k} q_{-k}^{(-)}+(1+\sin\theta_{-k})p_{-k}^{(+)}]/2
	=\tilde{p}_{-k}^{(+)}
\end{align}
and $\tilde{p}_r^{(+)}$ must be a real field. This fact suggests that one need to introduce
the two real conjugate pairs \cite{scalefreedom}:
\begin{eqnarray}
	\tilde{p}_k^{(+)} & = & [-iS\cos\theta_k q_k^{(-)}+(1+\sin\theta_k)p_k^{(+)}]/2,\quad
	\tilde{q}_k^{(+)}=q_k^{(+)}+\frac{i(1-\sin\theta_k)}{S\cos\theta_k}p_k^{(-)}    \\   \nonumber
	\tilde{q}_k^{(-)} & = & [(1+\sin\theta_k) q_k^{(-)}-\frac{i\cos\theta_k}{S} p_k^{(+)}]/2,\quad
	\tilde{p}_k^{(-)}=p_k^{(-)}+\frac{iS(1-\sin\theta_k)}{\cos\theta_k}q_k^{(+)}
\label{tildeset}
\end{eqnarray}
where $\tilde{p}_r^{(\pm)}$ and $\tilde{q}_r^{(\pm)}$ are all real and keep the commutation
relations $[\tilde{q}_k^{(s)}, \tilde{p}_{-k}^{(s')}]=i\delta_{ss'}$.
It is easy to check that in the C limit $ \theta_k \rightarrow \pi/2 $, the conjugate pairs recover those of the C-magnons:
$ ( \tilde{p}_k^{(\pm)}, \tilde{q}_k^{(\pm)} ) \rightarrow  ( p_k^{(\pm)}, q_k^{(\pm)} )  $.

  The corresponding  action can be written as:
\begin{equation}
    S_\text{low}=\frac{1}{2}\int d\tau\sum_k\{
	\tilde{p}_{k,\tau}^{(+)}\partial_\tau \tilde{q}_{k,\tau}^{(+)*}
	-\tilde{q}_{k,\tau}^{(+)}\partial_\tau \tilde{p}_{k,\tau}^{(+)*}
	+J[\frac{4(\lambda_k^--\chi_k)}{1+\sin\theta_k}\tilde{p}_{k,\tau}^{(+)}\tilde{p}_{k,\tau}^{(+)*}
	+S^2(1+\sin\theta_k)(\lambda_k^-+\chi_k)\tilde{q}_{k,\tau}^{(+)}\tilde{q}_{k,\tau}^{(+)*}]\}
\label{tildepqplusact}
\end{equation}
   which indeed reproduce $\omega_k^-$. The fact that only half of the degree of freedoms
   $ ( \tilde{p}_k^{(+)}, \tilde{q}_k^{(+)} ) $
 appear in the action means that this pair is controlled by the lower energy branch $ \omega_{k}^- $.
 While the other half of degree of freedoms $ ( \tilde{p}_k^{(-)}, \tilde{q}_k^{(-)} ) $  controlled by the higher energy branch $ \omega_{k}^+ $
 set are projected out, so only make sub-leading contributions ( which will
 still be discussed in the next section ).

 From the expressions of $ \tilde{p}_k^{(+)}, \tilde{q}_k^{(-)} $ in Eq.\ref{tildeset}, we obtain:
\begin{equation}
	p_k^{(+)}=\tilde{p}_k^{(+)}+\frac{iS\cos\theta_k}{1+\sin\theta_k}\tilde{q}_k^{(-)},\quad
	q_k^{(-)}=\tilde{q}_k^{(-)}+\frac{i\cos\theta_k}{S(1+\sin\theta_k)}\tilde{p}_k^{(+)}
\label{eq:-pq}
\end{equation}

 Because the IC magnons have two minima $ \pm Q $ which are coupled to each other.
 Expanding around the two IC- minima $ k= \pm Q + \delta k $ leads to:
\begin{align}
	S_\text{low}=\int d\tau\sum_{\delta k\ll\Lambda}\{
	&\tilde{p}_{k,\tau}^{(+)} \partial_\tau \tilde{q}_{-k,\tau}^{(+)}
	%-\tilde{q}_{k,\tau}^{(+)} \partial_\tau \tilde{p}_{-k,\tau}^{(+)}\\
	+\frac{J}{2}[\frac{4(\lambda_k^--\chi_k)}{1+\sin\theta_0}
	\tilde{p}_{k,\tau}^{(+)}\tilde{p}_{-k,\tau}^{(+)}
	+S^2(1+\sin\theta_0)(\lambda_k^-+\chi_k)
	\tilde{q}_{k,\tau}^{(+)}\tilde{q}_{-k,\tau}^{(+)}]\}
\label{tildepq}
\end{align}
where  $ \sin\theta_0 =\sin ( \pi - \theta_0) $ is even under $ k \rightarrow -k $ ( Fig. \ref{threemag} ) and
\begin{align}
	\theta_0=\theta_Q
	=\arcsin\Big(\frac{\sin^2\alpha}{\sin2\beta\sqrt{\sin^4\alpha+\sin^22\beta}}\Big)
\label{Qvalue}
\end{align}

   Since $\tilde{q}^{(+)}$ mode remains massive, integrating it out leads to
\begin{align}
	S_\text{eff}[\tilde{p}]=\frac{1}{2}\sum_{\delta k,\omega_n}
	\frac{(4JS\omega_{k}^-)^2+\omega_n^2}
		{2JS^2(1+\sin\theta_0)(\lambda_{k}^-+\chi_{k})}
	\tilde{p}^{(+)}(k,i\omega_n)\tilde{p}^{(+)}(-k,-i\omega_n)	
\label{onlytildep}
\end{align}
   which will be used to evaluate the SSCFs in the IC regime in the next section.
   In fact, it take a similar form as that for the C-magnons Eq.\ref{ppplusaction} after replacing
   $ \tilde{p}^{(+)}(k,i\omega_n) $ by $ p^{(+)}(k,i\omega_n) $. Indeed, taking the C-limit,
   it recovers Eq.\ref{ppplusaction}.

\subsection{ Effective actions at the phase boundary}

   By looking at the effective actions at the phase boundary, one can clearly distinguish the critical modes from the massive modes.

  At C$_0$ boundary, we have
\begin{align}
	\lambda_k^- & =\frac{1}{2}\sin^2\beta
	+\frac{1}{4}[\cos^2\beta(\delta k_x)^2+(1-6\sin^2\beta)(\delta k_y)^2]+\cdots    \\  \nonumber
	\chi_k & =\frac{1}{2}\sin^2\beta-\frac{1}{4}\sin^2\beta(\delta k_x)^2+\cdots
\end{align}
 which shows $ \lambda_k^- - \chi_k $ becomes gapless, while $ \lambda_k^- + \chi_k $ remains massive
\begin{align}
	H_\text{low}\approx J\sum_k[
		\frac{1}{2}((\delta k_x)^2+(1-6\sin^2\beta)(\delta k_y)^2)p_{k}^{(+)}p_{-k}^{(+)}
		+2\sin^2\beta S^2q_{k}^{(+)}q_{-k}^{(+)}]
\label{c0crit}
\end{align}
  which shows that $ p^{+} $ becomes the critical mode, while its conjugate variable $ q^{+} $ remains massive.
  Of course, as shown in Fig.1, it was pre-emptied by a 1st order transition anyway.

At C$_\pi$ boundary, we have
\begin{align}
	\lambda_k^- & =\frac{1}{2}\cos^2\beta
	+\frac{1}{4}[\cos^2\beta(\delta k_x)^2+(1-6\cos^2\beta)(\delta k_y)^2]+\cdots    \\  \nonumber
	\chi_k  & =\frac{1}{2}\cos^2\beta-\frac{1}{4}\cos^2\beta(\delta k_x)^2+\cdots
\end{align}
then
\begin{align}
	H_\text{low}\approx J\sum_k[
		\frac{1}{2}((\delta k_x)^2+(1-6\cos^2\beta)(\delta k_y)^2)p_{k}^{(+)}p_{-k}^{(+)}
		+2\cos^2\beta S^2q_{k}^{(+)}q_{-k}^{(+)}]
\label{cpicrit}
\end{align}
    which at the quadratic order,  can be obtained from Eq.\ref{c0crit} by the mirror transformation $ \beta \rightarrow \pi/2 - \beta $
    with the difference that C$_\pi$ magnons's minimum is at $ k^0_y=\pi $ ( Fig.\ref{threemag} ).
    It recovers the spurious Goldstone mode Eq.\ref{gapless} in Sec.II-A.

At the IC- boundary, we have $\alpha=\arcsin\frac{\sqrt{6}\sin2\beta}{\sqrt{9\sin^22\beta-1}}$
and $\sin\theta_0=\frac{6}{1+9\sin^22\beta}$, then
\begin{align}
    \lambda_k^- & =\frac{3\sin^22\beta-1}{2(9\sin^22\beta-1)}
	+\frac{9\sin^22\beta}{81\sin^42\beta-1}(\delta k_x)^2
	+\Big(\frac{1}{4}+\frac{9}{2+18\sin^22\beta}+\frac{4}{1-9\sin^2\beta}\Big)(\delta k_y)^2\cdots   \\  \nonumber
    \chi_k &  =\frac{3\sin^22\beta-1}{2(9\sin^22\beta-1)}
		-\frac{3\sin^22\beta-1}{4(9\sin^22\beta-1)}(\delta k_x)^2+\cdots
\end{align}
  which again shows $ \lambda_k^- - \chi_k $ becomes gapless, while $ \lambda_k^- + \chi_k $ remains massive,
\begin{align}
	H_\text{low}\approx & J\sum_k\Big\{
	4 \Big[
	\frac{1-12\sin^22\beta-27\sin^42\beta}{2(9\sin^22\beta-1)(9\sin^22\beta+7)}(\delta k_x)^2
	+\Big(\frac{1}{4}+\frac{9}{2+18\sin^22\beta}+\frac{4}{1-9\sin^2\beta}\Big)(\delta k_y)^2
	\Big]\tilde{p}_k\tilde{p}^{*}_{k}    \\  \nonumber
	& +S^2\frac{(3\sin^22\beta-1)(9\sin^22\beta+7)}{81\sin^42\beta-1}\tilde{q}_k\tilde{q}^{*}_{k}
	\Big\}
\label{ciccrit}
\end{align}
   which shows that $ \tilde{p}_k $ becomes the critical mode, while its conjugate variable $ \tilde{q}^{*}_k $ remains massive.

   In appendix F, we will push the actions beyond the boundary to study the nature of the transitions to
   commensurate or in-commensurate phases.

\section{ The spin-spin correlation functions in Y-x state via the low energy effective actions }

   We will use the effective low energy actions for the C- and IC-magnons in the Y-x phase in terms of
   the critical modes ( order parameters )
  to derive their  spin-spin correlation functions (SSCFs ) respectively which, as stressed in the main text,
  take very different forms in the two cases. They are also directly
  experimentally measurable physical quantities.

\subsection{The Commensurate magnons}

{\sl 1. Leading contributions }

   Eq.\ref{ppplusaction} leads to the only nonzero correlation function for the critical modes:
\begin{align}
	\langle p^{(+)}(k,i\omega_n)p^{(+)}(-k,-i\omega_m)\rangle
	=\frac{4JS^2(\lambda_k^-+\chi_k)}{(4JS\omega_k^-)^2+\omega_n^2}
\end{align}

   Because $ ( p^-,q^- ) $ are high energy modes which do not even appear in Eq.\ref{pqplusaction},
   so at the low energy sector \cite{takelimit},  imposing $p_k^{(-)}=0$,
   leads to $ p_k^A=p_k^B=p_k^{(+)}/\sqrt{2}$.
   Due to $p_i=S\delta\eta_i=\delta S_i^x$, then we conclude the relations between the
   quantum spin and the critical mode:
\begin{equation}
\delta S_A^x(k,i\omega_n)=\delta S_B^x(k,i\omega_n)=p^{(+)}(k,i\omega_n)/\sqrt{2}
\label{spinOP1}
\end{equation}
  which tells the two sublattices A and B play quite similar roles. Then we obtain
\begin{align}
	& \langle S_A^x(k,i\omega_n)S_A^x(-k,-i\omega_n)\rangle
	 =\langle S_B^x(k,i\omega_n)S_B^x(-k,-i\omega_n)\rangle   \\  \nonumber
	& =\langle S_A^x(k,i\omega_n)S_B^x(-k,-i\omega_n)\rangle
	=\langle S_B^x(k,i\omega_n)S_A^x(-k,-i\omega_n)\rangle   \\  \nonumber
	& =\frac{1}{2}\langle p^{(+)}(k,i\omega_n)p^{(+)}(-k,-i\omega_m)\rangle
\end{align}
    All the other SSCFs can be ignored at the leading order. For example, setting $ q^{+}=0 $ and $ q^{-}=0 $ lead to all the spin $ ZZ $ correlation functions vanish at the leading order. But they may still contribute at the sub-leading order to be discussed in the next subsection.

    Eq.\ref{ppplusaction} leads to the dynamic susceptibility function:
\begin{align}
	\chi_{pp}(k,i\omega_n)
	=\frac{4JS^2(\lambda_k^-+\chi_k)}{(4JS\omega_k^-)^2+\omega_n^2}
\end{align}
   Its analytical continuation lead to
\begin{align}
	\mathrm{Im}[\chi_{pp}(k,i\omega_n\to\omega+i0^+)]
	=\frac{S(\lambda_k^-+\chi_k)}{\omega_k^-}
	[-\pi\delta(4JS\omega_k^-+\omega)+\pi\delta(4JS\omega_k^--\omega)]
\end{align}
   thus the equal-time $pp $ correlation function is
\begin{align}
	S_{pp}(k)=\int\frac{d\omega}{2\pi}
	\frac{-2\mathrm{Im}[\chi_{pp}(k,\omega)]}{1-e^{-\omega/T}}
	=\Big(\frac{1}{2}-\frac{1}{e^{4JS\omega_k^-/T}-1}\Big)
	\frac{S(\lambda_k^-+\chi_k)}{\omega_k^-}
\end{align}
    we can conclude the equal-time SSCF ( structure factor ) are
\begin{align}
	S^{xx}_{AA}(k)=S^{xx}_{BB}(k)=S^{xx}_{AB}(k)
	=\Big(\frac{1}{2}-\frac{1}{e^{4JS\omega_k^-/T}-1}\Big)
	\frac{S(\lambda_k^-+\chi_k)}{\omega_k^-}
\label{XXstructure}
\end{align}
     At a very low temperature $ T \ll \Delta $, the second term in the braket can be dropped.
     The structure factor clearly peaks at $ (0,0) $ or $ (\pi,\pi) $ for the C$_0$ or C$_\pi$ respectively.
     The dominant SSCFs is only the XX component which can be contrasted with those XX, ZZ, XZ and ZX components
     corresponding to the IC- magnons listed in Eq.\ref{XXZZstructure}.

{\sl 2. Sub-Leading contributions: the hierarchy of energy scales }

  Because $ q^{+}=q^{-} =0 $, putting  $ \delta \xi_i =0 $ in Eq.\ref{spinYx} leads to
  the dominant quantum fluctuations in the $ XY $ plane:

\begin{align}
    \mathbf{S}_i=S(-\sin \delta \eta_i,  (-1)^{i_x}\cos \delta\eta_i,  0  ),
\label{spinYxdom}
\end{align}

  Setting Eq.\ref{spinOP1} $ p^+_i = \sqrt{2} S \delta \eta_i $ into Eq.\ref{ppplusaction}  leads to the effective action in terms of  $ \delta\eta_i $:
\begin{align}
	S_\text{eff}[  \delta \eta(k,i\omega_n)   ]=\frac{1}{2}\sum_{k,\omega_n}
	\frac{(4J\omega_k^-)^2+\omega_n^2}{2J(\lambda_k^-+\chi_k)} \delta \eta(k,i\omega_n) \delta \eta(-k,-i\omega_n)
\label{ppplusactioneta}
\end{align}
  where $ k $ is confined in the BZ: $  0 < k_x < \pi, -\pi < k_y  < \pi $.

   We can summarize the hierarchy of energy scales and the SSCFs as follows:

  (a) Eq.\ref{spinYxdom} and \ref{ppplusactioneta} can be used to evaluate the SSCFs in the $ XY $ plane beyond the linear approximation.
  Of course, its linear approximation reduces to the results in  Eq.\ref{XXstructure}.
  However, any SSCFs involving the $ Y $ component will be much smaller than the dominant $ XX$ SSCF ( or structure factor )
  in Eq.\ref{XXstructure}.

  When the spectrum becomes gapless such as the gapless phason mode in Eq.\ref{phason} as done in Sec.VI-C or $ U(1) $ Dicke model \cite{u1dicke}, one must use  this exponential form to remove the IR divergence due to this gapless mode.

  (b) In order to evaluate the SSCFs involving the $ Z $ component, then one need to use the Eq.\ref{spinYx}
      and Eq.\ref{pqplusaction} which still keeps the massive $ q^{+} $  mode, but
      setting $ q^{-}=0 $ and $ p^{-} =0$ \cite{takelimit}. Then  $  q^A=q^B=q^{+}/\sqrt{2} $
      and $ p^A=p^B=p^{+}/\sqrt{2} $ and also the definition $ q= \delta \xi, p = S \delta \eta $.
      They are even much smaller than those involving the $ Y $ component.

  (c) If putting back the higher branch $ \omega_k^+ $ in Eq.\ref{H2Yx}, then one need to use Eq.\ref{spinYx} and Eq.\ref{Yxpathk}
      to evaluate the complete SSCFs.
      However, it is hard to extract any physics from such a complete, but complicated expression.
      To explore the hierarchy of energy scales and SSCFs, one must perform the above  projection procedures step by step.

  Both (a) and (b) are still within the $ \omega_k^- $ manifold by taking $ \omega_k^+ \rightarrow \infty $ limit \cite{takelimit}.
  (c) is putting back $ \omega_k^+ $. All are listed below in the Table 1.

\vspace{0.25cm}
%\begin{table}
\begin{tabular}{ |c|c|c|c| }    \hline
 Leading Level   &   $ \omega_k^+ \rightarrow \infty  $:  $ p^-=0, q^{-}=0 $   & integrate out $ q^+$, then set $q^{+}=0 $ &  Linear Approximation Eq.\ref{spinOP1}    \\ \hline
 Level (a)   &   $ \omega_k^+ \rightarrow \infty  $:  $ p^-=0, q^{-}=0 $   & integrate out $ q^+$, then set $q^{+}=0 $ &  Exponential Eq.\ref{spinYxdom}  \\  \hline
 Level (b)  &    $ \omega_k^+ \rightarrow \infty  $:  $ p^-=0, q^{-}=0 $    & keep $ q^+$,  $q^{+} \neq 0 $ &  Exponential Eq.\ref{spinYx}      \\ \hline
 Level (c)  &   keep $ \omega_k^+ $,  $ p^-\neq 0, q^{-} \neq 0 $    &  $q^{+} \neq 0 $ &  Exponential Eq.\ref{spinYx}      \\ \hline
\end{tabular}
\par
\vspace{0.25cm}
{\footnotesize  {Table 1: The four  hierarchy of  energy scales in the C-magnon case. The linear approximation to (a)
in the spin operators leads to the leading contributions. Setting $ q^{+}=0 $ in (b) leads to (a).
Setting $ p^-\neq 0, q^{-} \neq 0 $ in (c) leads to (b) \cite{takelimit}.  } }
%\end{table}
\vspace{0.25cm}

\subsection{Incommensurate magnons}

{\sl 1. Leading contributions }

   Eq.\ref{onlytildep} leads to the only non-vanishing correlator for the critical field:
\begin{align}
	\langle\tilde{p}^{(+)}(k,i\omega_n)\tilde{p}^{(+)}(-k,-i\omega_n)\rangle
	&=\frac{2JS^2(1+\sin\theta_0)(\lambda_k^-+\chi_k)}{(4JS\omega_k^-)^2+\omega_n^2},~~~~~~ |k\pm Q|\ll\Lambda
\end{align}

Because $\tilde{q}_k^{(-)}, \tilde{p}_k^{(-)}$ are high energy modes which do not even appear in the low energy effective action Eq.\ref{tildepqplusact},
so at the low energy limit \cite{takelimit}, one can safely set $\tilde{q}_k^{(-)}=0, \tilde{p}_k^{(-)}=0$.
So setting $\tilde{q}_k^{(-)}=0 $ in  Eq.\eqref{eq:-pq} leads to:
\begin{align}
	\langle p^{(+)}(k,i\omega_n)p^{(+)}(-k,-i\omega_n)\rangle
	&=\langle\tilde{p}^{(+)}(k,i\omega_n)\tilde{p}^{(+)}(-k,-i\omega_n)\rangle,\\ \nonumber
	S^2\langle q^{(-)}(k,i\omega_n)q^{(-)}(-k,-i\omega_n)\rangle
	&=\frac{1-\sin\theta_0}{1+\sin\theta_0}
	\langle\tilde{p}^{(+)}(k,i\omega_n)\tilde{p}^{(+)}(-k,-i\omega_n)\rangle,\\  \nonumber
	S\langle p^{(+)}(k,i\omega_n)q^{(-)}(-k,-i\omega_n)\rangle
	&=\frac{-i\cos\theta_0}{1+\sin\theta_0}
	\langle\tilde{p}^{(+)}(k,i\omega_n)\tilde{p}^{(+)}(-k,-i\omega_n)\rangle,\\  \nonumber
	S\langle q^{(-)}(k,i\omega_n)p^{(+)}(-k,-i\omega_n)\rangle
	&=\frac{+i\cos\theta_0}{1+\sin\theta_0}
	\langle\tilde{p}^{(+)}(k,i\omega_n)\tilde{p}^{(+)}(-k,-i\omega_n)\rangle,
\end{align}
  where one can see the last two equation differs by a minus sign.

 Because the massive mode $\tilde{q}^{(+)}$ has been integrated out in Eq.\ref{onlytildep},
 so one can simply set $ \tilde{q}_k^{(+)}=0 $. Putting $\tilde{q}_k^{(+)}=0$  and $ \tilde{p}_k^{(-)}=0 $ in
 Eq.\ref{tildeset} lead to  $q_k^{(+)}=0$ and $p_k^{(-)}=0$.
 Then  $q_k^A=-q_k^B=q_k^{(-)}/\sqrt{2}$ and $p_k^A=p_k^B=p_k^{(+)}/\sqrt{2}$.
Due to $p_i=S\delta\eta_i=\delta S_i^x$ and $(-1)^{i_x}Sq_i=(-1)^{i_x}S\delta\xi_i=\delta S_i^z$, we find
the relation between the quantum spin and the critical fields:
\begin{eqnarray}
\delta S_A^x(k,i\omega_n) & = & \delta S_B^x(k,i\omega_n)=p^{(+)}(k,i\omega_n)/\sqrt{2}    \nonumber   \\
\delta S_A^z(k,i\omega_n) & = & \delta S_B^z(k,i\omega_n)=Sq^{(-)}(k,i\omega_n)/\sqrt{2}
\label{spinOP2}
\end{eqnarray}
which tells the two sublattices $ A $ and $ B $ play a similar role.

    So we obtain all the non-vanishing transverse ( relative to the Y-x state ) SSCFs for the sublattice $ A $:
\begin{align*}
	\langle S_A^x(k,i\omega_n)S_A^x(-k,-i\omega_n)\rangle
	&=\frac{1}{2}\langle\tilde{p}^{(+)}(k,i\omega_n)\tilde{p}^{(+)}(-k,-i\omega_n)\rangle\\  \nonumber
	\langle S_A^z(k,i\omega_n)S_A^z(-k,-i\omega_n)\rangle
	&=\frac{1-\sin\theta_0}{2(1+\sin\theta_0)}
	\langle\tilde{p}^{(+)}(k,i\omega_n)\tilde{p}^{(+)}(-k,-i\omega_n)\rangle\\  \nonumber
	\langle S_A^x(k,i\omega_n)S_A^z(-k,-i\omega_n)\rangle
	&=\frac{-i\cos\theta_0}{2(1+\sin\theta_0)}
	\langle\tilde{p}^{(+)}(k,i\omega_n)\tilde{p}^{(+)}(-k,-i\omega_n)\rangle\\  \nonumber
	\langle S_A^z(k,i\omega_n)S_A^x(-k,-i\omega_n)\rangle
	&=\frac{+i\cos\theta_0}{2(1+\sin\theta_0)}
	\langle\tilde{p}^{(+)}(k,i\omega_n)\tilde{p}^{(+)}(-k,-i\omega_n)\rangle
\label{connecttildepp}
\end{align*}
   where the $ XZ $ and $ ZX $ SSCF differ by a minus sign.

   Because the two sublattices $ A $ and $ B $ play a similar role, the other nonzero SSCFs follow as:
\begin{align*}
	\langle S_A^x(k,i\omega_n)S_A^x(-k,-i\omega_n)\rangle
	&=\langle S_B^x(k,i\omega_n)S_B^x(-k,-i\omega_n)\rangle
	=\langle S_A^x(k,i\omega_n)S_B^x(-k,-i\omega_n)\rangle
	=\langle S_B^x(k,i\omega_n)S_A^x(-k,-i\omega_n)\rangle\\
	\langle S_A^z(k,i\omega_n)S_A^z(-k,-i\omega_n)\rangle
	&=\langle S_B^z(k,i\omega_n)S_B^z(-k,-i\omega_n)\rangle
	=\langle S_A^z(k,i\omega_n)S_B^z(-k,-i\omega_n)\rangle
	= \langle S_B^z(k,i\omega_n)S_A^z(-k,-i\omega_n)\rangle\\
	\langle S_A^x(k,i\omega_n)S_A^z(-k,-i\omega_n)\rangle
	&=\langle S_B^x(k,i\omega_n)S_B^z(-k,-i\omega_n)\rangle
	=\langle S_A^x(k,i\omega_n)S_B^z(-k,-i\omega_n)\rangle
	=\langle S_B^x(k,i\omega_n)S_A^z(-k,-i\omega_n)\rangle\\
	\langle S_A^z(k,i\omega_n)S_A^x(-k,-i\omega_n)\rangle
	&=\langle S_B^z(k,i\omega_n)S_B^x(-k,-i\omega_n)\rangle
	=\langle S_A^z(k,i\omega_n)S_B^x(-k,-i\omega_n)\rangle
	=\langle S_B^z(k,i\omega_n)S_A^x(-k,-i\omega_n)\rangle
\end{align*}
   All other SSCFs can be ignored at this leading orders. However, they may still contribute to the subleading orders to be discussed in the
   next subsection.

Taking the susceptibility function
\begin{align}
	\chi_{\tilde{p}\tilde{p}}(k,i\omega_n)
	=\frac{2JS^2(1+\sin\theta_0)(\lambda_k^-+\chi_k)}{(4JS\omega_k^-)^2+\omega_n^2}
\end{align}
Its analytical continuation lead to
\begin{align}
	\mathrm{Im}[\chi_{\tilde{p}\tilde{p}}(k,i\omega_n\to\omega+i0^+)]
	=\frac{S(1+\sin\theta_0)(\lambda_k^-+\chi_k)}{2\omega_k^-}
	[-\pi\delta(4JS\omega_k^-+\omega)+\pi\delta(4JS\omega_k^--\omega)]
\end{align}
thus  the equal-time correlation function  is
\begin{align}
	S_{\tilde{p}\tilde{p}}(k)=\int\frac{d\omega}{2\pi}
	\frac{-2\mathrm{Im}[\chi_{\tilde{p}\tilde{p}}(k,\omega)]}{1-e^{-\omega/T}}
	=\Big(\frac{1}{2}-\frac{1}{e^{4JS\omega_k^-/T}-1}\Big)
	\frac{S(1+\sin\theta_0)(\lambda_k^-+\chi_k)}{2\omega_k^-}
\end{align}

 Eq.\ref{connecttildepp} leads to the equal-time SSCFs ( structure factor ):
\begin{eqnarray}
	S^{xx}_{AA}(k) &= & S^{xx}_{BB}(k)=S^{xx}_{AB}(k)
	=\Big(\frac{1}{2}-\frac{1}{e^{4JS\omega_k^-/T}-1}\Big)
	\frac{S(1+\sin\theta_0)(\lambda_k^-+\chi_k)}{4\omega_k^-}\\  \nonumber
	S^{zz}_{AA}(k) & = & S^{zz}_{BB}(k)=S^{zz}_{AB}(k)
	=\Big(\frac{1}{2}-\frac{1}{e^{4JS\omega_k^-/T}-1}\Big)
	\frac{S(1-\sin\theta_0)(\lambda_k^-+\chi_k)}{4\omega_k^-}\\  \nonumber
	S^{xz}_{AA}(k)& = & S^{xz}_{BB}(k)=S^{xz}_{AB}(k)
	=\Big(\frac{1}{2}-\frac{1}{e^{4JS\omega_k^-/T}-1}\Big)
	\frac{-iS\cos\theta_0(\lambda_k^-+\chi_k)}{4\omega_k^-}\\  \nonumber
	S^{zx}_{AA}(k) & = & S^{zx}_{BB}(k)=S^{zx}_{AB}(k)
	=\Big(\frac{1}{2}-\frac{1}{e^{4JS\omega_k^-/T}-1}\Big)
	\frac{+iS\cos\theta_0(\lambda_k^-+\chi_k)}{4\omega_k^-}
\label{XXZZstructure}
\end{eqnarray}
where $|k-Q|\ll\Lambda$ ( Fig.\ref{threemag} ). Again, the last two equations differ by a minus sign.

Because $ \cos\theta_k $ ( $ \sin \theta_k $ )  is an odd ( even ) function under  $ \vec{k} \rightarrow  - \vec{k} $,
the first two equations on the $ XX $ and $ZZ$ SSCF also hold for $|k + Q|\ll\Lambda$ ( Fig.\ref{threemag} ) with a ratio
 $ S^{zz}/S^{xx}= (1-\sin\theta_0)/(1+\sin\theta_0) $ which is tuned by the IC- angle $ \sin\theta_0 $.
% In the C- limit, $ \theta_0 \rightarrow \pi/2 $, the ratio gets to zero.
However, the last two exchange under $ \vec{k} \rightarrow  - \vec{k} $, namely,  $S^{xz}_{AA}(-Q + \delta k )=S^{zx}_{AA}( Q -\delta k)$.
 When approaching the C- boundary, $ \theta_0 \rightarrow \pi/2 $,
 $ S^{zz} $ and $ S^{zx}, S^{xz} $ all approach zero, only $ S^{xx} $ survives which recover the C limit in Eq.\ref{XXstructure}.

  At a very low temperature $ T \ll \Delta $, the second term in the braket can be dropped.
  The structure factor clearly peaks at $ (0, \pm k_0 ) $. So can be used to map out the minima contour of the IC- magnons.
  The dominant SSCFs contain the XX, ZZ, XZ and ZX components which can be contrasted with the only XX component
  corresponding to the C- magnons listed in Eq.\ref{XXstructure}.

{\sl 2. Sub-Leading contributions: the hierarchy of energy scales }

  We can summarize the hierarchy of energy scales and the SSCFs as follows:

  (a) By using  Eq.\ref{spinYx} and the action \ref{onlytildep}, one can evaluate the SSCFs in the $ XYZ $ beyond the
  linear approximation as follows:
  Using  $q_k^A=-q_k^B=q_k^{(-)}/\sqrt{2}, p_k^A=p_k^B=p_k^{(+)}/\sqrt{2}$ and
  $ p_k^{(+)}=\tilde{p}_k^{(+)}, q_k^{(-)}= \frac{i\cos\theta_k}{S(1+\sin\theta_k )}\tilde{p}_k^{(+)} $.
  Of course, its linear approximation reduces to the explicit results in  Eq.\ref{XXZZstructure}.
  However, any SSCFs involving the longitudinal $ Y $ component will be much smaller than the dominant
  $ XX, ZZ $ or $ XZ $ SSCF ( or structure factor ) in Eq.\ref{XXZZstructure}.

  (b) By using  Eq.\ref{spinYx} and the action Eq. \ref{tildepq} which still keeps $ \tilde{q}^{+} $,
   one can also evaluate the SSCFs in the $ XYZ $
   with a better accuracy than in (a) by setting $ \tilde{q}^{-}=0 $ and $ \tilde{p}^{-} =0$ .
   Then  all the angle variables $ \delta \xi_A,  $ and  $ \delta \eta_A,  $ can be expressed in terms of
  $ \tilde{p}_k^{(+)}, \tilde{q}_k^{(+)} $ appearing in the effective action Eq.\ref{tildepq} as follows:
  $ p_k^{(+)}=\tilde{p}_k^{(+)}, q_k^{(-)}= \frac{i\cos\theta_k}{S(1+\sin\theta_k )}\tilde{p}_k^{(+)} $
  and  $ q_k^{(+)}= \frac{1+ \sin \theta_k }{2} \tilde{q}_k^{(+)},
  p_k^{(-)}= -\frac{iS (1-\cos\theta_k)(1+ \sin\theta_k ) }{ 2 \cos \theta_k }\tilde{q}_k^{(+)} $.
  The practical use of this procedure is limited. Of course, setting  $ \tilde{q}_k^{(+)} =0 $ recovers (a).

  (c) If putting back the higher branch $ \omega_k^+ $ as in Eq.\ref{H2Yx}, then one need to use Eq.\ref{spinYx} and Eq.\ref{Yxpathk}
      to evaluate the complete SSCFs.
      However, it is hard to extract any physics from such a complete, but complicated expression.
      To explore the hierarchy of energy scales and SSCFs, one must perform the above  projection procedures step by step.

   Similar to the C-magnon case in the last subsection,
   both (a) and (b) are still within the $ \omega_k^- $ manifold by taking $ \omega_k^+ \rightarrow \infty $ limit \cite{takelimit}.
   (c) is putting back $ \omega_k^+ $. A counter-part of Table 1 for the IC-magnons can also be made.

\section{ The quantum Lifshitz action from the Y-x phase on the right to the IC-SkX phase  }

 Inside the Y-x phase, it is enough to use the effective actions Eq.\ref{ppplusaction} and Eq.\ref{onlytildep}
 at the Gaussian level to evaluate the SSCFs  with C- and IC- magnons respectively as did in the last section.
 However, when getting close to the magnon condensation boundary, it is important to consider the higher
 order terms represented by the $ \cdots $ in Eq.\ref{pathhigh}.
 The physical picture is presented in Fig.\ref{phasesarc}.

\subsection{ The transition driven by the condensation of  C$_{\pi}$ and  C$_{0}$  magnons  }

  Note that in the Sec.IV, we construct an effective action to study the quantum Lifshitz transition from the Y-x phase to the IC-XY-y phase
  along the diagonal line in the polar coordinate $ (\theta, \phi) $ using the  $ Z $ as the spin quantization axis.
  In the appendix C and D, we did the coherent state
  path integral in the polar coordinate $ (\eta, \xi) $ using the  $ X $ as the spin quantization axis.
  Both $X$ and $ Z $  spin quantization axis can be used to characterize the spin-orbital order in the XY plane.
  In fact after making the shift $ \phi \rightarrow \phi + \pi/2 $ mentioned above Eq.\ref{2times2order}, $ \phi $ and $ \eta $
  stand for the same angle.
  Indeed,  by comparing $ \delta S_i^x = S\delta\eta_i $  with Eq.\ref{phiofqd},  we can identify:
\begin{equation}
  \delta \eta_i = (-1)^{i_y}  \phi_i
\label{etaphi}
\end{equation}
   which  $ k= \pi + q $ takes care of the C$_\pi$ magnons's minimum at $ k^0_y=\pi $.
    It also corresponds to folding or un-folding between
   the 2-sublattices in the $ (\eta, \xi) $ and the 4-sublattices in the $ (\theta, \phi) $ polar coordinate.

   When plugging it into Eq.\ref{spinYxdom} leads to the spin in terms of the order parameter $ \phi $:
\begin{align}
    \mathbf{S}_i=S( -(-1)^{i_y} \sin \phi_i , (-1)^{i_x}\cos \phi_i ,  0  ),
\label{spinYxdomphi}
\end{align}
  which is identical to Eq.\ref{2times2order} in the main text achieved from
  the diagonal line $ \alpha=\beta $ by an order from quantum disorder analysis (OFQD).
  So we achieve the same result from the Y-x state on the right $ \beta < \alpha $.

  One can systematically expands Eq.\ref{pathhigh} to the cubic, quartic and higher orders.
  Following the procedures leading to Eq.\ref{ppplusactioneta}, one can express all these higher order terms
  in term of the complete sets $  ( p^{\pm}, q^{\pm} ) $. The $ \pm $ sets are decoupled at the quadratic orders, but are are coupled
  by higher order terms.
  When projecting out the higher energy branch  $ \omega_k^+ $,
  one can simply set $  p^{-}=0, q^{-}=0 $ in all these terms \cite{fermionout}, then integrating out the massive   $ q^{+}  $ mode to reach
  an effective action solely in terms of $ p^{+} $.
  From the symmetry point of view, one should expect a cubic term in  $ p^{+} $. By using the identification
  $ p^+_i = \sqrt{2} S \delta \eta_i $ and Eq.\ref{etaphi} which absorbs
  the C$_\pi$ magnons's minimum at $ k^0_y=\pi $, one reach the effective action driven by
  the condensation of the C$_{\pi}$ magnons ( See Fig.\ref{phasesarc} ):
\begin{align}
	S_{Y-x,C_{\pi} } [  \phi ]=\frac{1}{2}\sum_{k,\omega_n}
	[ \omega_n^2 + \Delta^2 + v^2_x q^2_x + v^2_y q^2_y  ] \phi(q,i\omega_n) \phi(-q,-i\omega_n) + \lambda \phi^3 + \kappa \phi^4 + \cdots
\label{ppphi3}
\end{align}
   where  $ 0 < q_x < \pi, 0 < q_y < \pi $  is within the RBZ due to the 4-sublattices in the $ (\theta, \phi) $ polar coordinate,
   in Eq.\ref{spinYxdomphi} and also listed below Eq.\ref{fournk}.
   The cubic term leads to a 1st order transition at $ \Delta^2_0= \lambda^2/2 \kappa > 0 $ which happens before
   the putative 2nd order transition $ \Delta^2=0 $.

   As shown in Sec.IV, along the diagonal line $ \alpha= \beta $, due to the  $ [C_4 \times C_4]_D $ symmetry at $ \alpha= \beta $, the cubic term vanishes  $ \lambda=0 $, but the gap remains $ \Delta > 0 $, so the system remains in the Y-x phase.
   So we expect $ \lambda $ changes sign along $ \alpha= \beta $. Setting  $ \lambda=0 $ along
   $ \alpha= \beta $ in Eq.\ref{ppphi3} recovers Eq.\ref{Yxp} in the main text with $ q_{ic}=0 $.
   The gap $ \Delta $ and the quartic term $ \kappa $ has been evaluated by the OFQD analysis
   in the paragraphs before Eq.\ref{Yxp}.

   The transition driven by C$_0$ magnons is given by Eq.\ref{eta33} where $ k $ is within the BZ.
   It is pre-emptied by the $ W=1/2 $ piece of the in-complete devil staircase.

\subsection{ The quantum Lifshitz transition from the Y-x to the IC-SkX-y driven by the condensation of IC-magnons }

  Plugging $ p^{+}= \sqrt{2} S \delta \eta =\tilde{p}^{+} $ into Eq.\ref{onlytildep} leads to
\begin{align}
	S_\text{eff}[\tilde{p}]=\frac{1}{2}\sum_{\delta k,\omega_n}
	\frac{(4JS\omega_{k}^-)^2+\omega_n^2}
		{J(1+\sin\theta_0)(\lambda_{k}^-+\chi_{k})}
	\delta \eta(k,i\omega_n)  \delta \eta (-k,-i\omega_n)	
\label{onlytildeptoeta}
\end{align}
where $ k $ is within the BZ.

  Again, one can systematically expand Eq.\ref{pathhigh} to the cubic, quartic and higher orders.
  Following the procedures leading to Eq.\ref{onlytildeptoeta}, one can express all these higher order terms
  in term of the twisted complete sets $  ( \tilde{p}^{\pm}, \tilde{q}^{\pm} ) $.
  The $ \pm $ sets are decoupled at the quadratic orders, but are are coupled
  by higher order terms.
  When projecting out the higher energy branch  $ \omega_k^+ $,
  one can simply set $  \tilde{p}^{-}=0, \tilde{q}^{-}=0 $ in all these terms \cite{fermionout}, then integrating out the massive   $ \tilde{q}^{+}  $ mode to reach
  an effective action solely in terms of $ \tilde{p}^{+} $.
  From the symmetry point of view, one should expect a cubic term in  $ \tilde{p}^{+} $. By using the identification
  $ \tilde{p}^{+}_i= p^+_i = \sqrt{2} S \delta \eta_i $, one reach the effective action Eq.\ref{etaeta3} driven by
  the condensation of the IC-magnons ( dropping $ \delta $ for the notational convenience ):
\begin{align}
	S[  \phi ]_{Y-x,IC}=\frac{1}{2}\sum_{k,\omega_n}
	[ \omega_n^2 + \Delta^2 + v^2_x k^2_x + v^2_y (k^2_y-Q^2)^2  ] \eta(k,i\omega_n) \eta(-k,-i\omega_n) + \lambda \eta^3 + \kappa \eta^4 + \cdots
\end{align}
  which was taken as Eq.\ref{etaeta3} in the main text and shown in Fig.\ref{phasesarc}.
  The spin is expressed  in terms of the order parameter in Eq.\ref{spinOP2}.

   As mentioned below Eq.\ref{ppphi3}, approaching to the diagonal line, $ \lambda \rightarrow 0 $, then using the identification
   Eq.\ref{etaphi}, one recovers  Eq.\ref{Yxp} in the main text.
\begin{equation}
{\cal L}[ \phi ]_{Y-x,D}  =  \phi(- \omega_n, -q_x, -q_y)
[\omega^2_{n}/A +  v^2_x q^2_x + u^2 ( q^2_y - q^2_{ic} )^2
  +   \Delta ] \phi( \omega_n, q_x, q_y) + \kappa \phi^4 + \cdots     \nonumber
\end{equation}
 where $ -\pi/2 < q_x, q_y < \pi/2 $ is in the Reduced Brillouin Zone (RBZ).
   The non-coplanar IC-SkX-y phase reduces to the co-planar IC-XY-y phase
   along the diagonal line ( See Fig.\ref{phasesarc} ).
   It is remarkable that one reach consistent results from the right and along the diagonal line.
   More detailed discussions are given in Sec.V-B.

   As said in Sec. VIII-C, it can be contrasted to the IC-SkX at $ h_{c1}  < h < h_{c2} $ discovered in \cite{rhh}
   where the $ U(1)_{soc} $ dictates $ S^x $ and $ S^{z} $ must have the same maximum magnitude and an associated
   Goldstone mode $ \phi $.  It will also be discussed in appendix G.
   As shown in Fig.1, there is a direct transition from the Y-x to the $ 3 \times 3 $ SkX along the counter $  k^{0}_{y}=\pi-\pi/3  $
   at the M point.
   Despite Eq.\ref{etaeta3} contains the two ordering wavevectors $ (0, \pm 2 \pi/3 ) $ of the  $ 3 \times 3 $ SkX phase,
   it remains a puzzle to understand the  $ 3 \times 3 $ SkX and  many other phases around the M point.

\section{The effective action of the  RFHM in a weak longitudinal field  $ h < h_{c1} $, complex order parameter
 and the role of the $ U(1)_{soc} $ symmetry }

 In the Y-x state which remains the exact ground state when $ h_{c1} < h < h_{c2} $,
 projecting out the higher branch $ \omega_{h,k}^+ $ in \cite{rhh}, the low-energy effective theory in
 the canonical quantization  takes the form
\begin{align}
	H_\text{low,h}=4JS\sum_k \omega_{h,k}^-\beta_k^\dagger \beta_k
\label{rhhlow}
\end{align}
which can be written in terms of original HP boson
\begin{align}
	H_\text{low}=4JS\sum_k
	    \begin{pmatrix}
		    a_k^\dagger &b_k^\dagger
	    \end{pmatrix}
	    \begin{pmatrix}
		    \omega_{h,k}^-(1+\cos\theta_{h,k}) &\omega_{h,k}^-\sin\theta_{h,k}\\
		    \omega_{h,k}^-\sin\theta_{h,k}	   &\omega_{h,k}^-(1-\cos\theta_{h,k})\\
	    \end{pmatrix}
	    \begin{pmatrix}
		    a_k\\
		    b_k\\
	    \end{pmatrix}
\label{earlyab}
\end{align}
In fact, one can reach these results from previous section by identifying
\begin{align}
	\theta_k\to\theta_{h,k},\quad
	\lambda_k^-\to\omega_{h,k}^-,\quad
	\chi_k\to0\>.
\label{chik0}
\end{align}
where
\begin{align}
	\omega_{h,k}^-&=1-\frac{1}{2}\cos2\beta\cos k_y
			-\frac{1}{2}\sqrt{\cos^2k_x+(\sin2\beta\sin k_y-h)^2}  \\  \nonumber
	\sin\theta_{h,k}&=\frac{\cos k_x}{\sqrt{\cos^2k_x+(\sin2\beta\sin k_y-h)^2}}   \\   \nonumber
	\cos\theta_{h,k}&=\frac{\sin2\beta\sin k_y-h}{\sqrt{\cos^2k_x+(\sin2\beta\sin k_y-h)^2}}
\end{align}
  In contrast to the $ h=0 $ case at any $ (\alpha, \beta ) $ in Eq.,  $ \sin\theta_{h,k} $ is not even,
  $ \cos\theta_{h,k} $   is not odd under $ \vec{k} \rightarrow - \vec{k} $ anymore.

 Now we shift the gear from the canonical quantization to the coherent path integral.
 If one use the same parametrization as Eq.\ref{spinYZ} in the polar coordinate along the X-axis:
\begin{align}
	S_i=S(\cos\eta_i,\sin\xi_i\sin\eta_i,\cos\xi_i\sin\eta_i)
\end{align}
then
\begin{align}
	p_i=S\delta\eta_i,\quad q_i=\delta\xi_i
\end{align}
and one still introduce new variables as
\begin{align}
    a_k=(p_{-k}^A+iSq_{-k}^A)/\sqrt{2S}, \quad b_k=(p_{-k}^B+iSq_{-k}^B)/\sqrt{2S}
\end{align}
then it will transform the low energy effective Hamiltonian as
\begin{align}
    H_\text{low}=J\sum_k\omega_{h,k}^-\{
		&[S^2(1-\sin\theta_{h,k})q_{k}^{(-)}q_{-k}^{(-)}
		+(1+\sin\theta_{h,k})p_{k}^{(+)}p_{-k}^{(+)}
		-iS\cos\theta_{h,k}(q_{k}^{(-)}p_{-k}^{(+)}-p_{k}^{(+)}q_{-k}^{(-)})]   \\  \nonumber
		+&[S^2(1+\sin\theta_{h,k})q_{k}^{(+)}q_{-k}^{(+)}
		+(1-\sin\theta_{h,k})p_{k}^{(-)}p_{-k}^{(-)}
		-iS\cos\theta_{h,k}(q_{k}^{(+)}p_{-k}^{(-)}-p_{k}^{(-)}q_{-k}^{(+)})]   \\  \nonumber
        +&i[S(1+\sin\theta_{h,k})p_{k}^{(+)}q_{-k}^{(+)}
        -S(1-\sin\theta_{h,k})q_{k}^{(-)}p_{-k}^{(-)}
        -i\cos\theta_{h,k}(p_{k}^{(+)}p_{-k}^{(-)}+S^2q_{k}^{(-)}q_{-k}^{(+)})]  \\  \nonumber
        +&i[S(1-\sin\theta_{h,k})p_{k}^{(-)}q_{-k}^{(-)}
        -S(1+\sin\theta_{h,k})q_{k}^{(+)}p_{-k}^{(+)}
        -i\cos\theta_{h,k}(p_{k}^{(-)}p_{-k}^{(+)}+S^2q_{k}^{(+)}q_{-k}^{(-)})]
		\}
\label{eq:RFHlow}
\end{align}
 where the first two lines formally become similar to Eq.\ref{hpqcomplex} after using the mapping in Eq.\ref{chik0}.
 The last two lines, in the $h\to0$ limit, cancel each other due to the fact that
 $\sin\theta_{0,k}=\sin\theta_{0,-k}, \cos\theta_{0,k}=-\cos\theta_{0,-k}$ and $\omega_{h,k}=\omega_{h,-k}$.

 In IC- regimes, $\theta_{h,k}$  takes the value $\theta_0$ at the only minimum $ k^0_y $ \cite{rhh}, thus $H_\text{low}$ can be approximated as
\begin{equation}
   H_\text{low}=J\sum_k \omega_{h,k}^-
	\Big[\frac{4}{1+\sin\theta_0}\tilde{p}_k\tilde{p}_{k}^*
	+S^2(1+\sin\theta_0)\tilde{q}_k\tilde{q}_{k}^*
    +i2S(\tilde{p}_k\tilde{q}_{k}^*-\tilde{q}_k\tilde{p}_{k}^*)\Big]
\label{allu1}
\end{equation}
where
\begin{align}
	\tilde{p}_k=[-iS\cos\theta_0q_k^{(-)}+(1+\sin\theta_0)p_k^{(+)}]/2,\quad
	\tilde{q}_k=q_k^{(+)}+\frac{i(1-\sin\theta_0)}{S\cos\theta_0}p_k^{(-)}
\end{align}
  where one can see the  crucial difference than Eq.\ref{hpqcomplex}
  is that due to $ \chi_k =0 $, both $ \tilde{p}_k $ and  $ \tilde{q}_k $ become critical at the same time,
  so must be treated at the same footings. This is due to the $ U(1)_{soc} $ symmetry relating $ \tilde{p}_k $ to $ \tilde{q}_k $.
  Another crucial difference is that  both $ \tilde{p}_k $ and  $ \tilde{q}_k $  are complex instead of being real,
  namely $ \tilde{p}^{*}_k \neq  \tilde{p}_{-k} $ and  $ \tilde{q}^{*}_k \neq  \tilde{q}_{-k} $.
  Therefore although only half appear in  Eq.\ref{hpqcomplex}, all the degree of freedoms appear in Eq.\ref{allu1}.

  In order to see the roles of the $ U(1)_{soc} $ and treat $ \tilde{p}_k $ and  $ \tilde{q}_k $ at the same footings, it is convenient to
  rescale away the prefactors $ 1+ \sin \theta_0 $ in  Eq.\ref{allu1}, namely,  $\tilde{p}_k$ by
  $\sqrt{2}/(\sin\theta_0/2+\cos\theta_0/2) =1/\cos(\theta_0/2-\pi/4) $
  and $\tilde{q}_k$ by $(\sin\theta_0/2+\cos\theta_0/2)/\sqrt{2}= \cos(\theta_0/2-\pi/4)$, which still keeps the commutation relation.
  For notational simplicity,  we still keep the same notation and re-write Eq.\ref{allu1} as:
\begin{align}
    H_\text{low}=2J\sum_k \omega_{h,k}^-
	\Big[\tilde{p}_k\tilde{p}_{k}^*
	+S^2\tilde{q}_k\tilde{q}_{k}^*
    +iS(\tilde{p}_k\tilde{q}_{k}^*-\tilde{q}_k\tilde{p}_{k}^*)\Big]
    =2J\sum_k \omega_{h,k}^-(\tilde{p}_k-iS\tilde{q}_k)(\tilde{p}_k^*+iS\tilde{q}_k^*)
\end{align}
where the rescaled $ \tilde{p}_k $ and  $ S\tilde{q}_k $ are more symmetrically written as \cite{alsouse}:

\begin{align}
	\tilde{p}_k=\cos(\tilde{\theta}_0/2) p_{k}^{(+)}+i \sin(\tilde{\theta}_0/2) Sq_{k}^{(-)},\quad
	S\tilde{q}_k=\cos(\tilde{\theta}_0/2) Sq_{k}^{(+)}- i \sin(\tilde{\theta}_0/2) p_{k}^{(-)}, \quad \tilde{\theta}_0=\theta_0-\pi/2
\label{piq}
\end{align}
 which still satisfy $ [\tilde{q}_k,\tilde{p}_k^*]=\cos^2(\tilde{\theta}_0/2)[q_{k}^{(+)},
  p_{-k}^{(+)}]-\sin^2(\tilde{\theta}_0/2)[p_{k}^{(-)},q_{-k}^{(-)}]=i $.

  In fact, from the very early relation Eq.\ref{earlyab}, one can find
\begin{align}
        \beta_k=\cos(\theta_{h,k}/2) a_k+\sin(\theta_{h,k}/2) b_k =-(\tilde{p}_k^*+iS\tilde{q}_k^*)/\sqrt{2S}
\end{align}
 which is clearly well defined and  also satisfy $ [ \beta_k, \beta^{\dagger}_k]=1 $. In fact, it resembles
 $ \sqrt{\omega_{k}^{-}}\beta_k $ for IC-magnons listed below Eq.\ref{fakebeta}.

 Inserting this relation \cite{shiftstar} to Eq.\ref{piq} leads back to the original Eq.\ref{rhhlow}.
 The critical field is nothing but the original $ \beta_k $ itself.
 This is in sharp contrast to Eq.\ref{betak} where the $ \beta_k $ is not the critical field, only $ \tilde{p}_k, S\tilde{q}_k $ are.
 From the Eq.\ref{HPab}, one can immediately
 see that it indeed transforms under the  $ U(1)_{soc}(\phi)=e^{ i \phi \sum_{i} (-1)^x S^{y}_i } $  symmetry as $ \beta_k \rightarrow \beta_k e^{i \phi} $ which in turn, immediately leads to the effective action:
\begin{equation}
   {\cal L}[\beta]= ( i \omega_n +  \Delta + \frac{ k^2_x}{2m_x}
   +\frac{ ( k_y-k^0_y)^2 }{2m_y} ) \beta( k, i \omega_n) \beta^{\dagger}( k, i \omega_n)
   + ( \beta^{\dagger} \beta )^2 + \cdots
\label{actionzeeman}
\end{equation}
  where  is formally the same as the zero-density superfluid to Mott transition
  with the dynamic exponent $ z=2 $, with the crucial difference of the IC- momentum at $ k^0_y $.
The crucial role of the $ k^0_y $, especially the mirror symmetry on the complete effective action Eq.\ref{actionzeeman} will be studied in
\cite{rhhrg}. From which we will study the IC-SkX phase Eq.\ref{icskxh}  at $ h_{c1} < h < h_{c2} $ which
breaks the $ U(1)_{soc} $ symmetry and leads to the peculiar Goldstone mode in Eq.\ref{minusGold} reached in \cite{rh}.

  The purpose of the whole cycle is really not necessary to explore Eq.\ref{rhhlow}, but very useful to clarify
  the crucial role of the  $ U(1)_{soc} $ symmetry which holds
  in the longitudinal Zeeman field, but absent in the SOC case in appendix A-F.  The two cases are dramatically different.
  Explicitly, despite Eq.\ref{effYx} and Eq.\ref{rhhlow} take formally the same form, they lead to completely different physics.

  Overall, the magnon condensations, especially in the presence of SOC, is a new class of problems which maybe
  quite different than the BEC studied in \cite{pifluxgold,pifluxqsl,SFnon,SFQAH}.

\end{widetext}


\begin{thebibliography}{99}
\bibitem{firstorder} The 1st order  transition line ( the solid line )
 connecting the $ M $ point to $ (\pi/2,0) $ of the boundary of the Y-x phase in Fig.1
 matches very precisely the line achieved from the previous works \cite{classdm1,classdm2} using classical Monte-Carlo simulations. However, the previous two works \cite{classdm1,classdm2} are purely numerical, so no physical insights on the global phase diagram Fig.1.
    In fact, as clarified in Sec.VI-D,  the solid line is just $ W=1/2 $ segment of the whole in-complete staircase instead of a 1st order transition line.
    In any numerical calculation on a finite size, it appears as a 1st order transition line.
\bibitem{spinICXYy} Just from the momenta where the magnon condenses, the state has the 4 orbital ordering wave-vectors $ ( 0, \pm ( \pi-q^{0}_y ) ) $ and $ ( \pi, \pm ( \pi-q^{0}_y ) ) $. However, it remains challenging to determine the associated spin structure. The most general spin structures for the three spin components  can be written as:
 $ S_{\alpha}= Re [ ( \psi_{1 \alpha} + \psi_{2 \alpha}(-1)^{x} ) e^{i (\pi-q^{0}_y ) y } ] $
 where $ \alpha=x,y,z $.   We used the effective action  Eq.\ref{Yxp} in Sec.IV, and  the effective action
  Eq.\ref{onlytildeptoeta} in appendix F to determine
 them to be the co-planar IC-XY-y and the non-planar IC-SkX-y state respectively which leads to the sepcific
 values of $ \psi_{1 \alpha} $ and $ \psi_{2 \alpha} $.
\bibitem{firstwinding} The physical meaning of this expresion is clear:
The first factor gives the winding number when $ x \rightarrow x + N $, the second factor is just periodic $ \tilde{\xi}(x+N)=\tilde{\xi}(x) $.
\bibitem{takelimit} It corresponds to take the $ \omega_k^+ \rightarrow \infty $ limit.
Similarly, in a QH system, one usually take the LLL projections by sending the Landau level spacing $ \hbar \omega_c \rightarrow \infty $,
then one may ignore the Landua level mixing, the system at $ \nu=1/2 $ has an approximate particle-hole symmetries \cite{hlr,zwang3}.
Putting back $ \omega_k^+ $ corresponds to the Landau level mixing.
\bibitem{differpm}  Note that this $ \pm $ sets stand for the uniform and staggered components, so
completely different from the $ \omega^{\pm}_k $ which stand for the high and low energy parts.
\bibitem{scalefreedom} There are always some scale degree of freedoms in choosing the two conjugate pairs to satisfy
the commutation relations, but any physical quantities such as SSCFs are independent of  the scale factors.
\bibitem{nonanalytic}
 It is important to observe that $ \omega^{-}_{k} $ in  Eq.\ref{effYx}, is non-analytic,
 therefore, as stressed below Eq.\ref{betak}, $ \beta_k $ may  not be a suitable field to construct an effective action.
 While the effective action Eq.\ref{ppplusaction}  for  C- and Eq.\ref{onlytildep}  for IC-magnons, the propagator becomes analytic.
 This kind of strategy to construct an analytic effective action to transform away the non-analyticity in an propagator
 also happens in the quantum rotor spin glass studied in \cite{glass1,glass2}.
\bibitem{dvm} In the dual vortex method to study the transitions from superfluid to various lattice symmetry breaking Mott insulating states
  in \cite{pq1,dual2,dual3}, in order to characterize the symmetry breaking patterns of the bosons of the Mott states
  in the direct lattice in terms of the vortex degree of freedoms in the dual lattice,
  one still need to identify the vortex fields in terms of the  eigen-modes ( order parameters ) of the dual vortex effective action
  as done in \cite{dual2,dual3} similar to Eq.\ref{betak} here.
  So the way to identity the patterns developed in the original paper \cite{pq1} which avoided such a procedure was
  shown to be in-sufficient in \cite{dual2,dual3}.
\bibitem{alsouse} One can also use this more symmetry set of definitions in all the previous appendices.
As stressed in \cite{scalefreedom}, the scale factor will not affect any physical quantities.
\bibitem{shiftstar} If one like to express $\beta$ in terms of $\tilde{p},\tilde{q}$ instead of $\tilde{p}^*,\tilde{q}^*$,
then one only need to switch the definitions between $\tilde{p},\tilde{q}$ and $\tilde{p}^*,\tilde{q}^*$.
\end{thebibliography}
\end{document}